\documentclass[10pt,journal,compsoc]{IEEEtran}

\usepackage{graphicx}
\usepackage{caption}
\usepackage{url}
\usepackage{amsmath,amsthm,amssymb}
\usepackage{mathtools}
\usepackage{mathrsfs}
\usepackage{bm}
\usepackage{acronym}
\usepackage{balance}
\usepackage{color}
\usepackage{float}
\usepackage{dblfloatfix}
\usepackage{subcaption}
\usepackage{booktabs, multicol, multirow}
\usepackage[utf8]{inputenc}
\usepackage[ruled,linesnumbered]{algorithm2e}
\usepackage{tabulary}
\usepackage{colortbl}
\usepackage{booktabs, cellspace, hhline}
\usepackage{cite}

\newcommand{\beq}{\begin{equation}}
\newcommand{\eeq}{\end{equation}}
\newcommand{\beqa}{\begin{eqnarray}}
\newcommand{\eeqa}{\end{eqnarray}}

\definecolor{newgreen}{rgb}{0.133,0.545,0.133}

\newcommand{\LineSep}[1][-0.2]{%
	\par\vspace*{\dimexpr-\baselineskip+7mm}%
}

\makeatletter
\makeatletter
\def\ps@IEEEtitlepagestyle{%
	\def\@oddfoot{\mycopyrightnotice}%
	\def\@oddhead{\hbox{}\@IEEEheaderstyle\leftmark\hfil\thepage}\relax
	\def\@evenhead{\@IEEEheaderstyle\thepage\hfil\leftmark\hbox{}}\relax
	\def\@evenfoot{}%
}
\def\mycopyrightnotice{%
	\begin{minipage}{\textwidth}
		\centering \scriptsize
		Copyright~\copyright~2021 IEEE. Personal use of this material is permitted. Permission from IEEE must be obtained for all other uses, in any current or future media, including reprinting/republishing this material for advertising or promotional purposes, creating new collective works, for resale or redistribution to servers or lists, or reuse of any copyrighted component of this work in other works by sending a request to pubs-permissions@ieee.org.
	\end{minipage}
}
\makeatother

\begin{document}


\title{Performability of Network Service Chains: \\ Stochastic Modeling and Assessment of \\ Softwarized IP Multimedia Subsystem}

\author{Mario~Di~Mauro,~\IEEEmembership{Member,~IEEE,}
	Giovanni~Galatro,
        Fabio~Postiglione, Marco Tambasco
\IEEEcompsocitemizethanks{\IEEEcompsocthanksitem M. Di Mauro, G. Galatro, F. Postiglione are with the Department of Information and Electrical Engineering and Applied Mathematics (DIEM), University of Salerno,  Italy. E-mails: \{mdimauro,fpostiglione\}@unisa.it, g.galatro1@studenti.unisa.it
	{\IEEEcompsocthanksitem M. Tambasco is with Ericsson Telecommunications Italy. E-mail: marco.tambasco@ericsson.it}
\protect\\

}
}

\IEEEtitleabstractindextext{%
\begin{abstract}
Service provisioning mechanisms implemented across $5$G infrastructures take broadly into use the network service chain concept. Typically, it is coupled with Network Function Virtualization (NFV) paradigm, and consists in defining a pre-determined path traversed by a set of softwarized network nodes to provide specific services. A well known chain-like framework is the IP Multimedia Subsystem (IMS), a key infrastructure of $5$G networks, that we characterize both by a performance and an availability perspective. Precisely, supported by a designed from scratch testbed realized through \textit{Clearwater} platform, we perform a stochastic assessment of a softwarized IMS (softIMS) architecture where two main stages stand out: $i)$ a performance analysis, where, exploiting the queueing network decomposition method, we formalize an optimization problem of resource allocation by modeling each softIMS node as an $M/G/c$ system; $ii)$ an availability assessment, where, adopting the Stochastic Reward Net methodology, we are able to characterize the behavior of softIMS in terms of failure/repair events, and to derive a set of optimal configurations satisfying a given availability requirement (e.g. five nines) while minimizing deployment costs. Two routines dubbed \textit{OptCNT} and \textit{OptSearchChain} have been devised to govern the performance and availability analyses, respectively.  
\end{abstract}

\begin{IEEEkeywords}
IP Multimedia Subsystem, Performance Analysis, Availability Analysis, Stochastic Reward Networks, Redundancy Optimization.
\end{IEEEkeywords}}

\maketitle

\IEEEdisplaynontitleabstractindextext

\IEEEpeerreviewmaketitle

\section{Introduction}\label{sec:intro}

\begin{figure}[t!]
	\centering
	\captionsetup{justification=centering}
	\includegraphics[scale=0.32,angle=90]{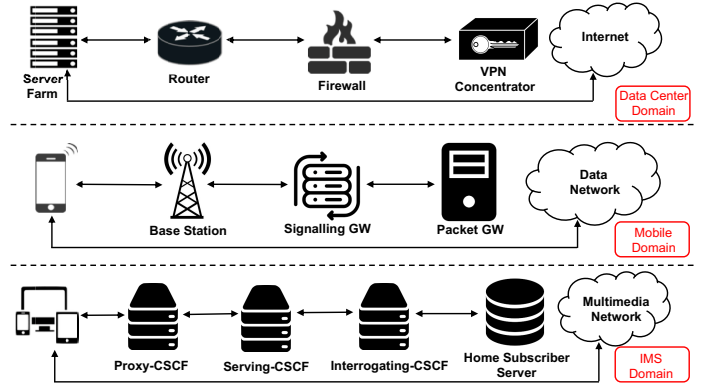}
	\caption{Service Chain Logic adopted by different domains: Data Center (uppermost panel), Mobile (middle panel), IP Multimedia Subsystem (downmost panel).}
	\label{fig:chain}
\end{figure}

\IEEEPARstart{N}{etwork} Function Virtualization (NFV) paradigm is a crucial technology enabler for $5$G \cite{NFV5g}. The main novelty introduced by NFV relies on the possibility of strongly decoupling network functions from the underlying physical infrastructure. It means that network elements such as routers, switches, firewalls, and many others, may be \textit{softwarized} so as to provide their functionalities with no need of being connected to special-purpose hardware. Such a new model brings undoubtedly many benefits in terms of: flexibility, concerning the scale-up/scale-down operations (adding/removing resources to achieve a desired state of performance), saving costs according to a pay-per-use model, maintenance optimization since the underlying physical resources can be shared among other network operators/tenants.  
Moreover, NFV introduces the concept of service chain, namely, a group of softwarized network elements aimed to provide specific services if traversed in a predetermined order, whose performance and availability are starting to be hot topics \cite{cerroni1,cerroni2,tola1,tola2,dim_tsc}. Some telco/networking domains which benefit from a service chain logic are pictorially sketched in Fig. \ref{fig:chain}: Data Center/Cloud domain (uppermost panel) where a chain of virtualized elements (router, firewall, VPN concentrator) is arranged to provide the access to data center resources \cite{eri2014}; Mobile network domain (middle panel) where paradigms such as software defined radio allows to softwarize radio access elements (e.g. Base Station) as well \cite{nfvchain18}; IP Multimedia Subsystem (IMS) domain (downmost panel) in charge of managing multimedia content within novel telco architectures \cite{nec2015}. The increasing importance of this latter as a crucial part of $5$G networks is confirmed either by standardization technical groups \cite{ts_1,ts_2,ts_3}, and by big industry players \cite{industry-eri,industry-huawei,5gbook}. Remarkably, according to GSMA statistics \cite{gsma}, $138$ carriers launched around the world new IMS-based networks by May 2018.
Taking inspiration from such novel tendencies, we propose a performability assessment of softwarized service chain structures, where an IMS architecture has been elected as a valuable use case, and where a softwarized IMS (henceforth softIMS) testbed has been deployed to derive realistic parameters. The softwarized infrastructure has been realized through the container technology, the most effective virtualization concept which allows to deploy each softIMS node functionality as a lighweight virtualized entity (namely, a container). 
Two key issues are identified and tackled at the same time: $i)$ the performance of softIMS, intended as the ability of correctly guaranteeing the service under a delay constraint (Call Setup Delay - CSD), $ii)$ the availability of softIMS, intended as the ability of ensuring a given resiliency in presence of network faults (e.g. caused by flow congestion, nodes malfunctioning, hacker attacks) through ad-hoc redundancy strategies. Obviously, such an assessment can be easily adapted to other chained-like systems, provided that performance and availability metrics are accessible.    
\newpage
The main contributions are summarized in the following.
\begin{itemize}
\item We characterize the performance of softIMS chain through the queueing network decomposition method, allowing to analyze the whole chain by modeling each node as an $M/G/c$ queue; during this stage we formalize an optimization problem of resource (containers) allocation, where a CSD-based constraint has been taken into account. Such a characterization appears to be the first attempt in the literature.

\item We face an availability analysis and evaluation of softIMS by exploiting the Stochastic Reward Networks (SRN) technique to deal either with common mode failures (due to a nested layered structure of softwarized nodes), and with single point of failures (due to the series structure of softIMS); by means of SRN we design an optimal redundancy strategy aimed to satisfy given steady-state availability requirements at the minimal cost, where two softIMS deployments are compared and discussed (homogeneous and co-located).

\item We devise two algorithms working in cascade: \textit{OptCNT} useful to find an approximately optimal strategy for container allocation in terms of the adopted performance metric (CSD); \textit{OptSearchChain} to pinpoint the set of feasible (jointly satisfying delay and availability requirements) softIMS configurations through a heuristic search with pruning. The two algorithms are jointly exploited to derive the optimal chains achieving, at the same time, a desired performance level and a given availability target value. 

\end{itemize}

The whole assessment is supported by a designed from scratch testbed based on Clearwater \cite{clearwater}, a container-based IMS opensource platform which allows us to derive some realistic parameters through an intensive campaign of workload stress tests.  
 
A set of novelties emerge from our analysis.
First, our formulation leads to non-trivial findings due to the subtle interplay existing among the quantities at stake (availability, delay, costs). Second, we automate the whole chain optimization process by exploiting two algorithms going into the direction of the $6$G paradigm, characterized by highly automated processes in the network management field, and not yet faced at this level of detail. Finally, the integration between experimental results and the queueing network model allows us to deal with realistic deployments embodying a solid mathematical formulation as well. Such a novel approach lies in the middle between pure theoretical advances which are too far from the realistic world, and pure practical frameworks which do not allow any analytical characterization. 
 
The paper has the following structure: in Section \ref{sec:RW} we highlight the main novelties introduced in this work, w.r.t. the most relevant affine literature; Section \ref{sec:ims_arch} contains a description of a softIMS architecture with focus on the five-layered structure characterizing the container-based environment; in Section \ref{sec:performance} we detail the softIMS queueing model and the connected optimization problem useful to face the performance assessment; in Section \ref{sec:availability} we present the availability characterization relying on the Stochastic Reward Network technique used to evaluate two deployment schemes (\textit{Homogeneous} and \textit{Co-located}) of softIMS; in Section \ref{sec:autom} we introduce \textit{OptCNT} and \textit{OptSearchChain} algorithms supporting, respectively, the performance and the availability assessments; in Section \ref{sec:results} we provide details about the testbed and the performed experimental trials, along with a critical analysis of the resulting outcomes; finally, Section \ref{sec:conclusions} concludes the work along with some ideas about future developments.

\section{Related Work}
\label{sec:RW}

Redundancy mechanisms to guarantee availability requirements of service chains are brought to the attention both of  industrial and academic research. As a matter of facts, the European Telecommunications Standards Institute (ETSI) has released some best practices aimed at deploying high-availability service chains patterns \cite{etsirel}. A number of use cases is presented, but, obviously, the choice of implementing specific redundancy strategies is left to the network designer. 
On the other hand, the technical literature is focused on the methodological aspects concerning the availability issues of softwarized infrastructures. Accordingly, in this section we present a roundup of most relevant works along with differences and similarities with our proposal.

To better emphasize the offered contribution, it is useful to highlight the differences among our assessment and affine works along three lines, concerning: $i)$ the availability analysis, $ii)$ the optimal resource allocation problem, $iii)$ the queueing model. 

As regards the former aspect, the most common choice in literature is to characterize the availability of softwarized environments through state-space formalisms such as the Continuous-Time-Markov-Chains (CTMC). Valuable examples include: \cite{bezerra14} where the availability analysis concerns a cloud-based video streaming service; \cite{kim09} where the authors characterize a virtualized system (including hardware, hypervisor, and application layer) through a combination of fault trees and CTMCs; \cite{matos17} which proposes an availability evaluation of private cloud environments, where reliability block diagrams and Markov chains are hierarchically assembled; \cite{rahme17} where the authors introduce a scheme for developing reliable cloud-based platforms using multiple software spare components, and where the CTMCs are used to characterize the reliability model. 
Unfortunately, a known drawback of CTMC-based approaches is the difficulty of modeling complex systems due to the ungovernable state-space growth. In contrast, the characterization offered by techniques such as the Stochastic Reward Networks adopted in this work helps to prevent this issue since they offer the possibility to automate the generation of the underlying Markov chain directly starting from a compact description of the system.
A limitation of other availability-related works pertains instead the fact that failure and/or repair events are partially considered.   
For instance, in \cite{liu2016} the authors propose an availability assessment of softwarized broadband network chains, where only failure actions are considered; the repair modeling is neglected in \cite{fan2017} as well, where the focus is on the minimum number of redundant Virtual Network Functions (VNFs) to deploy for guaranteeing service chains availability. Again, failure/repair models lack in \cite{kong2017}, where the authors propose an algorithm to improve the service chains availability by solving an optimal VNFs allocation problem. Conversely, we take into account both failure and repair events characterizing: $i)$ each single node (with its internal nested structure) and $ii)$ the whole softIMS architecture whose chained structure raises the single point of failure problem.   

As regards the optimal resource allocation problem in softwarized environments, it is worth noting that we take into account a delay-sensitive infrastructure like IMS, where managing audio/video communications requires to deal with latency constraints. Conversely, typical solutions available for allocation problems in softwarized networks \cite{static1ok,static2ok,static3,static4} do not investigate in details realistic models and metrics to characterize the network latency.
Moreover, most works typically deal with the standard case of functions which do not require any convexity proof. Some examples include: \cite{optpb1}, where an optimal resource allocation problem is considered in the context of a publish/subscribe system deployed in a cloud-based environment; \cite{optpb2}, where an open queueing network model is adopted to characterize a data stream framework in terms of optimal resource allocation; \cite{static1ok} and \cite{static_opt}, where the problem of an optimal load distribution across cloud infrastructures is faced. In contrast, dealing with a time-related quantity coming from the realistic Cosmetatos approximation, we have to numerically prove the pertinent convexity, since no similar results exist in the literature (as the best knowledge of the authors). Finally, it is worth noting that the optimal resource allocation problem in our work is a part of a more complex process involving an availability evaluation, whereas in the aforementioned works such a problem is typically treated as a stand-alone procedure.
	
Finally, as concerns the service chain queueing model, we want to highlight that many approaches in the literature assume  exponential  service times, which is unrealistic in most situations. It is the case of \cite{mdmtnsm,tdsc1}, where the product-form property of Jackson networks (requiring an assumption of exponential service times) is exploited to model virtualized IMS-based systems. Jackson networks are again used in \cite{mahmood15} to model the interconnections between the controller and switches in an SDN-based environment, and in \cite{zhang17} to characterize a chain of VNFs across a datacenter. 
Other works consider the more realistic case of non-exponential service times but with several limitations. For instance, both in \cite{mg1_1} and in \cite{mg1_2}, a simple $M/G/1$ model is used to characterize a Software Defined Network architecture, where the presence of multiple instances (as occurs in more generic $M/G/c$ models) is neglected. Even when multiple instances are considered (see \cite{mgc_1,mgc_2}), the queueing analysis is limited to a single network node, by neglecting the interconnections existing among the nodes. 
	

Differently from aforementioned works, we exploit the more realistic framework of non-product-form queueing networks where the assumption of exponential service times is removed, and the interconnections among nodes are considered. Thanks to our IMS-based testbed, we perform automated and repeated trials of IMS requests, and, then, we empirically estimate the processing time spent by each interconnected node to handle such requests. The result is a generic-shape probability distribution of service times, whose evaluation of the first two moments (mean and variance) leads to the adoption of approximations for $M/G/c$ systems.

\begin{figure}[t!]
	\centering
	\captionsetup{justification=centering}
	\includegraphics[scale=0.3,angle=90]{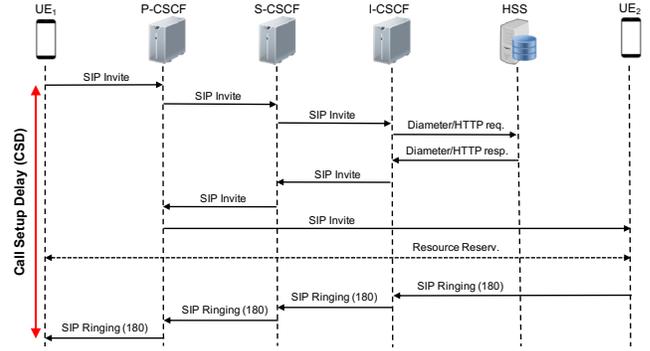}
	\caption{Call Setup Delay (CSD) in a simplified IMS scenario.}
	\label{fig:csd}
\end{figure} 

\section {Architectural perspective of softwarized IP Multimedia Subsystem}
\label{sec:ims_arch}

As a pivotal architecture of $5$G networks, IMS has been designed to manage multimedia contents (e.g. HD Voice/Video, presence, gaming, etc.) and to act as a broker with legacy networks, by guaranteeing backward compatibility with previous technologies (e.g. Long Term Evolution - LTE). 
From a topology viewpoint, IMS can be seen as a network chain composed of $3$ Call Session Control Function (CSCF) nodes and a database server element \cite{camarillobook}, as following detailed.  

\textit{Proxy CSCF (P-CSCF \textnormal{or} P \textnormal{for brevity})}: the first access point to the IMS domain, in charge to manage and route the SIP (Session Initiation Protocol) incoming requests (e.g. Register, Invite) received by subscriber devices. From a security viewpoint, such a node prevents unauthorized accesses to IMS infrastructure.

\textit{Serving CSCF (S-CSCF \textnormal{or} S \textnormal{for brevity})}: a crucial node aimed at controlling the status on each IMS session (including instant messaging, voice, multimedia transfer, etc.). Such a node acts as a SIP registrar, since it is responsible for authenticating the subscribers attempting an IMS registration.

\textit{Interrogating CSCF (I-CSCF \textnormal{or} I \textnormal{for brevity})}: this node queries the HSS (see below) to retrieve the user location through Diameter or HTTP protocols. Then, it routes the SIP request to the assigned S-CSCF.

\textit{Home Subscriber Server (HSS \textnormal{or} H \textnormal{for brevity})}: it represents the place where subscriber data are stored. Such data include: public and private identities, authentication keys, profiling information. Such a node can be accessed only by the S-CSCF belonging to the same IMS domain through the so-called Diameter protocol.

The Figure \ref{fig:csd} reports a simplified IMS scenario involving two User Equipments: the caller (UE$_1$) and the callee  (UE$_2$) attached to the same IMS domain. The call setup stage starts with the SIP invite message from UE$_1$ to P-CSCF (the contact point of the IMS domain). Such a message traverses the S-CSCF and I-CSCF nodes and arrive to the HSS in charge of returning the address of the S-CSCF that will manage the SIP session. Such an information is backpropagated to the P-CSCF that can now send a SIP Invite message (including a list of all the involved nodes) directly to UE$_2$. Finally, the SIP ringing message (message code $180$) encodes the signaling status that the UE$_2$ terminal is ringing.   

\begin{figure}[t!]
	\centering
	\captionsetup{justification=centering}
	\includegraphics[scale=0.3,angle=90]{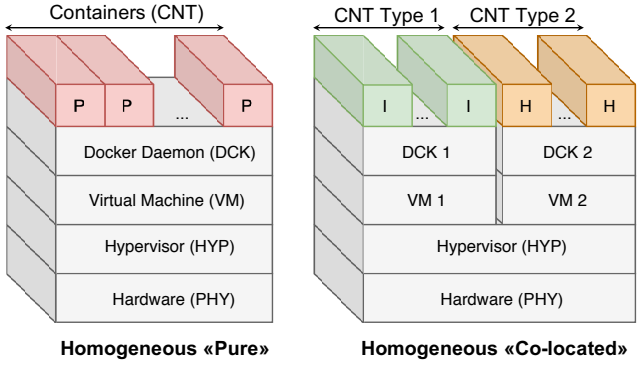}
	\caption{Network Replica deployments: Homogeneous (on the left), and Co-located (on the right).}
	\label{fig:nr}
\end{figure} 

\subsection{IMS softwarization and connected implications}

We remark that, from a technological point of view, the softwarization process can be implemented through paradigms such as virtualization or containerization. The latter offers better guarantees in terms of resource usage efficiency, since, differently from classic virtual machines, containers do not need a separate operating system to work. In contrast, classic virtualized settings exhibit a higher isolation degree, resulting in better managing security issues. 
Actually, it is possible to smartly combining both of technologies by implementing containers on top of virtual machines \cite{combe2016}. This is the solution adopted by Google Container Engine \cite{google_container} and by Amazon Web Services \cite{amazon_container}. By adhering to such a trend, each node of softIMS architecture is deployed through a so-called Network Replica (NR), a five-layered structure (see Fig. \ref{fig:nr}) that can be easily replicated for availability purposes. Precisely, an NR consists of:
\begin{itemize}
	\item {\em Containers (CNT)}: the softwarized network functionality to be provided across the IMS domain (e.g. Proxy, Interrogating, Serving, etc.);
	\item {\em Docker Daemon (DCK)}: the most popular engine \cite{docker} to manage the whole life-cycle of a container;
	\item {\em Virtual Machine (VM)}: the ecosystem providing isolation to the NR; 
	\item {\em Hypervisor (HYP)}: it represents an interface between the underlying hardware and the upper software-based layers;
	\item {\em Physical Layer (PHY)}: it embeds all the hardware equipment (Power supplies, CPU, RAM, etc.). 
\end{itemize} 
In the considered softIMS environment, we focus on the most common container-based deployment often known as {\em Homogeneous} \cite{sebastio18}, meaning that one and the same type of container can be hosted on top of the NR. Moreover, we introduce an additional taxonomy: $i)$ {\em Homogeneous Pure} (simply {\em Homogeneous} in the following) deployment, where containers share the whole underlying infrastructure (Fig. \ref{fig:nr} - left); $ii)$ {\em Homogeneous Co-located} (simply {\em Co-located} in the following), where different types of containers can coexist on the same NR by sharing only hypervisor and hardware layer (Fig. \ref{fig:nr} - right).   
Typically, homogeneous deployments are preferable in public cloud scenarios, where instances are kept separated for security issues or pricing concerns. Conversely, co-located deployments are suitable for private cloud, where multiple container instances (pertaining to different users) share part of the same infrastructure.

Such considerations also hold true for real IMS architectures, where HSS and I-CSCF are often co-located on the same infrastructure \cite{ims_syed}. 

It is also important to highlight that a single softIMS node can be realized by deploying one or more NRs.

\section{Performance of softIMS: the Queueing Network model}
\label{sec:performance}

As occurs in many telco systems, also IMS manages the arriving requests in an ordered way, so that the whole infrastructure is traversed node by node. Since each node has its own peculiarity (underlying technology, resource usage, software on-board, etc.), IMS requests are processed with different service times, thus, an accumulation of such requests is possible at specific nodes if they are not adequately designed/tuned. From a probabilistic point of view, this behavior can be captured through the queueing network (QN) formalism, where departures from a node become arrivals for the next node. As already stated, such a formalism (e.g. Jackson networks \cite{jackson}) typically requires the assumption of exponential service times distributions. In contrast, empirical measurements performed across our softIMS testbed reveal generic-shape distributions of nodes service times, thus, the exponential times assumption is violated. 
As a result, we assume an {\em M/G/c} model for each node, where: requests enter the softIMS system according to a Poisson random process {\em M}, service times follow an arbitrary distribution {\em G}, and a number of finite containers {\em c} is in charge of managing the requests.  
Then, the whole softIMS chain is treated as an open queueing network, being requests not reinserted in the system once processed. 
Due to the generic distribution of service times, we settle on an approximate performance analysis based on the QN decomposition method \cite{trivedibook}, which requires the chain to be broken into subsystems independently analyzed, and  involves two steps: $i)$ for each node, estimate mean $\mathbb{E}(\cdot)$ and variance $\mathbb{V}(\cdot)$ of service times to derive the coefficient of variation $v=\sqrt{\mathbb{V}(\cdot)}/\mathbb{E}(\cdot)$; $ii)$ evaluate performance measures (e.g. mean waiting times) through approximating formulas.   

In an open network with $N$ nodes, we have:
\beq
\alpha_n=\alpha_{ext} + \sum_{m=1}^{N}\alpha_m \cdot p_{mn}^{(r)},
\label{eq:balance}
\eeq
which is the so-called balance equation. The arrival rate $\alpha_n$ at node $n$ (with $n=1 \dots N$) is obtained as the sum of external and internal contributions, where:  
\begin{itemize}
\item $\alpha_{ext}$ is the mean external arrival rate of requests at P-CSCF node (the only node qualified to manage outside traffic flows);
\item $\alpha_m$ ($m \in \textnormal{S-CSCF, I-CSCF, HSS}$) is the mean internal arrival rate at the node $m$;
\item $p_{mn}^{(r)}$ is the probability that a request is routed to node $n$, once the process at the node $m$ is finished. 
\end{itemize}
Let us start by considering an $M/M/c_n$ queueing model. Let $1/\beta_n$ be the mean service time of node $n$, $c_n$ the number of containers at the node $n$, and $\rho_n=\alpha_n/c_n\beta_n$ be the utilization factor at node $n$, where the ergodic condition $\rho_n<1$ must be satisfied for the stability of the queueing system. 
According to the classic queueing theory \cite{bgbook}, the mean waiting time at node $n$ for an $M/M/c_n$ model is:
\beq
\mathbb{E}[W_n]_{M/M/c_n}= \frac{\rho_n}{\alpha_n(1-\rho_n)} \cdot \pi_n,
\label{eq:mmc}
\eeq
being $\pi_n$ the steady-state probability that an arriving request has to wait in queue. 
Among the existing approximations to derive performance measures of our $M/G/c_n$ system, we use the one provided by Cosmetatos \cite{cosme}, which, in case of medium/heavy-traffic condition ($\rho_n \geq 0.6$) and for a small number of $c_n$ (typically, $c_n \leq 10$), is a very good approximation as also highlighted in \cite{kimura} and in \cite{bolchbook}. Such assumptions are perfectly in line with the carried experiments and with realistic deployments. The Cosmetatos approximated formula allows to express the mean waiting time of requests at the node $n$ as

\beq
\mathbb{E}[W_n] \approx v^2_{\beta_n} \cdot \mathbb{E}[W_n]_{M/M/c_n} + (1- v^2_{\beta_n}) \cdot \mathbb{E}[W_n]_{M/D/c_n},
\label{eq:cosme}
\eeq
where $v_{\beta_n}$ is the coefficient of variation of service time at node $n$, and where $ \mathbb{E}[W_n]_{M/M/c_n}$ comes from (\ref{eq:mmc}). 
As part of Cosmetatos approximating formula, it is possible to express the main waiting time at node $n$ having deterministic (D) service times as
\beq
\mathbb{E}[W_n]_{M/D/c_n} \approx \frac{\mathbb{E}[W_n]_{M/M/c_n}}{2 \cdot \phi_n},
\label{eq:mdm}
\eeq
being $\phi_n$ the Cosmetatos approximation factor (for the node $n$) amounting to

\beq
\phi_n = \frac{1}{1+(1-\rho_n)(c_n-1) \frac{\sqrt{4+5c_n}-2}{16 \rho_n c_n} }.
\eeq
From the classic queueing theory \cite{bgbook}, we know that the mean response time (namely the mean time that a {\em job} spends in the queueing system) is the sum of the mean waiting time (time that a job spends in a queue, waiting to be serviced) and the service time itself.
	
Thus, by taking into account (\ref{eq:cosme}), the mean response time to process an IMS request at node $n$ can be expressed as
\beq
\mathbb{E}[T_n] = \frac{1}{\beta_n}+\mathbb{E}[W_n].
\label{eq:tot_wait_time}
\eeq
Finally, since the mean CSD (simply indicated by CSD in the following) can be interpreted as the total average time that IMS requests spend in the whole system, we can write: 
\beq
CSD \approx \sum_{n=1}^{N} \mathbb{E}[T_n]
\label{eq:csd}
\eeq
From queueing theory, we have that as the number of containers hosted by the node $n$ grows, $\mathbb{E}[W_n]$ decreases, and, in turn, $\mathbb{E}[T_n]$ will be dominated by the service time $1/\beta_n$ as pointed in (\ref{eq:tot_wait_time}). Actually, two connected drawbacks emerge: first, the underlying infrastructure could host only a limited number of containers; then, more containers imply more resource consuming, and, then, more costs. 
Accordingly, we are interested in minimizing the overall number of containers deployed across the softIMS chain, thus, the following optimization problem arises:

\beqa
\label{eq:optprob}
&\textnormal{minimize}& \sum_{n=1}^{N} c_n   \\  \nonumber   \\
&\textnormal{subject to}& 
\left\{
\begin{array}{l} 
	{\begin{array}{ll} 
			\hspace{-0.2cm} c_{n0} \leq c_n \leq c_{max} , ~~ c_n \in \mathbb{N}, \nonumber
		\end{array}}
		\\
		\\
		\sum_{n=1}^{N} \mathbb{E}[T_n] \leq CSD^\dagger.
	\end{array}
	\right.
\eeqa

The first constraint in (\ref{eq:optprob}) accounts for the queue stability condition to be achieved per node, where $c_{n0}=\lfloor \alpha_n/\beta_n \rfloor$ +1, being $\lfloor \cdot \rfloor$ the integer round-down operation, and where $c_{max}$ accounts for the maximum number of containers to preserve the convexity\footnote{We are safely guaranteed of lying in the convexity region, since $c_{max}$ never exceeds the value of $10$.}; the second constraint is aimed at preventing that the whole softIMS chain could introduce a delay exceeding a target CSD (CSD$^\dagger$) imposed by networking/telco standards.  

We note in passing that, in this formulation, we have neglected the propagation delays that are typically related to the geographic location of nodes (in cloud-based environments such delays amount approximately to zero). If needed, such delays can be considered as a fixed quantity that can be included into the R.H.S. of the second constraint in (\ref{eq:optprob}) with no lack of generality. 	

 \begin{figure*}[t!]
 	\setcounter{figure}{4}
 	\centering
 	\captionsetup{justification=centering}
 	\includegraphics[scale=0.63,angle=90]{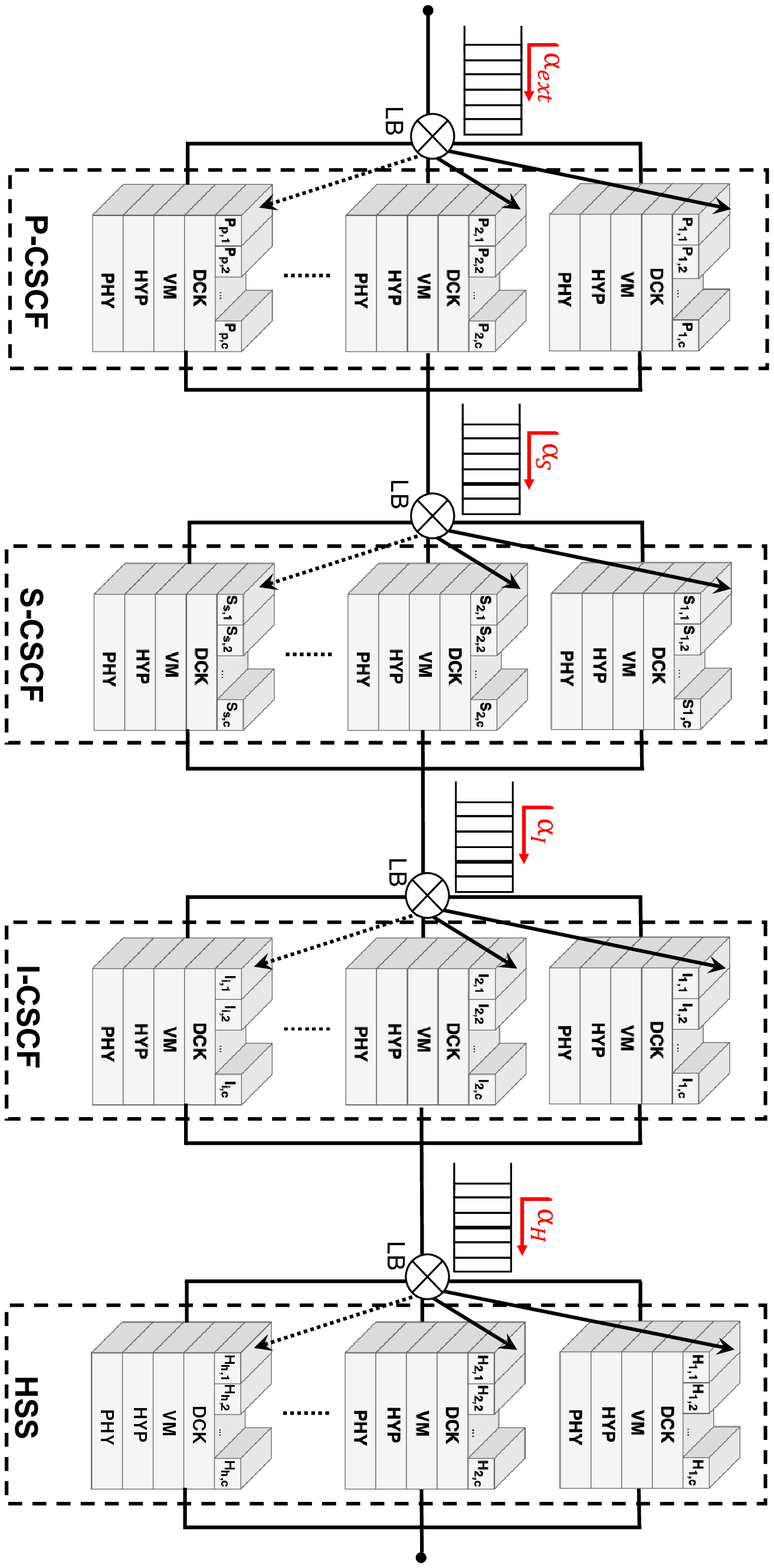}
 	\caption{High-level perspective of softIMS including series/parallel interconnections and queueing network structure.}
 	\label{fig:rbd}
 \end{figure*} 
 
 \begin{figure}[t!]
 	\setcounter{figure}{3}
 	\centering
 	\captionsetup{justification=centering}
 	\includegraphics[scale=0.35]{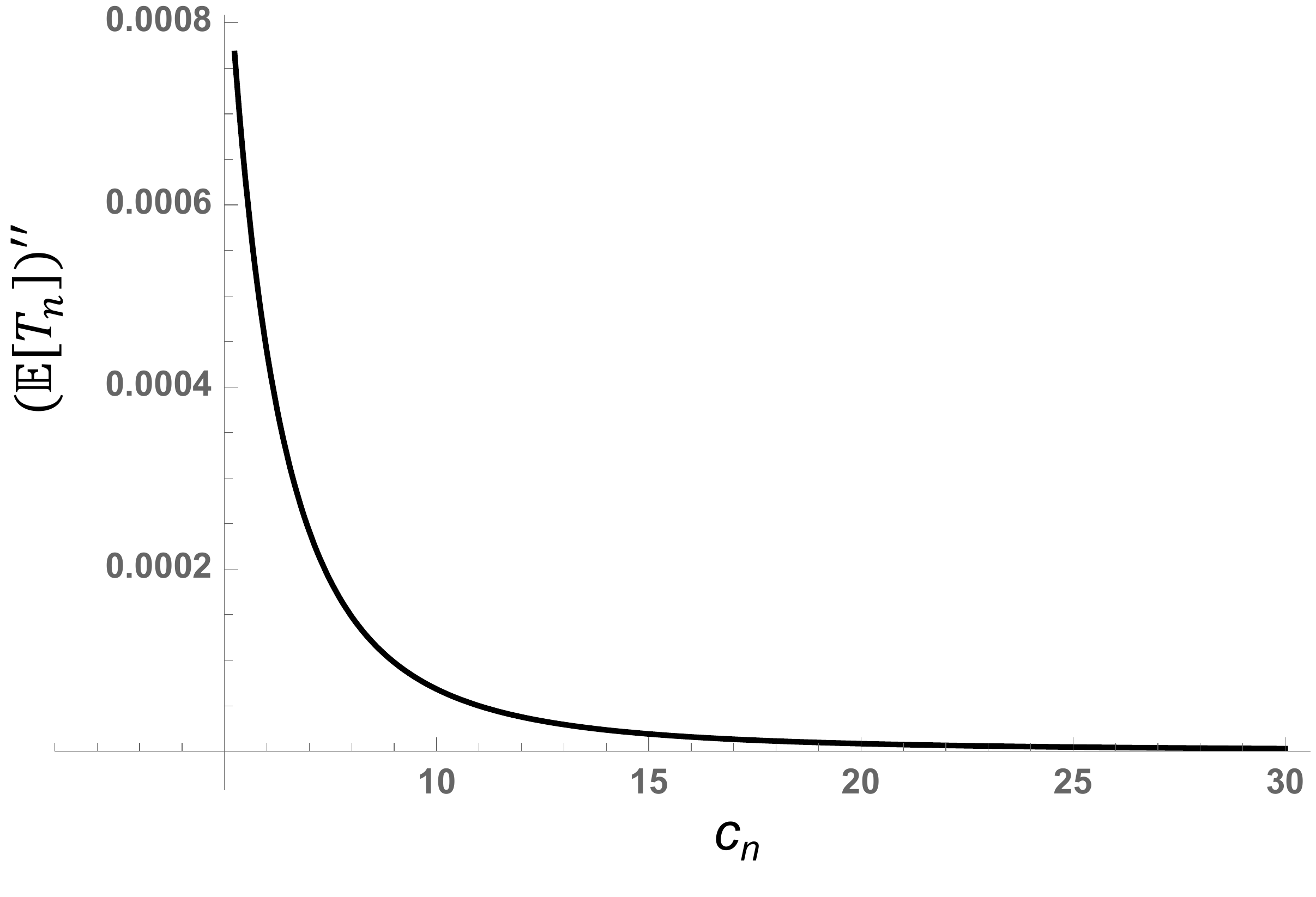}
 	\caption{Convexity (numerical proof) of $\mathbb{E}[T_n]$ at the P-CSCF node.}
 	\label{fig:cosme_convex}
 \end{figure} 

\begin{figure*}[t!]
	\setcounter{figure}{5}
	\centering
	\captionsetup{justification=centering}
	\includegraphics[scale=0.62]{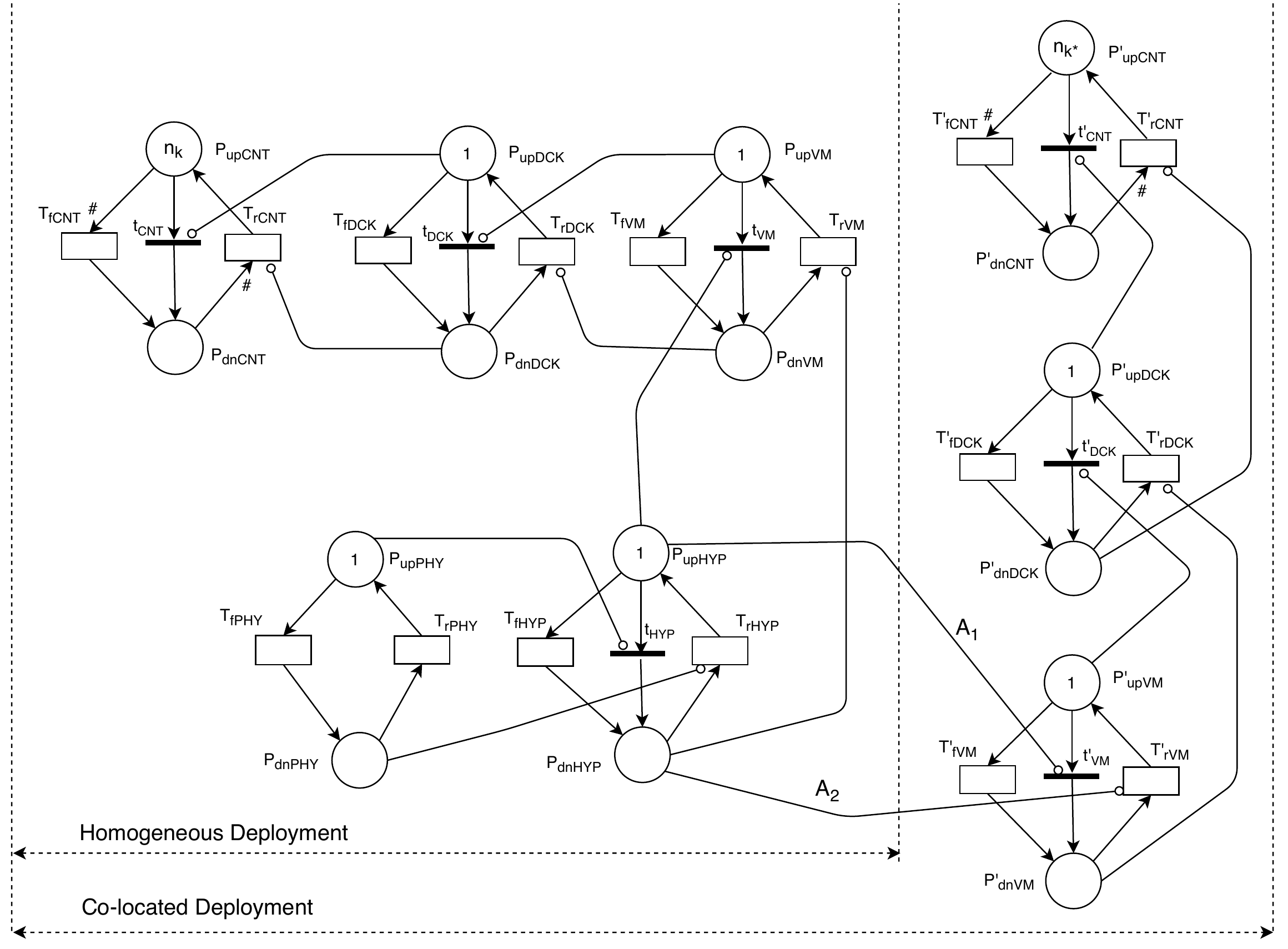}
	\caption{SRN representation of two NR deployments: Homogeneous and Co-located.}
	\label{fig:srn}
\end{figure*} 

The solution of (\ref{eq:optprob}) takes inspiration from some works dealing with optimal resource allocation in manufacturing networks \cite{manuf1,manuf2}, where the main problem is to minimize the Work-In-Process (WIP), namely the number of goods partially finished which wait for further processing at a station.

 It is interesting to notice that the optimization problem in (\ref{eq:optprob}) exhibits the structure of the well-known \textit{knapsack problem} which is NP-hard, thus heuristic approaches can be exploited to find the approximately optimal solutions. 
A greedy strategy for solving (\ref{eq:optprob}) is to start with the smallest possible allocation of containers per node where the stability condition is satisfied. At each iteration, a container can be added to that node where the ratio between the increase of $c_n$ ($+\Delta(c_n)$) and the decrease of $\mathbb{E}[T_n]$ ($-\Delta(\mathbb{E}[T_n]$)) is the smallest, namely

\beq
n~=~ \textnormal{argmin} \left(   \frac{\Delta(c_n)}{- \Delta(\mathbb{E}[T_n])}   \right).
\eeq

Such allocation problems are proved to be undominated \cite{manuf1}, provided that the involved functions are convex. 

Convexity of $\sum c_n$ holds true since it is a non-decreasing non-negative weighted sum, but, a convexity ``check'' must be performed for $\mathbb{E}[T_n]$. As rightly observed in \cite{manuf3}, for many queueing systems approximating formulas it is very hard, if not impossible, to prove analytically their convexity. Accordingly, we face this issue by a numerical point of view. By exploiting the fact that a function is convex \textit{iff} its second derivative is non-negative, in Fig. \ref{fig:cosme_convex} we verify the behavior of second derivative of $\mathbb{E}[T_n]$ for the P-CSCF node, with values: $\alpha_{ext}=200$ s$^{-1}$ (highest external load), $\beta_P=8 \cdot 10^{-3}$ being the mean service time empirically measured for P-CSCF (see forthcoming Table \ref{tab:params} for values pertaining to other nodes), and $v_{\beta_P}=1.25$ the measured coefficient of variation.
Similar behaviors can be observed for other nodes (not reported here). 

Hence, the result of (\ref{eq:optprob}) is an optimal vector of containers whose components, say $c^*_n$, are assigned to each node $n$ in order to satisfy the desired CSD-based performance constraint. Such a greedy procedure is a part of OptCNT algorithm presented in Sect. \ref{sec:autom}.
In summary, the proposed queueing network model coupled with the softwarized arrangement of each node can be captured by the high-level representation of Fig. \ref{fig:rbd}. Each softIMS node can be made of more NRs and each NR can be made of more containers. As requests enter a softIMS node, a load balancer (LB) pass them to a container. The minimum number of containers per node to satisfy the CSD constraint comes from (\ref{eq:optprob}), but such a number can be not enough to guarantee certain availability requirements. This issue will be tackled in Section \ref{sec:availability}. 
 
Furthermore, it is useful to give prominence to an important connection between queueing and failure/repair models in terms of time scales. As pointed in \cite{whittbook}, in the field of communication networks we can distinguish various and different time scales, such as the service time scales (STS) and the failure time scales (FTS). The former rules the typical queueing metrics (e.g. the service times), whereas the latter rules the failure/repair processes. 
In case a time scale completely dominates another one, we can neglect the transient effects pertaining to the dominated time scale. In this work, experimental trials clearly show that STS $\ll$ FTS (see numerical values derived in the forthcoming Table \ref{tab:params}, with the service times in the order of milliseconds and failures in the order of hours), thus, a decoupling between FTS and STS can be reasonably assumed, and the queue can be analyzed in its steady-state. 
 
\section{Availability characterization of softIMS}
\label{sec:availability}

From an availability modeling perspective, we adopt the SRN formalism \cite{muppala94} to capture the probabilistic behavior (in terms of failure and repair events) of each NR deployment (both homogeneous and co-located), and, consequently, of each node. Then, we derive the availability of the whole softIMS by combining the availability results obtained for each node.
Figure \ref{fig:srn} depicts the SRN-based model of the two considered NR deployments, where the classic Petri nets symbolism has been adopted as following detailed.
\begin{itemize}
\item {\em Places} (circles): denote particular conditions (e.g. an NR layer up or down) specified by an internal {\em token} represented with a number or a letter; 
\item {\em Timed Transitions} (unfilled thick rectangles): indicate actions (e.g. a layer fails or gets repaired), whose  times are supposed to be exponentially distributed with parameters $\lambda$ (failure rate) or $\mu$ (repair rate);
\item {\em Immediate Transitions} (filled thin rectangles): denote instantaneous actions occurring in a near-zero time interval (e.g. the DCK layer fault implying immediately the CNT layer fault);
\item {\em Inhibitory Arcs} (segments with a circle): indicate forbidden actions (e.g. repair a layer having a failed underlying layer). 
\end{itemize}
The solution of SRN proposed in Fig. \ref{fig:srn} is equivalent to the evaluation of the pertinent {\em reward function} $R(t)$. It is defined as a non-negative random process amounting to $1$ when the NR is working at time $t$ and to $0$ otherwise. The instantaneous availability $A(t)$ is the probability that the node is working at time $t$, and can be expressed in terms of $R(t)$ \cite{muppala94} as follows: 

\begin{equation}
\label{maineq}
\displaystyle { A(t) =  \mathbb{E}[R(t)] = \sum_{s \in S } {r_s \cdot p_s(t) },}
\end{equation}
being: $S$ the set of feasible tokens distributions (a.k.a. {\em markings}), $r_s$ the so-called {\em reward rate}, namely the value of $R(t)$ in marking $s$, and $p_s(t)$ the pertinent probability. 
We want now to solve, separately, the SRNs corresponding to the homogeneous and co-located deployments, respectively. 

\subsection{Homogeneous deployment: availability model}

Let us consider the SRN model for the homogeneous deployment represented in Fig. \ref{fig:srn}. Places $P_{upCNT}$, $P_{upDCK}$, $P_{upVM}$, $P_{upHYP}$, and $P_{upPHY}$ refer to the working conditions of the five layers of a homogeneous-type NR. In contrast, places $P_{dnCNT}$, $P_{dnDCK}$, $P_{dnVM}$, $P_{dnHYP}$, and $P_{dnPHY}$ take into account the counterposed failure conditions. 
Similarly, timed transitions $T_{fCNT}$, $T_{fDCK}$, $T_{fVM}$, $T_{fHYP}$, and $T_{fPHY}$ indicate failure actions for the five layers, whereas $T_{rCNT}$, $T_{rDCK}$, $T_{rVM}$, $T_{rHYP}$, and $T_{rPHY}$ are the counterposed repair actions. Transitions with $\#$ symbol adjacent, are said to be ``marking-dependent'', namely, their rates are multiplied by the number of tokens in the corresponding place.
In the initial working condition, all the token are contained in $P_{up}$ places. Remarkably, the only place containing more than one token (precisely, $n_k$) is $P_{upCNT}$, due to the possibility of considering more replicated container instances. In case of a container fault, $T_{fCNT}$ is ``fired" and $P_{upCNT}$ loses one token. Such a token is transferred to $P_{dnCNT}$, and will remain there until a repair action intervenes. 
Similarly, in case of Docker daemon fault, $T_{fDCK}$ is fired, and the transfer of the token from $P_{upDCK}$ to $P_{dnDCK}$ occurs. This action implies that all the containers running on top of docker layer are no longer working. As a result, the inhibitory arc between $P_{upDCK}$ and the immediate transition $t_{CNT}$ forces this latter to be fired, thus, all the token in $P_{upCNT}$ move to $P_{dnCNT}$. Once repaired the docker daemon, the same inhibitory arc is deactivated and $T_{rCNT}$. Likewise, hypervisor and physical layers admit a similar behavior in terms of failure and repair events. 
The reward rate pertaining the node $n$ in marking $s$ can now be expressed as
\begin{equation}
\label{eq:reward_hom}
\centering
r_{s}(n)=
\left\{
\begin{array}{l}
{\begin{array}{ll}
	\hspace{-0.2cm} 1 \;\;\;\;\; \text{if } \;\;\;\;\; \sum_{\ell=1}^{L} \ {\odot P_{upCNT}^{(\ell)} \ge  c^*_n,}
	\end{array}}
\\
\\
0 \;\;\;\;\;\text{otherwise,}
\end{array}
\right.
\end{equation}  
where: $\ell$ is the number of NRs composing the node $n$, $``\odot"$ refers to the number of tokens, and the threshold $c_n^*$ is the minimum number of containers hosted by the $n$-th node guaranteeing the CSD performance condition as derived from (\ref{eq:optprob}). It is useful to remark that the unique condition to guarantee on $c^*_n$ in (\ref{eq:reward_hom}) is that such a quantity is non-negative, being the number of tokens a non-negative quantity. Such a condition is surely statisfied from the problem formulation in (\ref{eq:optprob}), since we are guaranteed that $c_n \geq c_{n0}$, being $c_{n0}$ non-negative.

By virtue of (\ref{maineq}) and for $t \rightarrow \infty$, the steady-state availability of the $n$-th node, i.e. the probability that the node $n$ is working for $t\rightarrow \infty$, is:
\begin{equation}
\label{eq:ss}
\hspace{15mm}   \displaystyle  {  A^{(n)}= \lim_{t \to +\infty}   A^{(n)}(t) = \sum_{s \in S } {r_{s}(n) \cdot p_{sn}}, }
\end{equation}
where $r_{s}(n)$ stems from (\ref{eq:reward_hom}), and $p_{sn} = \lim_{t \to +\infty}  p_{sn}(t)$ is the corresponding steady-state probability. 
Now, the whole steady-state availability for the homogeneous deployement can obtained by starting from single nodes availability in (\ref{eq:ss}), viz.

\begin{equation}
\label{eq:sscims}
A^{(hom)}=\prod_{n=1}^{4} A^{(n)}.
\end{equation}

\noindent The product in (\ref{eq:sscims}) reflects the chained structure of Fig. \ref{fig:rbd} where the series connection implies that each node must be available to make the whole softIMS available.

\subsection{Co-located deployment: availability model}

We now consider the availability characterization of the co-located deployment represented by the ``largest" SRN in Fig. \ref{fig:srn}. From a modeling perspective, the co-located SRN can be obtained by adding a new part to the homogeneous SRN. In particular, we consider a co-located deployment having two different docker and VM layers (see Fig. \ref{fig:nr}), thus, two different types of containers can be hosted on top of the co-located NR. The new added elements (places, transitions, etc.) are denoted by a prime superscript ($P_{upCNT}'$, $P_{dnCNT}'$, and so forth). It is also useful to highlight the presence of two new inhibitory arcs: $A_1$ which is enabled when the hypervisor fails, thus $t_{VM}'$ gets fired, and $A_2$ which prevents the token to be transferred from $P_{dnVM}'$ to $P_{upVM}'$ until hypervisor gets restored. 
Let us now consider the reward rate for co-located case:

\begin{align}
\label{eq:reward_hom_col}
r'_{s}(n_1, n_2)=
\left\{
\begin{array}{l}
{\begin{array}{ll}
	\hspace{-0.2cm} 1 \;\;\;\;\; \text{if } 
	&\left( \sum_{\ell=1}^{L} \ {\odot P_{upCNT}^{(\ell)} \ge c^*_{n_1}}\right) \\   \\
	 &	\;\;\;\;\;\;\;\;\;\;\;\;\;\;\;\;\;\;\;\;   \wedge  \\    \\
	&\left( \sum_{\ell=1}^{L} \ {\odot P_{upCNT}^{'(\ell)} \ge c^*_{n_2}}\right),
	\end{array}}
\\
\\
0 \;\;\;\;\;\text{otherwise},
\end{array}
\right.
\end{align}
where ``$\wedge$''  indicates the logical {\em AND} operator. Relation (\ref{eq:reward_hom_col}) derives from (\ref{eq:reward_hom}) with threshold values $c^*_{n_1}$ and $c^*_{n_2}$ corresponding to the $n_1$ and $n_2$ co-located nodes, respectively (I-CSCF and HSS). 
In the end, the corresponding steady-state availability for the couple of co-located nodes is
\begin{equation}
\label{eq:ss_colocated}
\hspace{15mm}   \displaystyle  {  A^{(n_1,n_2)}= \lim_{t \to +\infty}   A^{(n_1,n_2)}(t) = \sum_{s \in S } {r'_{s}(n_1, n_2) \cdot p'_{ns} }, }
\end{equation}
being $r'_s(n_1, n_2)$ derived by (\ref{eq:reward_hom_col}), and $p'_{ns}$ the pertinent steady-state probability.

This leads to evaluate the overall softIMS steady-state availability for the co-located deployment as:
\begin{equation}
\label{eq:sscims_col}
A^{(col)}= A^{(n_1,n_2)} \cdot \prod_{n\neq n_1,n_2}A^{(n)},
\end{equation}
In (\ref{eq:sscims_col}), we exploit the fact that the first factor on R.H.S. accounts for the presence of $n_1$ and $n_2$ co-located nodes, whereas, the second factor considers the two remaining nodes.

\section{Performability Evaluation}
\label{sec:autom}

We remark that the main purpose of our analysis is to derive an optimal set of softIMS configurations guaranteeing, at the same time, a given performance level and a desired steady-state availability with the minimum deployment cost. At this aim, we designed two algorithms nicknamed OptCNT and OptSearchChain\footnote{The code of the two algorithms is available upon request.} to automate performance and availability assessments, respectively. OptCNT is intended to numerically solve the optimization problem (\ref{eq:optprob}), and returns the minimum number of containers per node (c$_n^*$) useful to guarantee the performance metric CSD$^\dagger$. By starting from such a result, OptSearchChain will be able to: $i)$ automatically build and evaluate SRN models of performance-compliant softIMS configurations, $ii)$ select the configurations satisfying the desired steady-state availability requirement with a minimal deployment cost.   

	\begin{algorithm}[t!]
		\caption{OptCNT}
		\label{myalg}
		\SetAlgoLined   
	\KwIn{$\alpha_n$, $\beta_n$, CSD$^\dagger$}
		\For{n=1 \dots N}{ 
 $c_{n0}=\lfloor \alpha_n/\beta_n \rfloor +1$ \\
 $c_n=c_{n0}$
		}
		  \While{$\sum_{n=1}^{N} \mathbb{E}[T_n]  \geq \textnormal{CSD}^\dagger $} 
		  {
		  	$c_n \leftarrow (c_n +1)$ $\textnormal{where}$ n~=~argmin $\left(   \frac{\Delta(c_n)}{- \Delta(\mathbb{E}[T_n])}   \right)$
		  } 
		  \% \textbf{Output}: c$_n^* \leftarrow c_n$ (opt. num. of containers per node) 
			\end{algorithm}
			
\vspace{10pt}
\noindent\textbf{OptCNT Algorithm:}
Let us start to analyze the OptCNT algorithm, whose pseudo-code is reported below. Lines $1-3$ refer to the initialization phase of the minimal number of containers so as to respect the stability queue condition as clarified in Sect. \ref{sec:performance}. In case the initialization value makes the performance metric to be satisfied, the algorithm directly returns the c$_n^*$ value. Otherwise, OptCNT enters the \textit{while-do} loop (lines $5-7$) and increases the number of containers per node up to reaching the desired performance condition. 			
			
	\begin{algorithm}[ht!]			
		\caption{OptSearchChain}
		\KwIn{$\textbf{K}$=SRNEval($\lambda$, $\mu$, A$^\dagger$, c$_n^*$, c$_{max}$, \textit{depl\_type})} 
		C$_{min} \leftarrow inf$ \\ 
		\For{p $\in$ {$\textbf{K$_{pcscf}$}$}}
		{{evaluate $Cost(p)$ }  \\
			\If{Cost(p) $>$ $0.5$ $\cdot$ C$_{min}$
				}  
			{ continue; }
			\For{s $\in$ {$\textbf{K$_{scscf}$}$}}
			{{evaluate $Cost(s)$}  \\
				\If{$\sum_{n=p,s} Cost(n)$ $>$  $0.75$ $\cdot$ $C_{min}$ \\ \textbf{OR} (A$^{(p)}$ $\cdot$ A$^{(s)}$) $<$A$^\dagger$} 
				{ continue; }
					\If{\textnormal{\textit{depl\_type}=homogeneous}}
					{
					\For{i $\in$ {$\textbf{K$_{icscf}$}$}}
					{
						{evaluate $Cost(i)$} \\
						\If{ $\sum_{n=p,s,i} Cost(n)$  $>$  C$_{min}$ \\ \textbf{OR} $\prod_{n=p,s,i}{A^{(n)}}$ $<$A$^\dagger$} 
						{ continue; }
						\For{h $\in$ {$\textbf{K$_{hss}$}$}}
						{evaluate $Cost(h)$ \\
							\If{  $\prod_{n=p,s,i,h}{A^{(n)}}$ $<$A$^\dagger$} 
							{ continue;  \\
							}   		 	
						     C$_{min}$ $ \leftarrow $ min\{C$_{min}$, TotCost\} \\
							{save [softIMS, A$^{(hom)}$, TotCost]}}
						}
					}
						\ElseIf{\textnormal{\textit{depl\_type}=co-located}}{
							\For{$\overline{h,i}$ $\in$ {$\textbf{K$_{col}$}$}}
							{
								evaluate $Cost(\overline{h,i})$ \\
								\If{ $Cost(p)$ + $Cost(s)$ + $Cost(\overline{h,i})$ $>$ C$_{min}$ \\ \textbf{OR} ${A^{(p)}} \cdot {A^{(s)}} \cdot {A^{(\overline{h,i})}}$ $<$A$^\dagger$} 
								{ continue; 
								}
								C$_{min}$ $\leftarrow $ min\{C$_{min}$, TotCost\} \\
								{save [softIMS, A$^{(col)}$, TotCost]}
							}
						}
			}
		}
	\end{algorithm}

\vspace{10pt}
\noindent\textbf{OptSearchChain Algorithm:}
It is a more sophisticated algorithm (see the pseudo-code on the right) than the previous one, and relies on a two-stage procedure. The first one includes a call to a sub-routine named SRNEval which interacts with TimeNET (Timed Petri Net Evaluation Tool) \cite{timenet}, a framework allowing to evaluate SRN models, and freely available for research purposes. SNREval admits as inputs: failure rate $\lambda$ and repair rate $\mu$ for each NR layer (whose values are specified in the forthcoming Table \ref{tab:params}); the steady-state availability target A$^\dagger$ that softIMS chain has to achieve (e.g. four/five/six nines); the minimum number (lower bound) of container per node c$^*_n$ as returned by OptCNT algorithm; the maximum number (upper bound) of sustainable containers per node c$_{max}$ due to the technological limits; the type of deployment \textit{depl\_type} allowing to specify if we are interested in evaluating homogeneous or co-located SRN models. 
As output, SRNEval returns a list \textbf{K} of feasible softIMS configurations where each node satisfies (separately) the $A^{(n)} \geq A^\dagger$ requirement. Within such a list we find configurations that: $i)$ may not satisfy A$^\dagger$ (namely, $\prod_{n}A^{(n)} < A^\dagger$), or $ii)$ satisfy A$^\dagger$ but, due to the enormous number of such configurations, it is needed to select few ones according to a cost-based criterion. For the sake of simplicity, the list \textbf{K} is organized in sub-lists. For instance, sublist \textbf{K}$_{pcscf}$ contains all feasible combinations of NRs/containers pertaining to P-CSCF node.

The second stage of OptSearchChain tackles the two aforementioned issues by performing a smart exhaustive search with pruning. Before delving into pseudo-code details, we clarify the adopted heuristic cost criterion. We assume that the cost of a node is made of two contributions: the infrastructure cost (including all NR's layers except for container layer) and the container layer cost that are then summed (see the detailed discussion on costs in Sect. \ref{sec:results}). 

Let us now detail the crucial steps of OptSearchChain pseudo-code. A cost initialization is performed at line $1$, where $C_{min}$ represents the whole softIMS cost calculated/updated within a cycle. Then, in the first block of instructions (lines $2-6$), OptSearchChain prunes all those softIMS configurations whose P-CSCF cost exceeds $0.5$ times the cost of the whole chain (line $4$). Such a smart (and obviously customizable) multiplier allows to discard in advance many expensive softIMS configurations (in other words, all configurations with a too over-priced P-CSCF will be pruned), guaranteeing the best trade off between the number of configurations to retain and the time spent to their availability evaluation. 
In the second block of instructions (lines $7-12$), the algorithm prunes the configurations whose P-CSCF+S-CSCF cost exceeds $0.75$ times the cost of the overall softIMS chain (line $9$) or whose joint availability (line $10$) does not satisfy the target requirement. Similarly to the previous case, the chosen multiplier is aimed at pruning in advance expensive configurations exhibiting a high cumulative cost of P-CSCF and S-CSCF. 
At this stage, the OptSearchChain is forked in two parts depending on the \textit{depl\_type} flag: homogeneous (lines $13-29$) in case I-CSCF and HSS are deployed on separate NRs, or co-located (lines $30-40$) in case I-CSCF and HSS share the same NR (within pseudo-code, such a case is accounted by $\overline{h,i}$ notation). For both homogeneous and co-located cases, the pruning procedure follows the same logic before described for P-CSCF and S-CSCF nodes. 
After updating the cost parameter (line $25$ for the homogeneous case, and line $37$ for the co-located case), the algorithm returns a "vector" including a subset of softIMS configurations each of which having a certain availability value and a total cost obtained by the sum of costs of single nodes (line $26$ for the homogeneous case, and line $38$ for the co-located case). 

OptCNT and OptSearchChain have ran on a laptop equipped with Intel Core $i5-7200U$@$2.50$ GHz (quadcore) and with a RAM of $16$ GB. OptCNT runs in less than $5$ seconds, whereas OptSearchChain requires about $45$ seconds (excluding the external call to the SRNEval sub-routine) considering a steady-state availability requirement amounting to $0.99999$, and a maximum number of containers hosted on top of an NR equal to $6$ (technological limit). The time required by SRNEval highly depends on the number of containers (ranging from few minutes for $2$ containers per NR up to 2 hours for $6$ containers).
Remarkably, OptSearchChain allows to obtain a number of softIMS configurations in the order of $10^{4}$ starting from a quasi-intractable number of configurations in the order of $10^{12}$.    
Some resulting configurations will be shown and analyzed in the next section.

\section{Numerical Results}
\label{sec:results}

\begin{figure}[t!]
	\centering
	\captionsetup{justification=centering}
	\includegraphics[scale=0.32,angle=90]{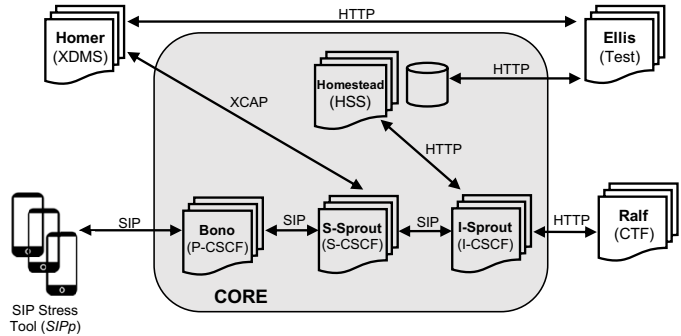}
	\caption{Schematic representation of the testbed based on the Clearwater platform.}
	\label{fig:clw}
\end{figure}

This section describes the experimental testbed and  discusses  the numerical results thereof. The whole softIMS architecture relies on the open-source Clearwater platform, whose main nodes are pictorially represented in Fig. \ref{fig:clw}. We deploy on separate VMs (2-core virtual CPU and 8 GB of RAM) the mandatory core nodes (depicted on a gray background in Fig. \ref{fig:clw}): {\em Bono} (P-CSCF), {\em Sprout} (a macro-node including S-CSCF and I-CSCF), and {\em Homestead} (HSS), this latter equipped with {\em Cassandra}, an evolved database used for keeping user information (profiles, private and public identities, etc.). The remaining ancillary nodes reported for the sake of completeness are: {\em Homer}, an XML document management server; {\em Ellis}, a web-based management GUI; {\em Ralf}, a charging/billing system. 
Moreover, we deploy a fourth Linux-based VM hosting a SIP stress tool (SIPp) \cite{sipp} useful to automatically emulate some SIP-based workload.  
Precisely, on the stress VM, we customize a script to automatically inject SIP flows having a Busy Hour Call Attempts (BHCA) value amounting to $2.6$ per user, derived from Voice over LTE (VoLTE) literature \cite{tonse}. All VMs communicate on a LAN environment through a Gigabit Ethernet switch.   
Our testbed allows to gather two types of results: the first one is available in terms of cumulative logs, directly provided by Clearwater, reporting the distribution of SIP calls within pre-defined CSD intervals. Such values are in the order of few hundreds of milliseconds, namely, an order of magnitude lesser than CSD experimented in real environments \cite{dim_lte}. This is absolutely reasonable due to the fact that we operate on a local infrastructure, where no propagation delays, noisy communication lines, nor congestion are present. 

\begin{table}[t!]
	\caption {Parameters values. Note that repair times of Container and Docker have to be interpreted as times needed for a software reboot.}
	\label{tab:params}
	\resizebox{.48\textwidth}{!}{
		\begin{tabular}{c|c|c}
			\hline
			Parameter & Description & Value\\
			\hline
			\\[-8pt]
			$1/\lambda_{CNT}$ & Container MTTF (h) & $500$  \\ \hline
			$1/\lambda_{DCK}$ & Docker daemon MTTF (h) & $1000$  \\ \hline
			$1/\lambda_{VM}$ & Virtual machine MTTF (h) & $2880$  \\ \hline
			$1/\lambda_{HPV}$ & Hypervisor MTTF (h) & $2880$  \\ \hline  
			$1/\lambda_{HW}$ & Hardware MTTF (h) & $60000$  \\ \hline 
			$1/\mu_{CNT}$ & Container MTTR (s) & $2$ \\ \hline
			$1/\mu_{DCK}$ & Docker daemon MTTR (s) & $5$  \\ \hline
			$1/\mu_{VM}$ & Virtual machine MTTR (h) & $1$ \\ \hline
			$1/\mu_{HPV}$ & Hypervisor MTTR (h) & $2$ \\ \hline  
			$1/\mu_{HW}$ & Hardware MTTR (h) & $8$ \\ \hline 
			$A^\dagger$ & Steady-state availability target & $0.9999 \leq A^\dagger \leq 0.999999$ \\ \hline 
			$CSD^\dagger$ & Call Setup Delay target (s) & $0.3$ \\ \hline 
			$\beta_P$  & Avg. service time for P-CSCF (s) & $8.0\cdot10^{-3}$  \\ \hline  
			$\beta_S$  & Avg. service time for S-CSCF (s) & $6.8\cdot10^{-3}$ \\ \hline  
			$\beta_I$  & Avg. service time for I-CSCF (s) & $5.4\cdot10^{-3}$ \\ \hline  
			$\beta_H$  & Avg. service time for HSS (s) & $9.0\cdot10^{-3}$  \\ \hline  
			$\alpha_{ext}$  & Max value of external arrival rate (s$^{-1}$) & $200$  \\ \hline  
			\hline
		\end{tabular}}
	\end{table}

\begin{table*}[t!]
	\caption {A subset of resulting Homogeneous (H) Configurations} \label{tab:homotab}
	\centering
	\resizebox{.99\textwidth}{!}{
		\begin{tabular}{|@{  }Sc|c|c|c|c|c|c|c|}
			\hline
			\textbf{Config.} & \textbf{P-CSCF} & \textbf{S-CSCF}  & \textbf{I-CSCF} & \textbf{HSS} & A$^{(hom)}$ & CSD (s) & Cost  \\
			\hline
			\\[-8pt]
			
			C$_{H}^*$  & NR$_1^{(1)}$, NR$_2^{(1)}$ & NR$_1^{(1)}$, NR$_2^{(1)}$ & NR$_1^{(1)}$, NR$_2^{(1)}$ & NR$_1^{(1)}$, NR$_2^{(2)}$ & {\textcolor{red}{0.99}}  & $0.257$ & $25$ \\ \hline
			
			C$_{1H}$  & NR$_1^{(2)}$, NR$_2^{(2)}$ & NR$_1^{(2)}$, NR$_2^{(2)}$ & NR$_1^{(2)}$, NR$_2^{(2)}$ & NR$_1^{(3)}$, NR$_2^{(3)}$ & $0.99999$ & $0.0493$ & $34$ \\ \hline
			
			C$_{2H}$  & NR$_1^{(2)}$, NR$_2^{(3)}$ & NR$_1^{(2)}$, NR$_2^{(3)}$ & NR$_1^{(2)}$, NR$_2^{(2)}$ & NR$_1^{(4)}$, NR$_2^{(4)}$  & $0.99999$ & $0.0384$ & $38$  \\ \hline
			
			C$_{3H}$ & NR$_1^{(2)}$, NR$_2^{(3)}$ & NR$_1^{(2)}$, NR$_2^{(3)}$ & NR$_1^{(2)}$, NR$_2^{(2)}$ & NR$_1^{(2)}$, NR$_2^{(2)}$, NR$_3^{(3)}$ & $\textcolor{red}{0.9999}$ & $0.0387$ & $39$  \\ \hline
			
			C$_{4H}$  & NR$_1^{(2)}$, NR$_2^{(2)}$ & NR$_1^{(3)}$, NR$_2^{(4)}$ & NR$_1^{(2)}$, NR$_2^{(3)}$ & NR$_1^{(3)}$, NR$_2^{(4)}$  & $0.99999$ & $0.0392$ & $39$ \\ \hline
			
			C$_{5H}$  & NR$_1^{(1)}$, NR$_2^{(1)}$, NR$_3^{(1)}$ & NR$_1^{(1)}$, NR$_2^{(1)}$, NR$_3^{(1)}$ & NR$_1^{(1)}$, NR$_2^{(1)}$, NR$_3^{(2)}$ & NR$_1^{(1)}$, NR$_2^{(1)}$, NR$_3^{(1)}$,  NR$_4^{(2)}$  &  $\textcolor{red}{0.9999}$ & $0.0787$ & $41$  \\ \hline
			
			C$_{6H}$  & NR$_1^{(2)}$, NR$_2^{(2)}$ & NR$_1^{(3)}$, NR$_2^{(3)}$ & NR$_1^{(2)}$, NR$_2^{(2)}$, NR$_3^{(2)}$ & NR$_1^{(2)}$, NR$_2^{(2)}$, NR$_3^{(3)}$   & $0.999999$ & $0.0387$ & $41$ \\ \hline
			
			C$_{7H}$  & NR$_1^{(3)}$, NR$_2^{(3)}$ & NR$_1^{(2)}$, NR$_2^{(3)}$ & NR$_1^{(1)}$, NR$_2^{(1)}$, NR$_3^{(2)}$ & NR$_1^{(2)}$, NR$_2^{(2)}$, NR$_3^{(3)}$   & $\textcolor{red}{0.9999}$ & $0.0368$ & $42$  \\ \hline
			
		\end{tabular}}
	\end{table*}
	
	\begin{table*}[t!]
		\caption {A subset of resulting Co-located (C) Configurations} \label{tab:colotab}
		\centering
		\resizebox{.99\textwidth}{!}{
			\begin{tabular}{|@{  }Sc|c|c|c|c|c|c|c|}
				\hline
				\textbf{Config.} & \textbf{P-CSCF} & \textbf{S-CSCF}  & \textbf{I-CSCF/HSS} (NR sharing) & A$^{(col)}$ & CSD (s) & Cost   \\
				\hline
				\\[-8pt]
				
				C$_{C}^*$  & NR$_1^{(1)}$, NR$_2^{(1)}$ & NR$_1^{(1)}$, NR$_2^{(1)}$ & NR$_1^{(2\textnormal{I},3\textnormal{H})}$ & {\textcolor{red}{0.99}} & $0.257$ & $19$ \\  \hline
				
				C$_{1C}$  & NR$_1^{(2)}$, NR$_2^{(2)}$ & NR$_1^{(2)}$, NR$_2^{(2)}$ & NR$_1^{(2\textnormal{I},3\textnormal{H})}$, NR$_2^{(2\textnormal{I},3\textnormal{H})}$  & $0.99999$ & $0.0493$ & $30$ \\  \hline
				
				C$_{2C}$  & NR$_1^{(2)}$, NR$_2^{(3)}$ & NR$_1^{(2)}$, NR$_2^{(3)}$ & NR$_1^{(2\textnormal{H})}$, NR$_2^{(1\textnormal{I},3\textnormal{H})}$, NR$_3^{(2\textnormal{I},3\textnormal{H})}$  & $0.99999$ & $0.0384$ & $36$ \\  \hline
				
				C$_{3C}$  & NR$_1^{(2)}$, NR$_2^{(2)}$ & NR$_1^{(1)}$, NR$_2^{(1)}$, NR$_3^{(1)}$ & NR$_1^{(1\textnormal{H})}$, NR$_2^{(2\textnormal{H})}$, NR$_3^{(2\textnormal{I},2\textnormal{H})}$, NR$_4^{(3\textnormal{I},3\textnormal{H})}$ & $0.99999$ & $0.0568$ & $38$  \\  \hline
				
				C$_{4C}$  & NR$_1^{(3)}$, NR$_2^{(3)}$ & NR$_1^{(3)}$, NR$_2^{(3)}$ & NR$_1^{(1\textnormal{I},2\textnormal{H})}$, NR$_2^{(1\textnormal{I},3\textnormal{H})}$, NR$_3^{(2\textnormal{I},3\textnormal{H})}$  & $\textcolor{red}{0.9999}$ & $0.0351$ & $38$ \\  \hline
				
				C$_{5C}$  & NR$_1^{(2)}$, NR$_2^{(2)}$ & NR$_1^{(4)}$, NR$_2^{(4)}$ & NR$_1^{(1\textnormal{H})}$, NR$_2^{(2\textnormal{H})}$, NR$_3^{(2\textnormal{I},2\textnormal{H})}$, NR$_4^{(2\textnormal{I},3\textnormal{H})}$ & $0.99999$ & $0.0423$ & $40$ \\  \hline
				
				C$_{6C}$  & NR$_1^{(2)}$, NR$_2^{(2)}$, NR$_3^{(4)}$ & NR$_1^{(2)}$, NR$_2^{(2)}$ & NR$_1^{(1\textnormal{H})}$, NR$_2^{(2\textnormal{H})}$, NR$_3^{(2\textnormal{I},2\textnormal{H})}$, NR$_4^{(2\textnormal{I},3\textnormal{H})}$ & $0.99999$ & $0.0402$ & $42$  \\  \hline
				
				C$_{7C}$  & NR$_1^{(1)}$, NR$_2^{(1)}$, NR$_3^{(1)}$, NR$_4^{(2)}$ & NR$_1^{(1)}$, NR$_2^{(1)}$, NR$_3^{(1)}$, NR$_4^{(1)}$ & NR$_1^{(2\textnormal{I},1\textnormal{H})}$, NR$_2^{(2\textnormal{I},2\textnormal{H})}$, NR$_3^{(2\textnormal{I},2\textnormal{H})}$  & $0.999999$ & $0.0389$ & $42$  \\  \hline
				
			\end{tabular}}
		\end{table*} 

The second set of results concerns an estimate of the service times per node, that we empirically derive by isolating the traffic flows at single nodes through the network sniffer Wireshark. 
In Table \ref{tab:params} we summarize all the parameters values where: parameters concerning failure/repair of various layers, such as the Mean-Time-to-Failure (MTTF) and the Mean-Time-to-Repair (MTTR) come both from expert hints and technical literature \cite{trivedi2012,sebastio18}, whereas, parameters related to service times are directly estimated from our experimental testbed. Parameter $\alpha_{ext}=200$ s$^{-1}$ has been chosen so to guarantee the queueing stability condition for a maximum external load (worst  case), whereas the $CSD^\dagger$ amounting to $0.3$ s has been derived by performing workload stress tests. 

Before delving into the numerical analysis, let us clarify how the costs are calculated. Being in practice impossible to estimate the cost of an application running on top of a container (including license costs, designing/coding costs, maintenance costs, etc.), we adopt the assumption that a container is worth half the remaining layers composing an NR. In a sense, we consider that the expenses due to the underlying NR infrastructure are amortized being in the cloud. Such an assumption is supported by information gathered from the Microsoft Azure platform in some realistic service scenarios, where a pricing simulation revealed that a licensed containerized instance costs about $1.5$ (namely $1+0.5$) the price of a non-licensed containerized instance. 
Just for example, let us consider a generic homogeneous softIMS configuration where: $i)$ both the P-CSCF and the S-CSCF nodes are composed of 3 NRs (each one with 2 containers on top); $ii)$ the I-CSCF node is composed of 4 NRs (each one with 3 containers on top); $iii)$ the	HSS node is composed of 3 NRs (each one with 2 containers on top). We have that: Cost(P-CSCF)$=$Cost(S-CSCF)$=1 \cdot 3+0.5\cdot3\cdot2 = 6$, Cost(I-CSCF)$=1\cdot4+0.5\cdot4\cdot3 =10$, Cost(HSS)$=1\cdot3+0.5\cdot3\cdot2 = 6$. Accordingly, the cost for this specific softIMS configuration amounts to $6+6+10+6=28$, in line with the pricing proposed by Azure.

\begin{figure*}[t!]
	\centering
	\begin{minipage}[t]{0.23\textwidth}
		\includegraphics[width=\textwidth]{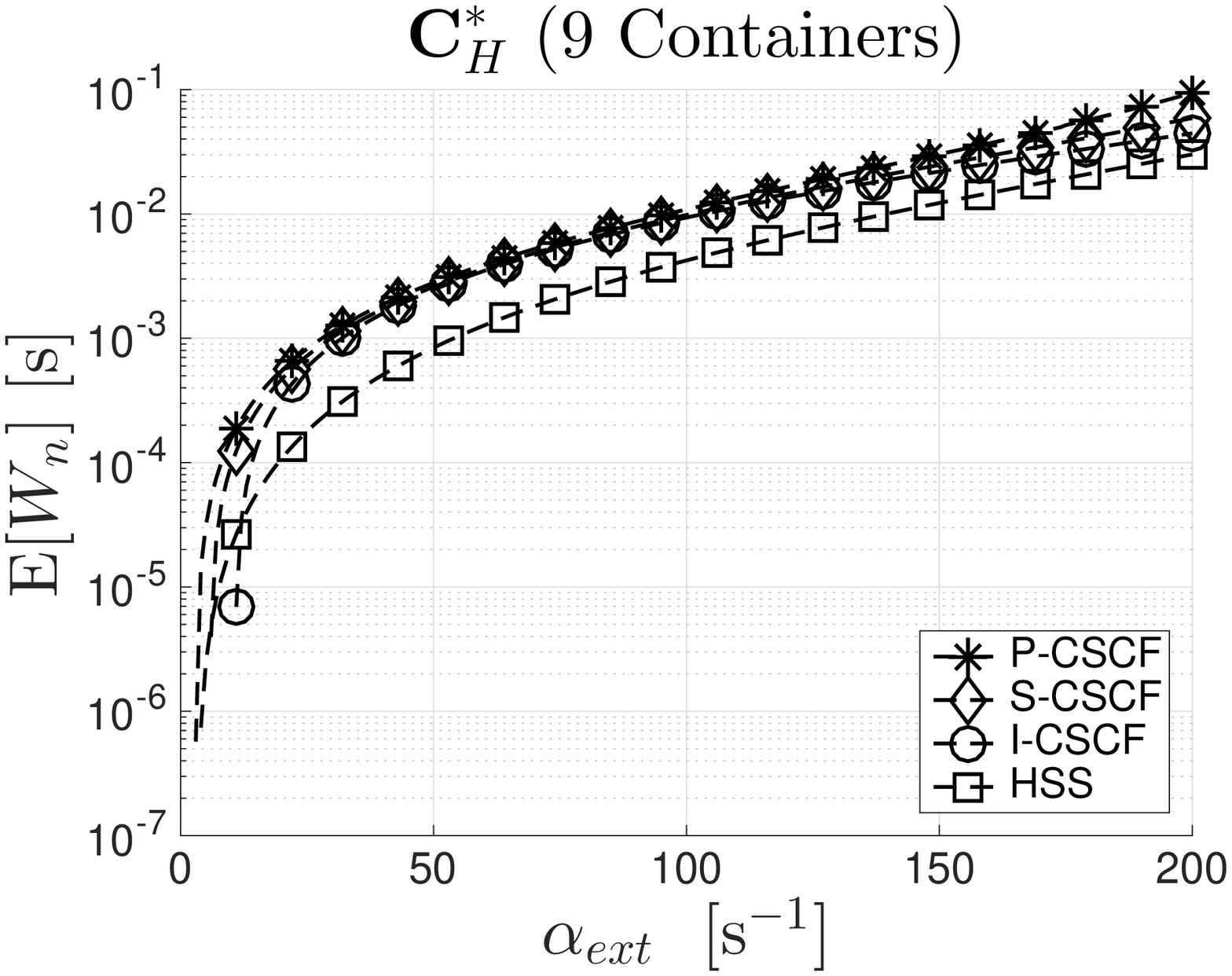}
	\end{minipage} \hspace{2pt}
	\begin{minipage}[t]{0.23\textwidth}
		\includegraphics[width=\textwidth]{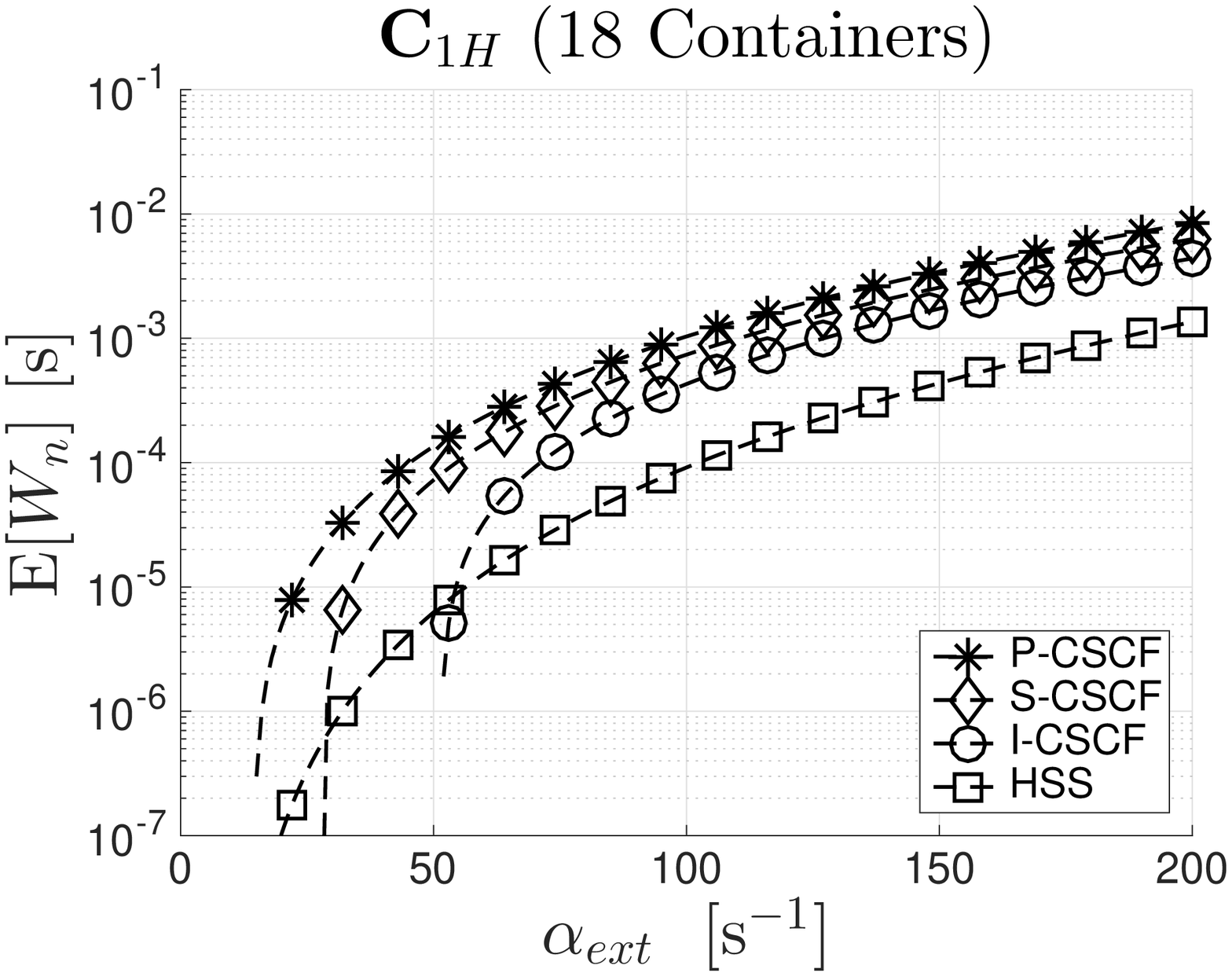}
	\end{minipage} \hspace{2pt}
	\begin{minipage}[t]{0.23\textwidth}
		\includegraphics[width=\textwidth]{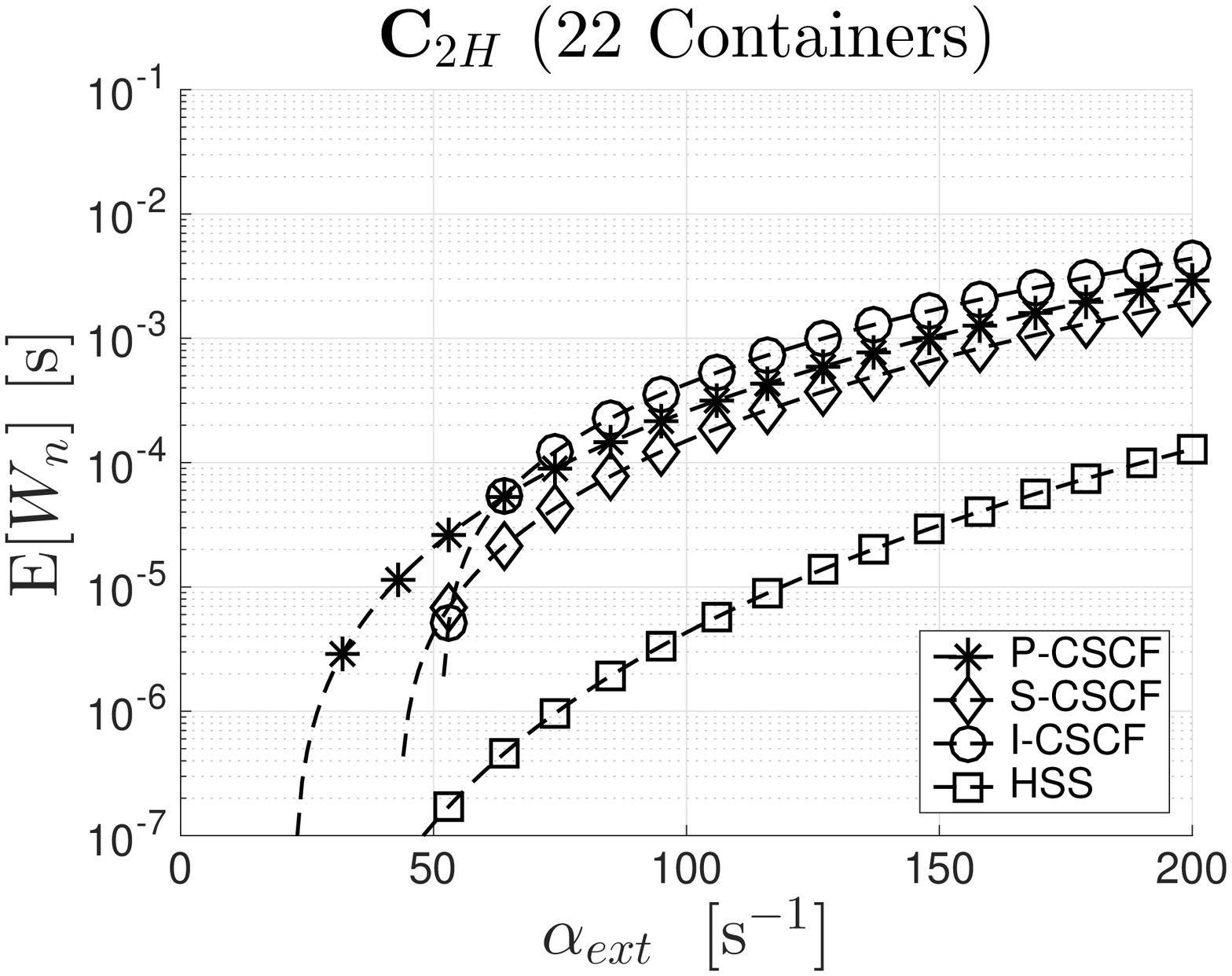}
	\end{minipage} \hspace{2pt}
	\begin{minipage}[t]{0.23\textwidth}
		\includegraphics[width=\textwidth]{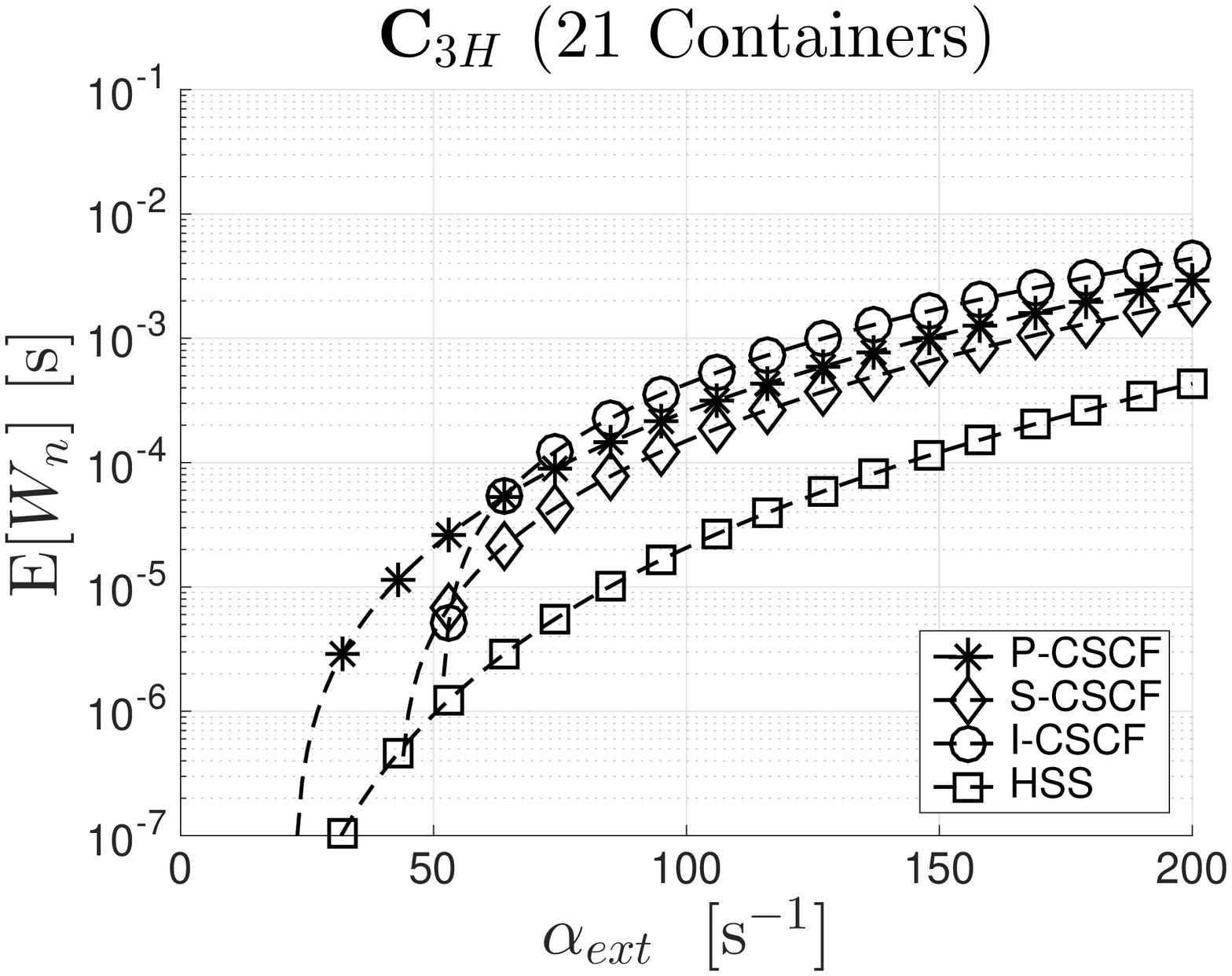}
	\end{minipage} \hspace{2pt}
	\LineSep
	\begin{minipage}[t]{0.23\textwidth}
		\includegraphics[width=\textwidth]{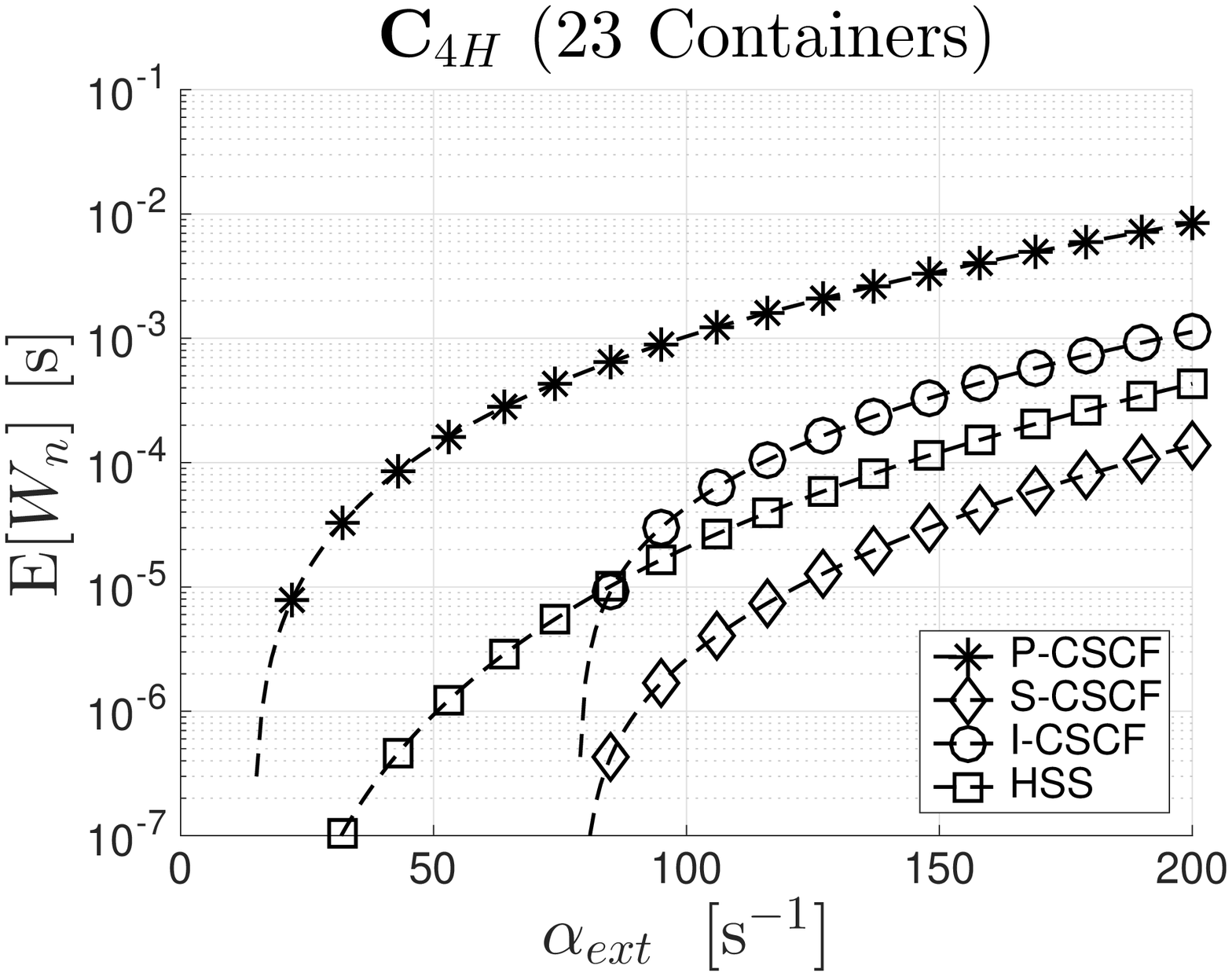}
	\end{minipage} \hspace{2pt}
	\begin{minipage}[t]{0.23\textwidth}
		\includegraphics[width=\textwidth]{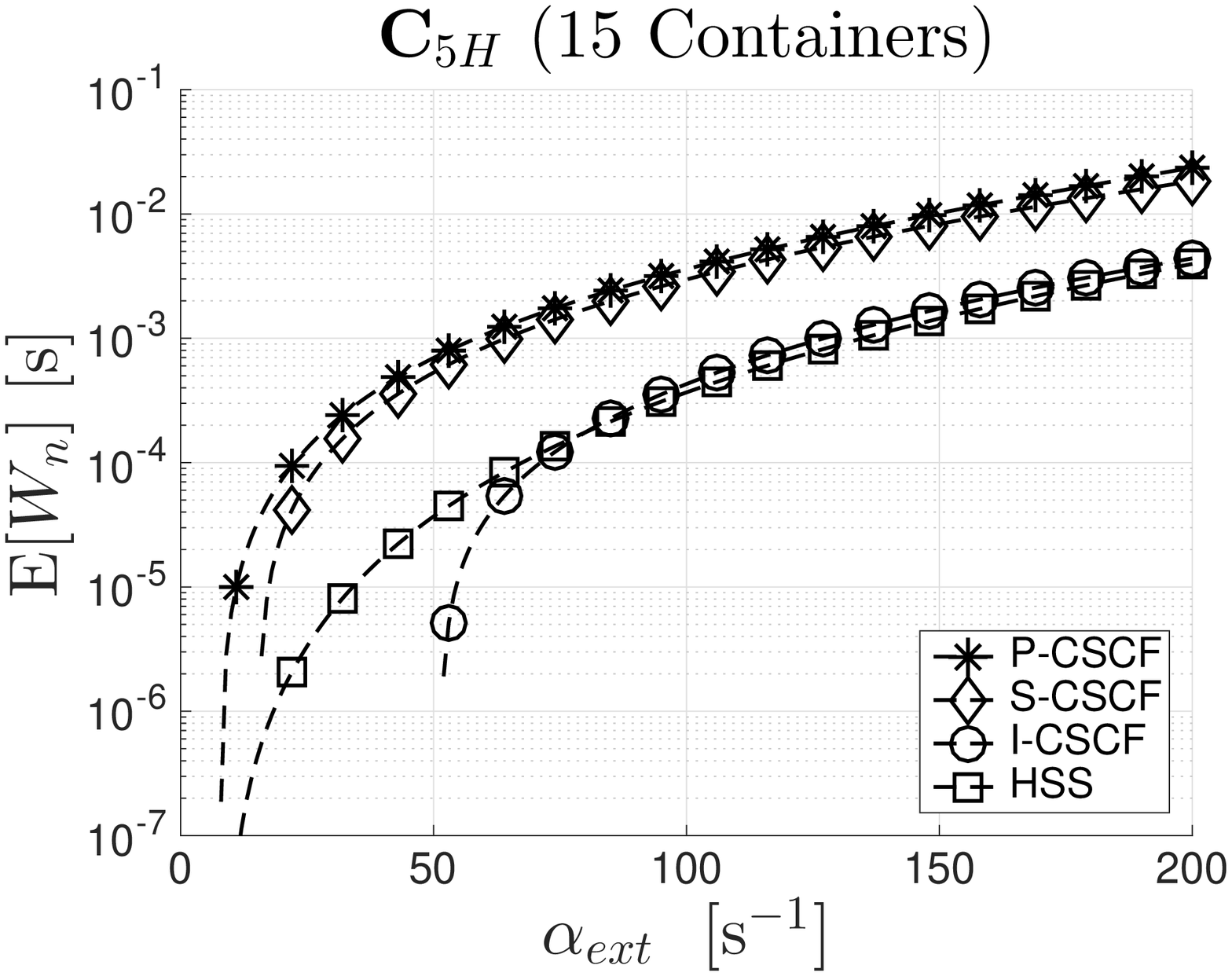}
	\end{minipage} \hspace{2pt}
	\begin{minipage}[t]{0.23\textwidth}
		\includegraphics[width=\textwidth]{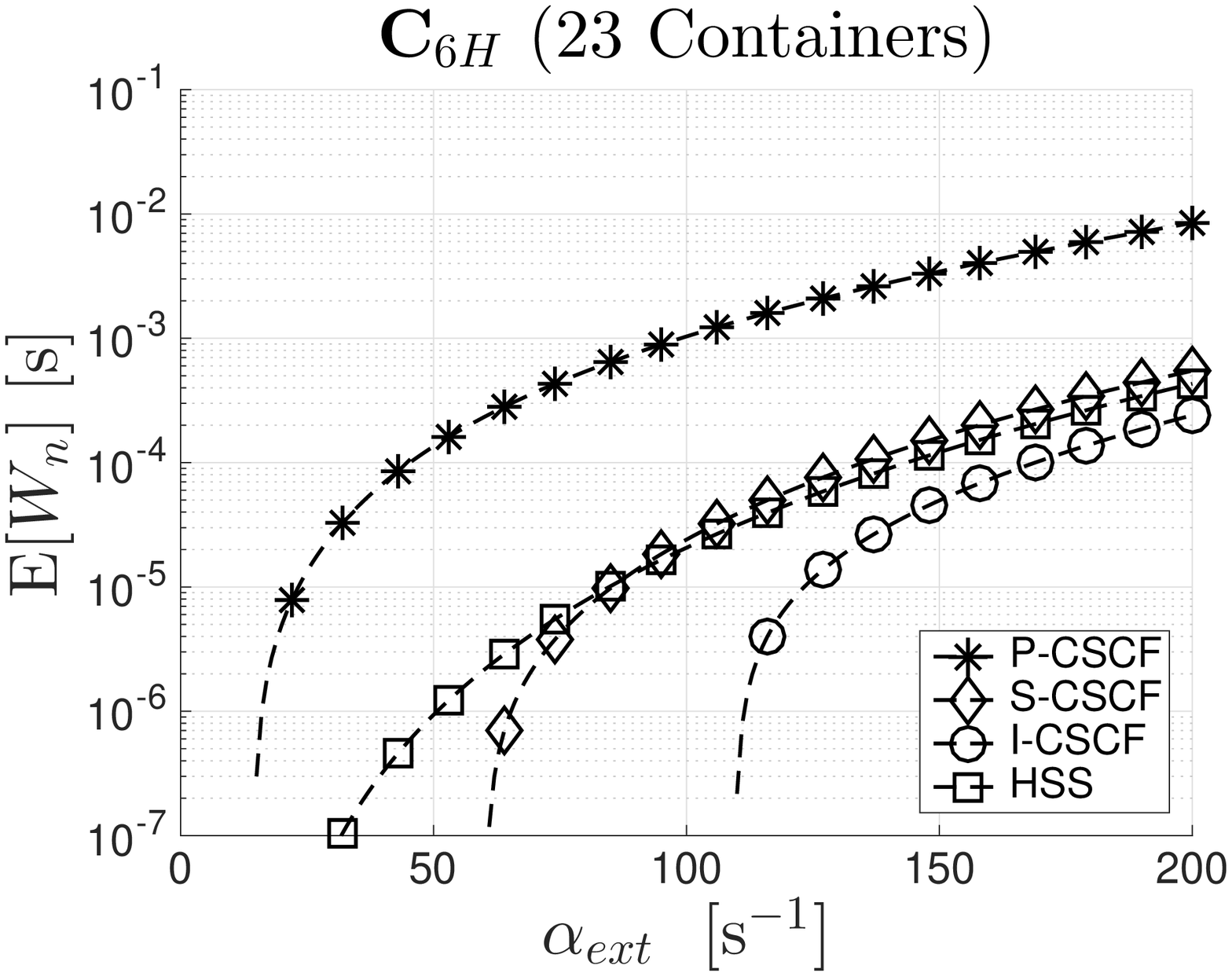}
	\end{minipage} \hspace{2pt}
	\begin{minipage}[t]{0.23\textwidth}
		\includegraphics[width=\textwidth]{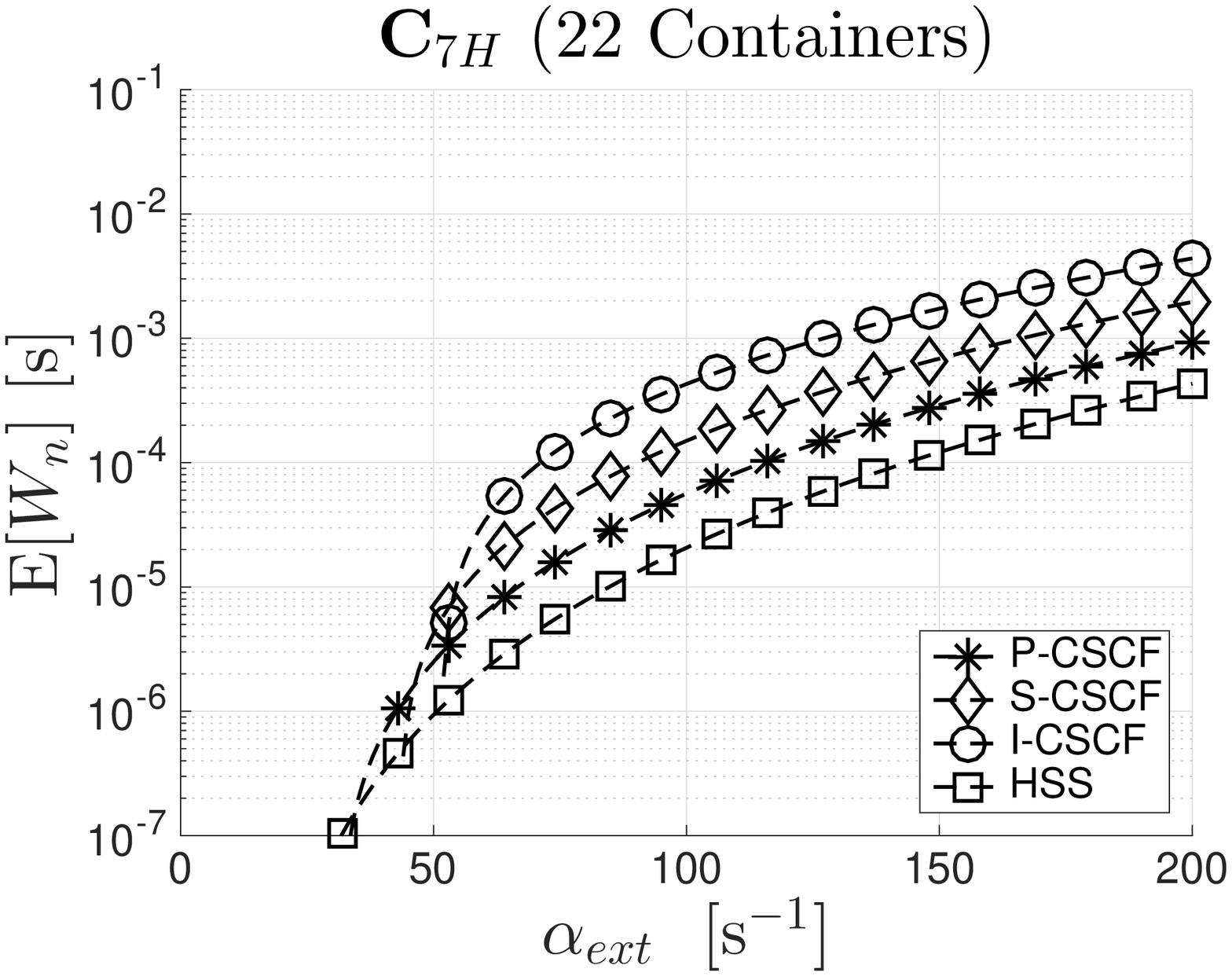}
	\end{minipage} \hspace{2pt}
	\caption{Homogeneous configurations: Mean waiting times per node $\mathbb{E}[W_n]$ for configurations from $C^*_H$ (top left panel) to $C_{7H}$ (bottom right panel).}
	\label{fig:all_w}
\end{figure*}
\begin{figure*}[t!]
	\centering
	\begin{minipage}[t]{0.23\textwidth}
		\includegraphics[width=\textwidth]{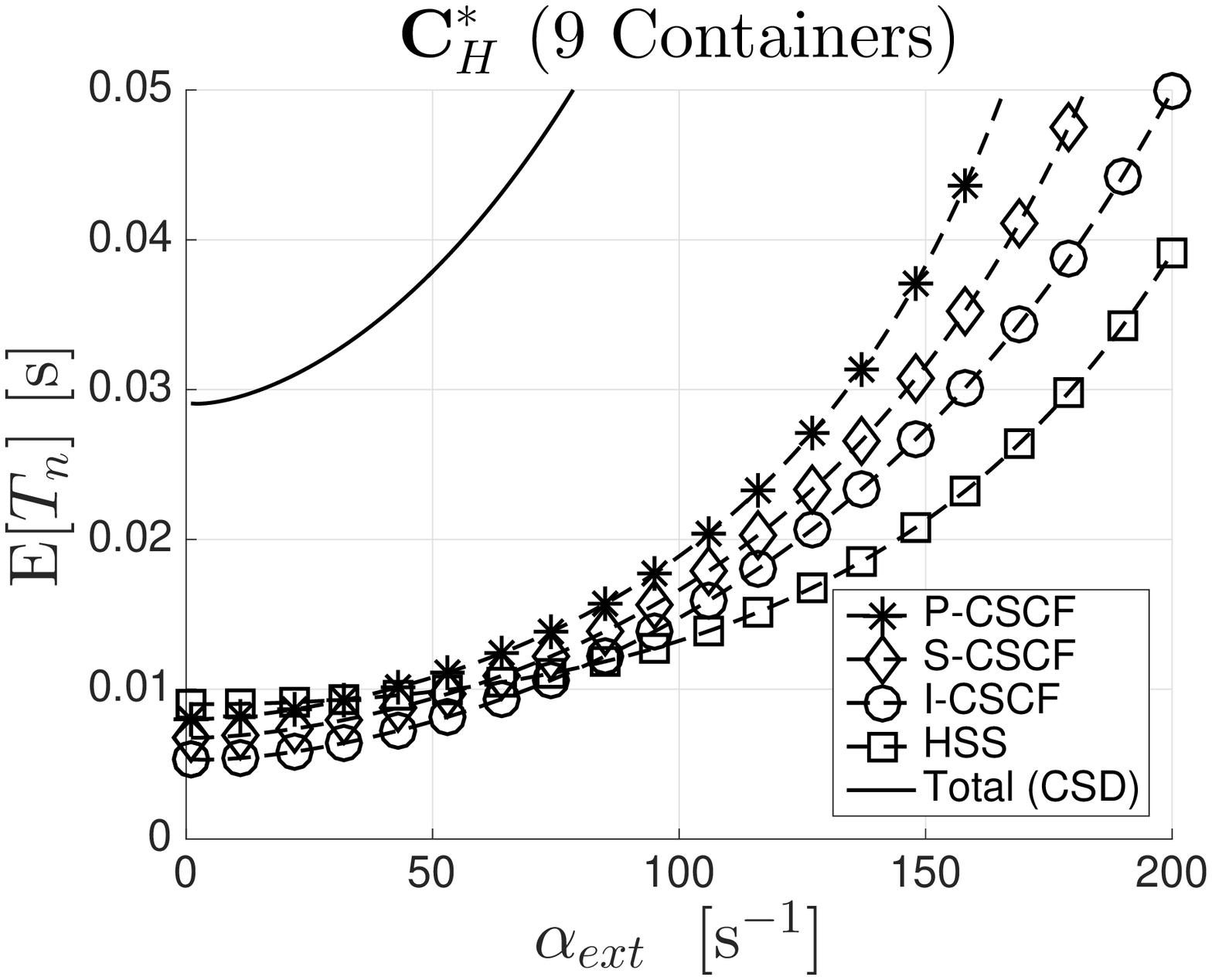}
	\end{minipage} \hspace{2pt}
	\begin{minipage}[t]{0.23\textwidth}
		\includegraphics[width=\textwidth]{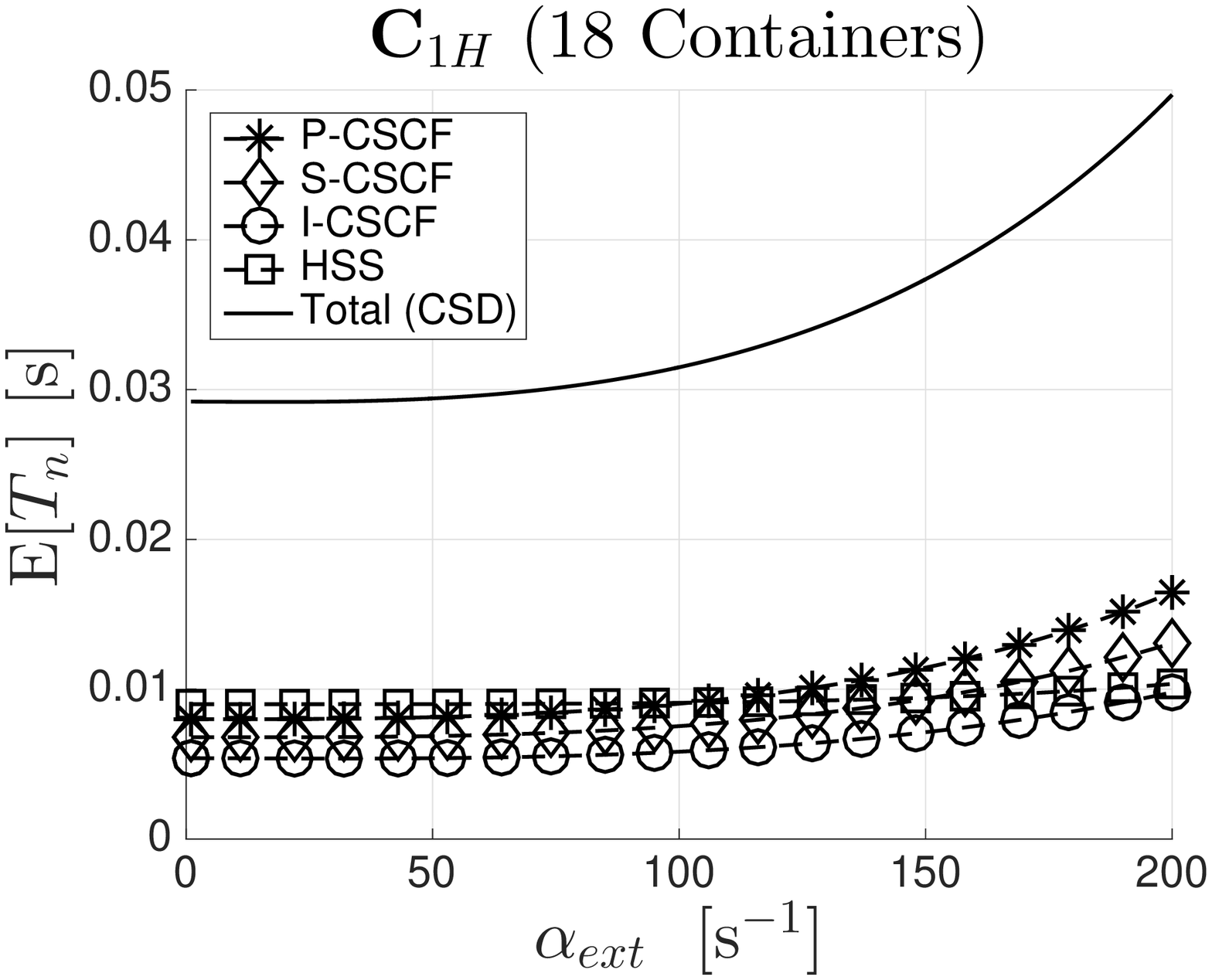}
	\end{minipage} \hspace{2pt}
	\begin{minipage}[t]{0.23\textwidth}
		\includegraphics[width=\textwidth]{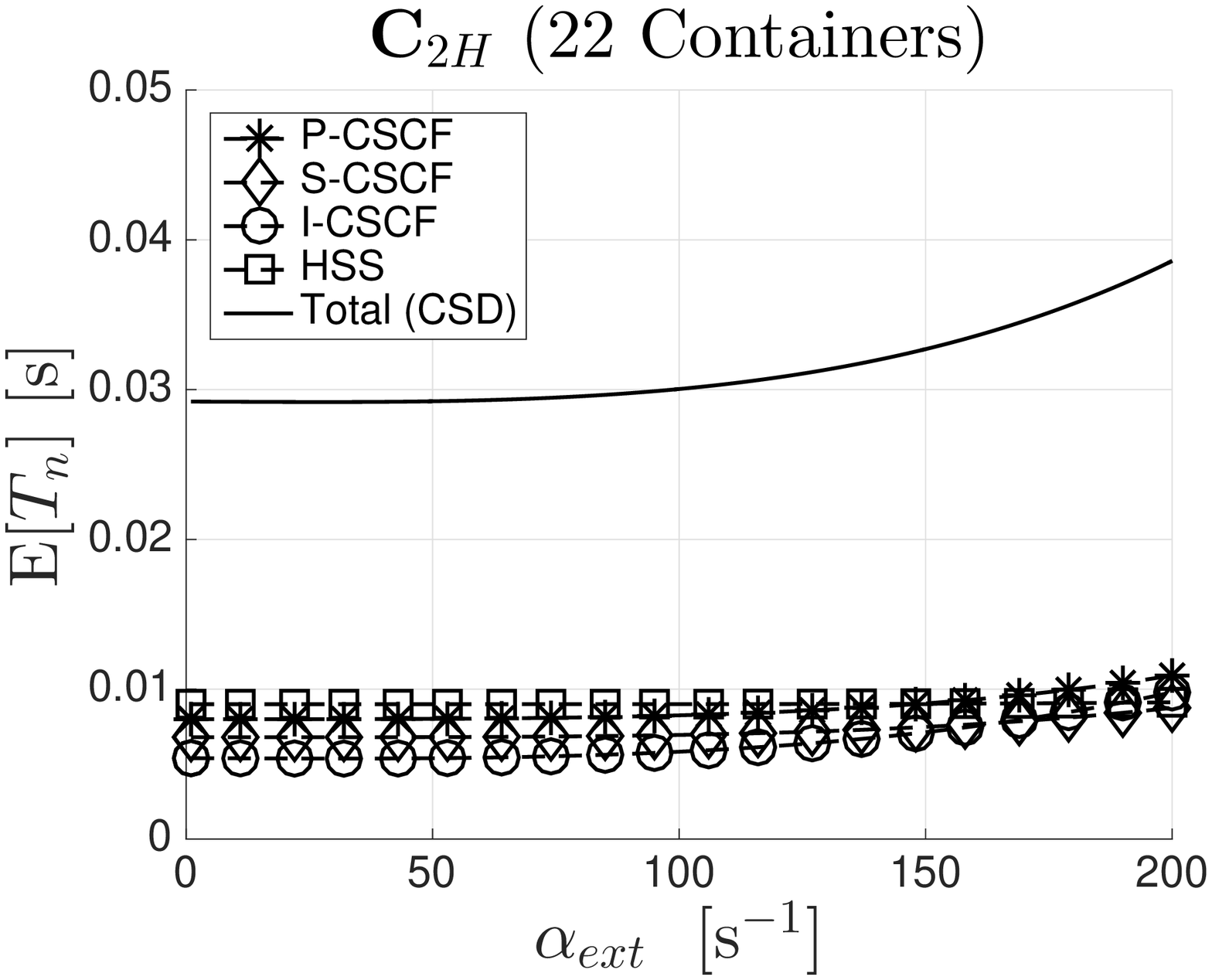}
	\end{minipage} \hspace{2pt}
	\begin{minipage}[t]{0.23\textwidth}
		\includegraphics[width=\textwidth]{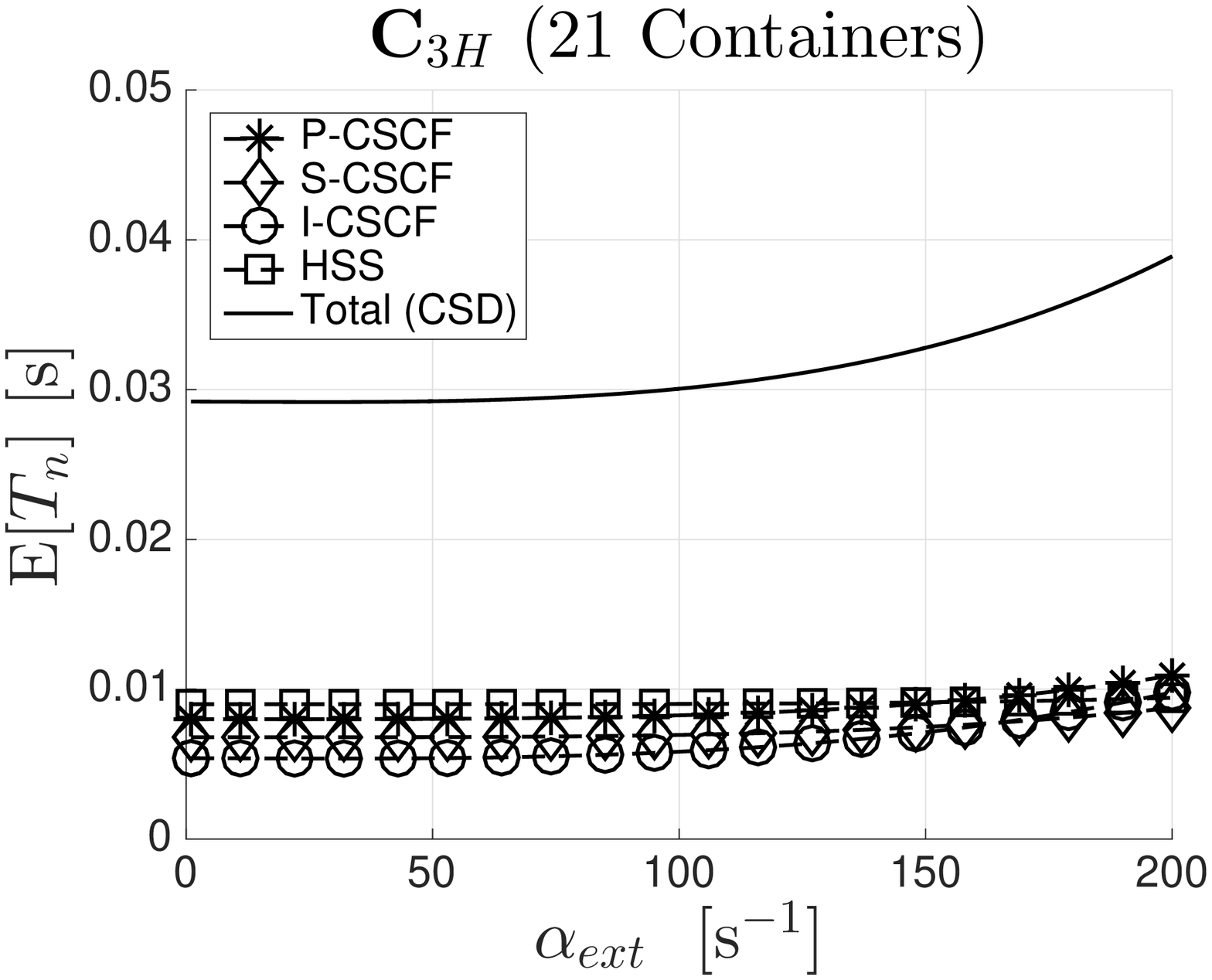}
	\end{minipage} \hspace{2pt}
	\LineSep
	\begin{minipage}[t]{0.23\textwidth}
		\includegraphics[width=\textwidth]{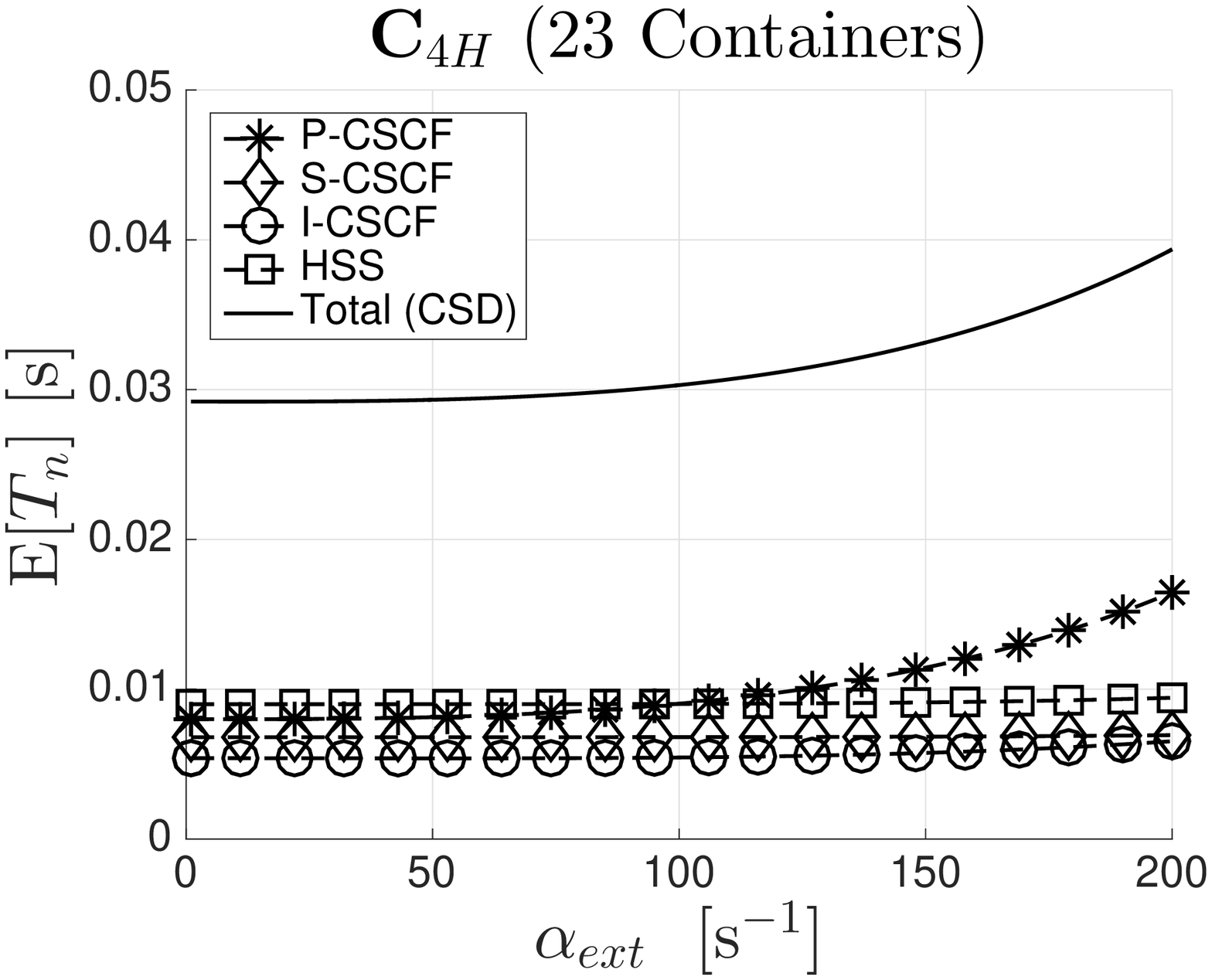}
	\end{minipage} \hspace{2pt}
	\begin{minipage}[t]{0.23\textwidth}
		\includegraphics[width=\textwidth]{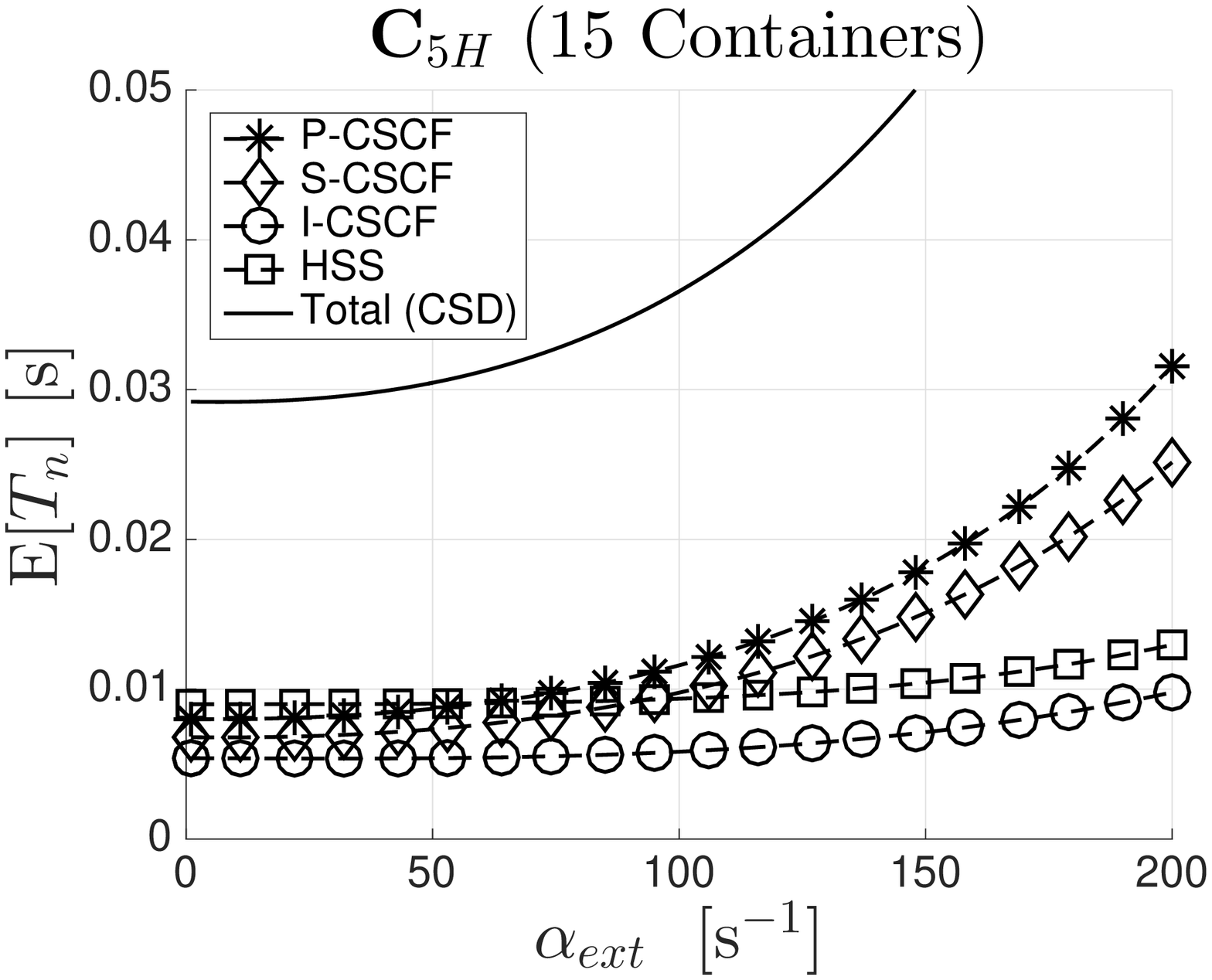}
	\end{minipage} \hspace{2pt}
	\begin{minipage}[t]{0.23\textwidth}
		\includegraphics[width=\textwidth]{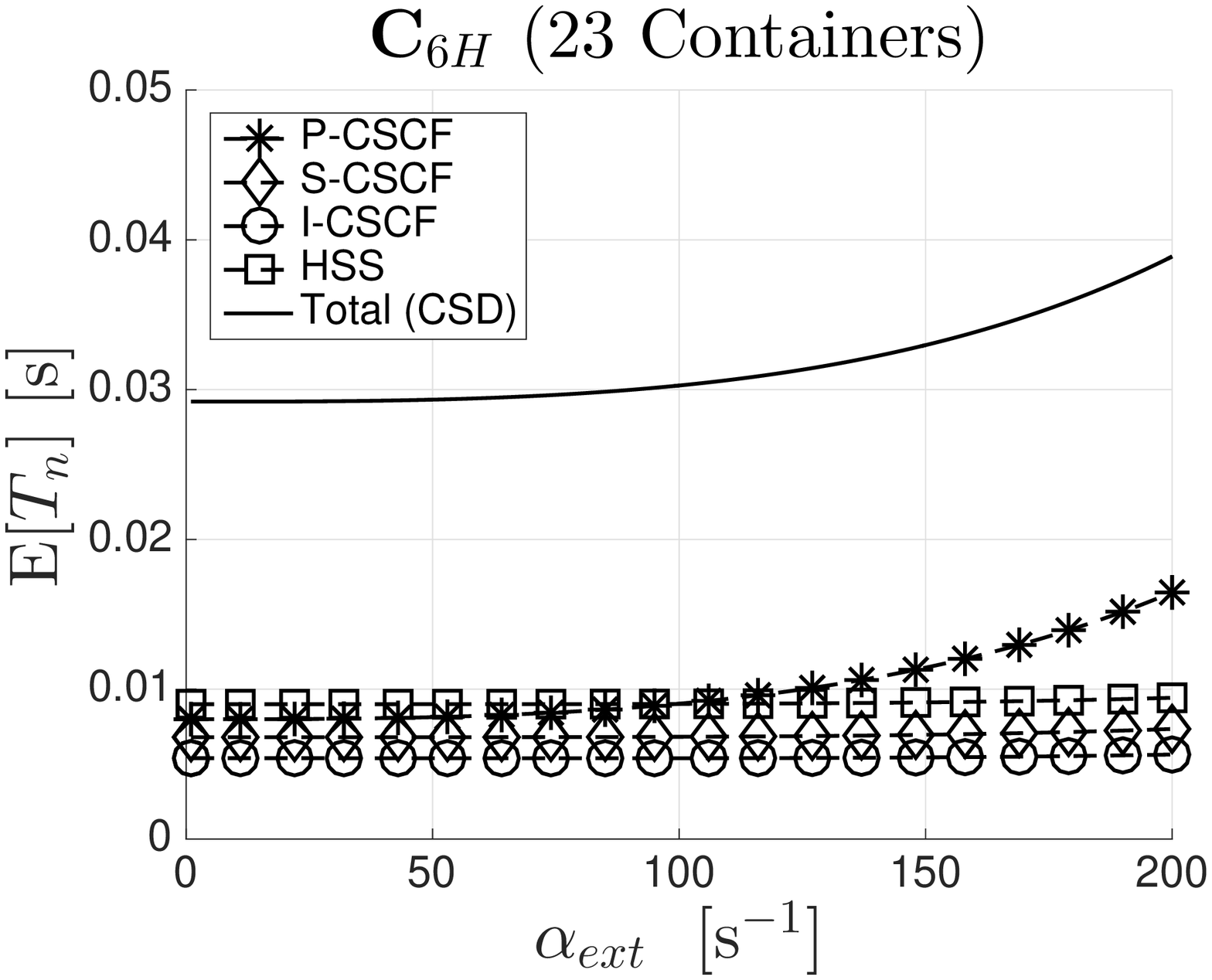}
	\end{minipage} \hspace{2pt}
	\begin{minipage}[t]{0.23\textwidth}
		\includegraphics[width=\textwidth]{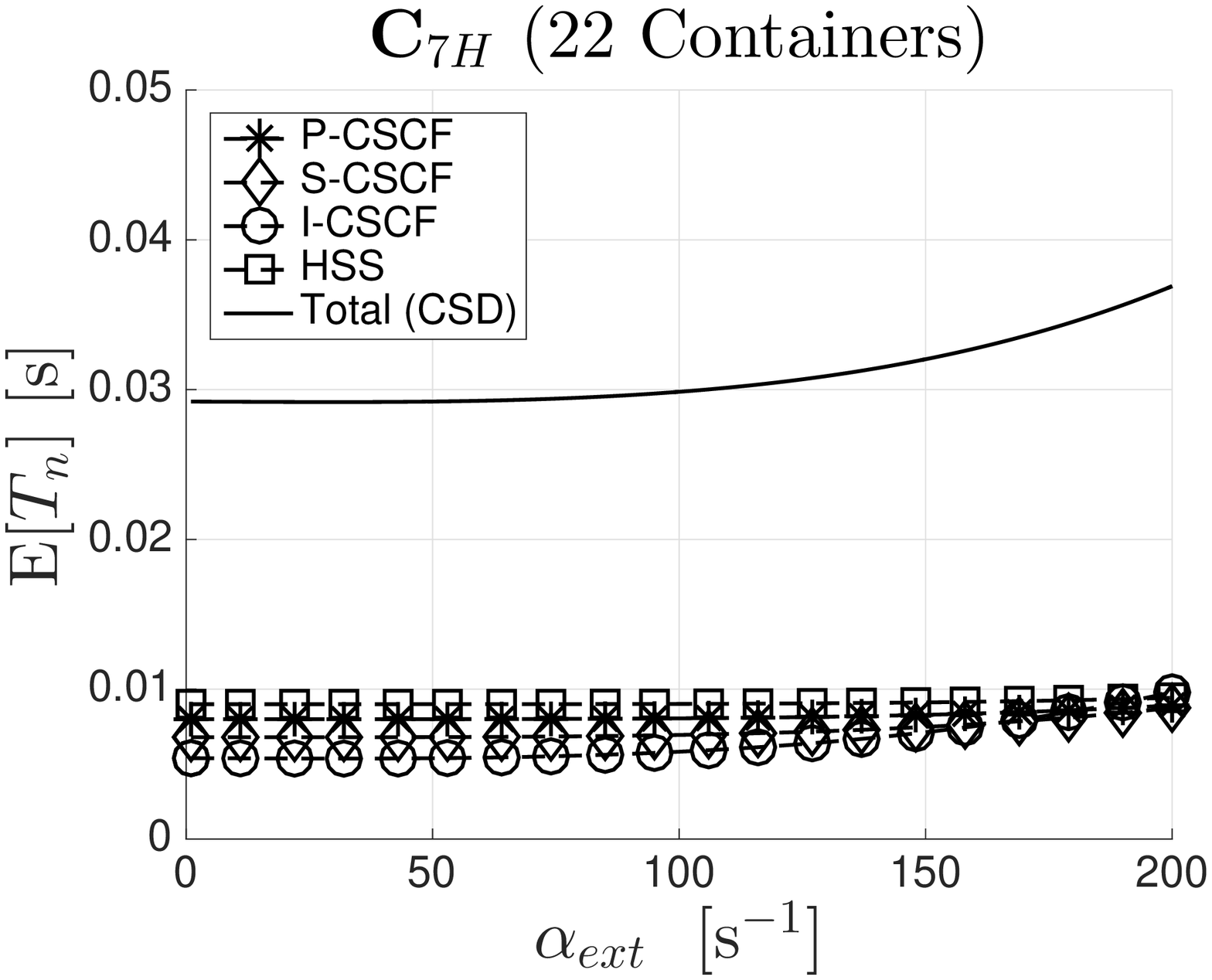}
	\end{minipage} \hspace{2pt}
	\caption{Homogeneous configurations: Mean response times per node $\mathbb{E}[T_n]$ for configurations from $C^*_H$ (top left panel) to $C_{7H}$ (bottom right panel).}
	\label{fig:all_t}
\end{figure*}

Let us now analyze some numerical results. Table \ref{tab:homotab} and \ref{tab:colotab} report, respectively, a selected subset of softIMS configurations for homogeneous and co-located deployments (ordered by increasing costs), obtained by applying in cascade OptCNT and OptSearchchain algorithms. For each table, the first column includes the configuration identifier; columns from $2$ to $5$ report the number of NRs per node, and the number of containers hosted on top of each NR (denoted by a superscript); column $6$ contains the availability value; column $7$ reports the CSD evaluated in case of high external load ($\alpha_{ext}=200$ s$^{-1}$); column $8$ includes the cost per configuration. It is worth highlighting that, in such analysis, we consider a variable availability steady-state target (from four nines to six nines) in order to have a broader set of configurations to compare for revealing some interesting facts. An exception is given by the two configurations $C^*_H$ and $C^*_C$ that we insert on top of Tables \ref{tab:homotab} and \ref{tab:colotab}, respectively. Such configurations have been obtained by only accounting for the output of OptCNT algorithm (namely $c^*_n$), thus they satisfy the CSD target constraint but exhibit an unacceptable availability value of $0.99$ (note that all the values below the five nines are highlighted in red).  
By analyzing remaining configurations, it is possible to notice that an increasing total number of containers or NRs do not necessarily imply better availability values. This is due to the fact that not all redundancy strategies (combination of redundant NRs and containers) are suitable. As an extreme example, let us consider configurations C$_{6H}$ and C$_{7H}$ in Table \ref{tab:homotab}. With the same number of NRs allocated per node, and with only one more container ($23$ total containers for C$_{6H}$, and $22$ total containers for C$_{7H}$ ), C$_{6H}$ exhibits a challenging $0.999999$ availability value (six nines), w.r.t. C$_{7H}$ whose availability value amounts to $0.9999$ (four nines). Such a surprising availability ``jump'' is the consequence of a bad allocation strategy of containers in C$_{7H}$ configuration, where more redundant containers have been assigned to P-CSCF by depriving of containers the I-CSCF node. Another example of bad allocation strategy is given by configuration C$_{5H}$, where a more robust infrastructure layer (namely, more NRs) brings the only effect of increasing cost, but not the availability (which gets stuck at $0.9999$).

Among the remaining homogeneous configurations, the most appealing is C$_{1H}$ which exhibits, a five nines availability value at the minimum cost. Similar considerations can be derived by the analysis of co-located configurations (Table \ref{tab:colotab}).
Moreover, comparing similar homogeneous and co-located configurations, it is quite easy to verify that the latter are cheaper due to the presence of a shared infrastructural part. For instance, in the comparison between C$_{1H}$ and C$_{1C}$ and between C$_{2H}$ and C$_{2C}$, the two co-located configurations come out as winners by saving $2$ NRs and $1$ NR, respectively. Obviously, more sophisticated criteria (than those merely based on costs) can be adopted to select optimal configurations, provided that the desired availability requirements are fulfilled. Among additional criteria we mention: $i)$ selecting configurations whose availability value is fully (and not barely) satisfied, $ii)$ choosing configurations whose distribution of NRs/containers is well-balanced so to simplify the application of automatic procedures; $iii)$ preferring configurations whose CSD value is quite far from the target constraint.  
If the containers distribution affects the availability results in the homogeneous/co-located comparison, no global effects are visible in the CSD variation. This is absolutely reasonable since, from a queueing system perspective, each node can be seen as a black box, thus, the increasing number of containers (it does not matter if they belong to separate or shared NRs) helps to reduce the whole CSD value. This latter, as expected, always satisfy the desired performance requirement (CSD $<0.3$ s) as correctly guaranteed by the OptCNT routine.

\begin{figure*}[t!]
	\centering
	\begin{minipage}[t]{0.23\textwidth}
		\includegraphics[width=\textwidth]{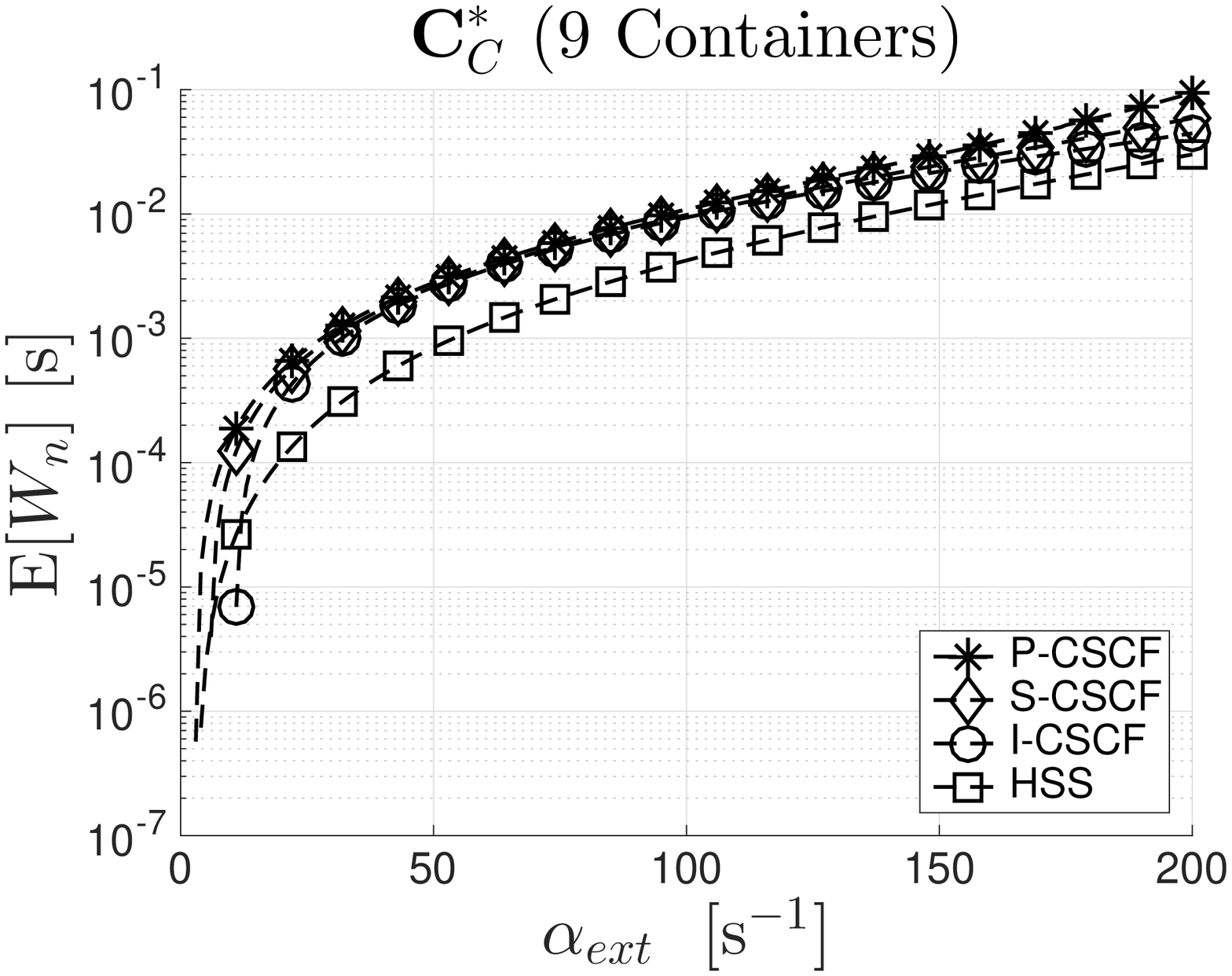}
	\end{minipage} \hspace{2pt}
	\begin{minipage}[t]{0.23\textwidth}
		\includegraphics[width=\textwidth]{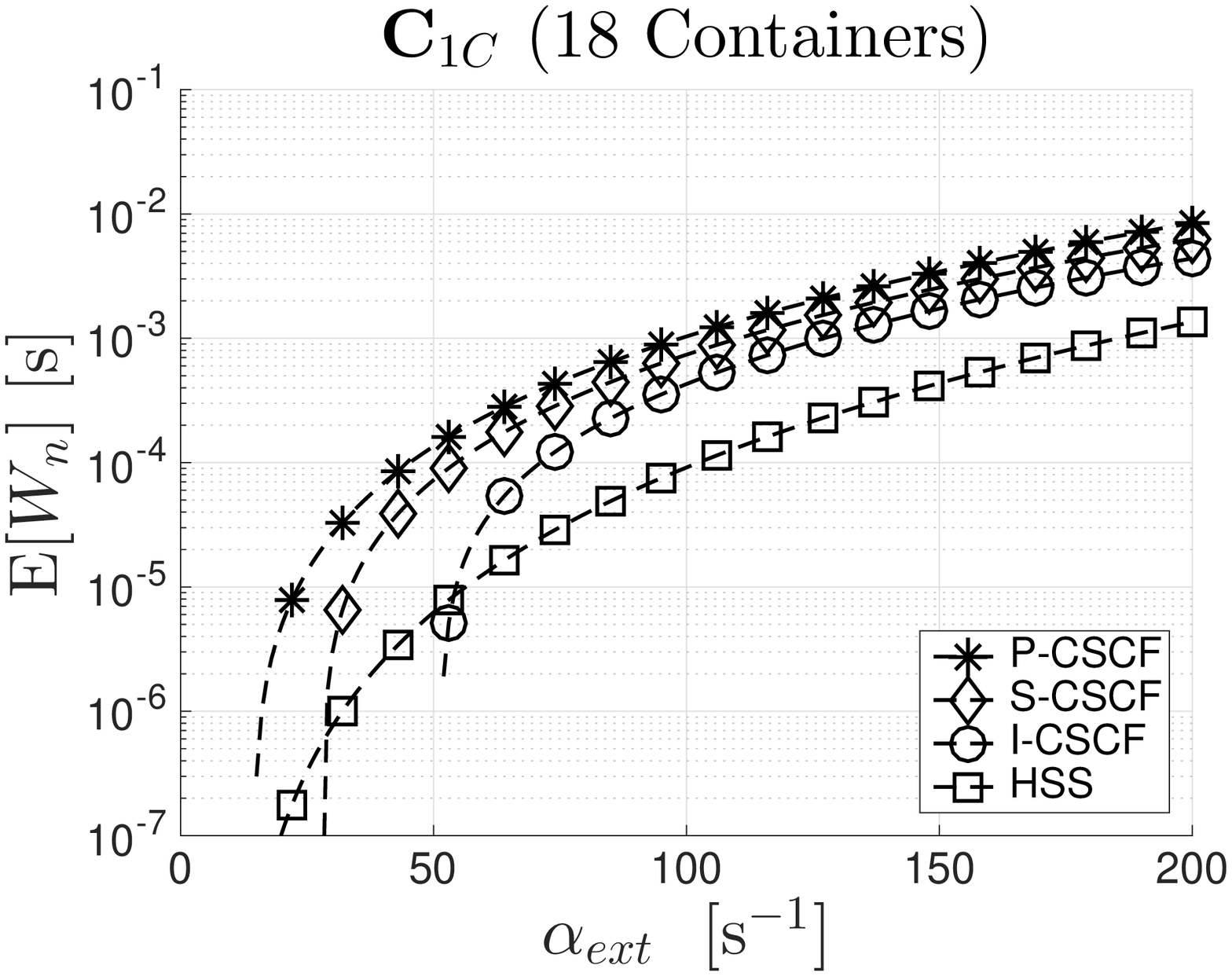}
	\end{minipage} \hspace{2pt}
	\begin{minipage}[t]{0.23\textwidth}
		\includegraphics[width=\textwidth]{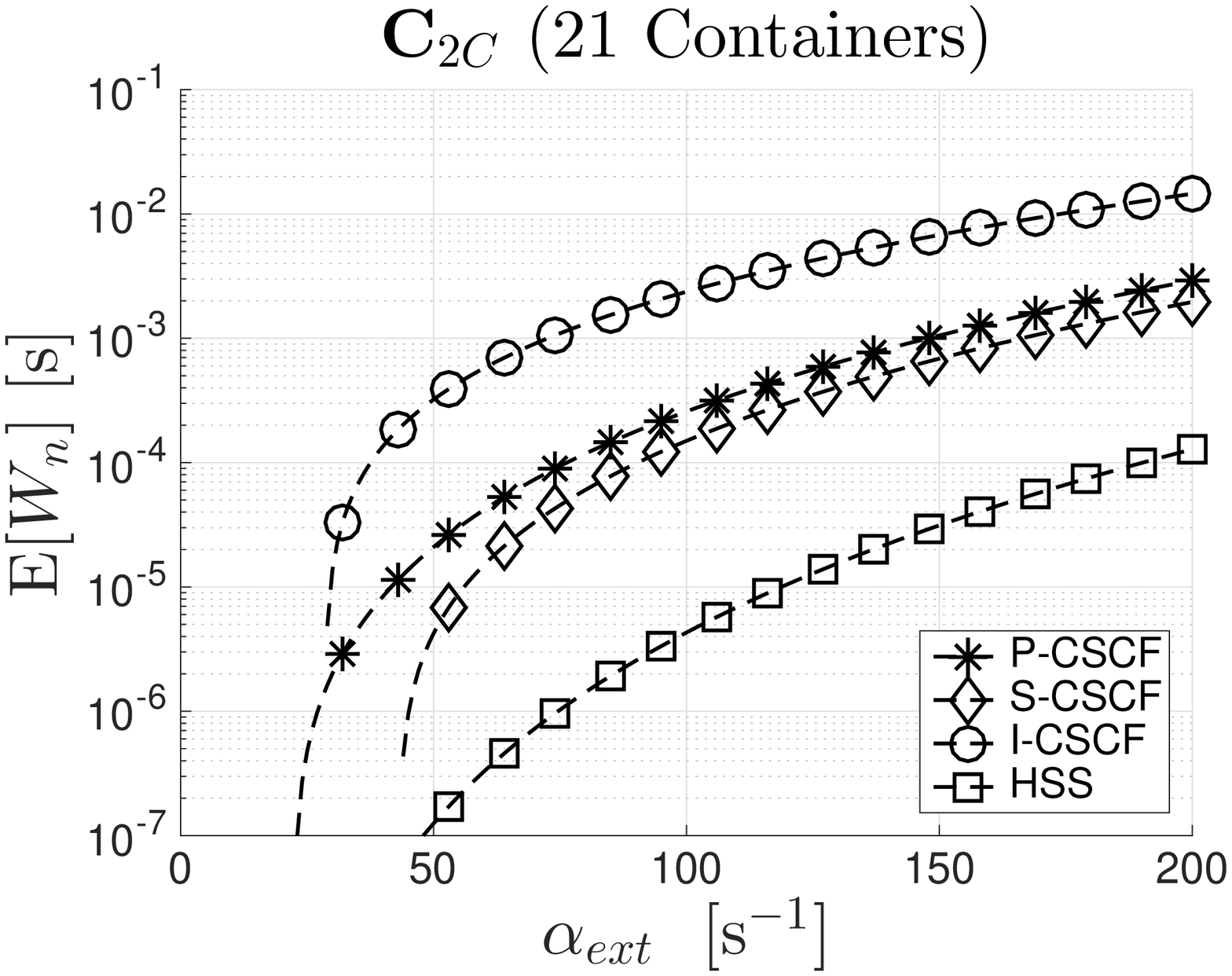}
	\end{minipage} \hspace{2pt}
	\begin{minipage}[t]{0.23\textwidth}
		\includegraphics[width=\textwidth]{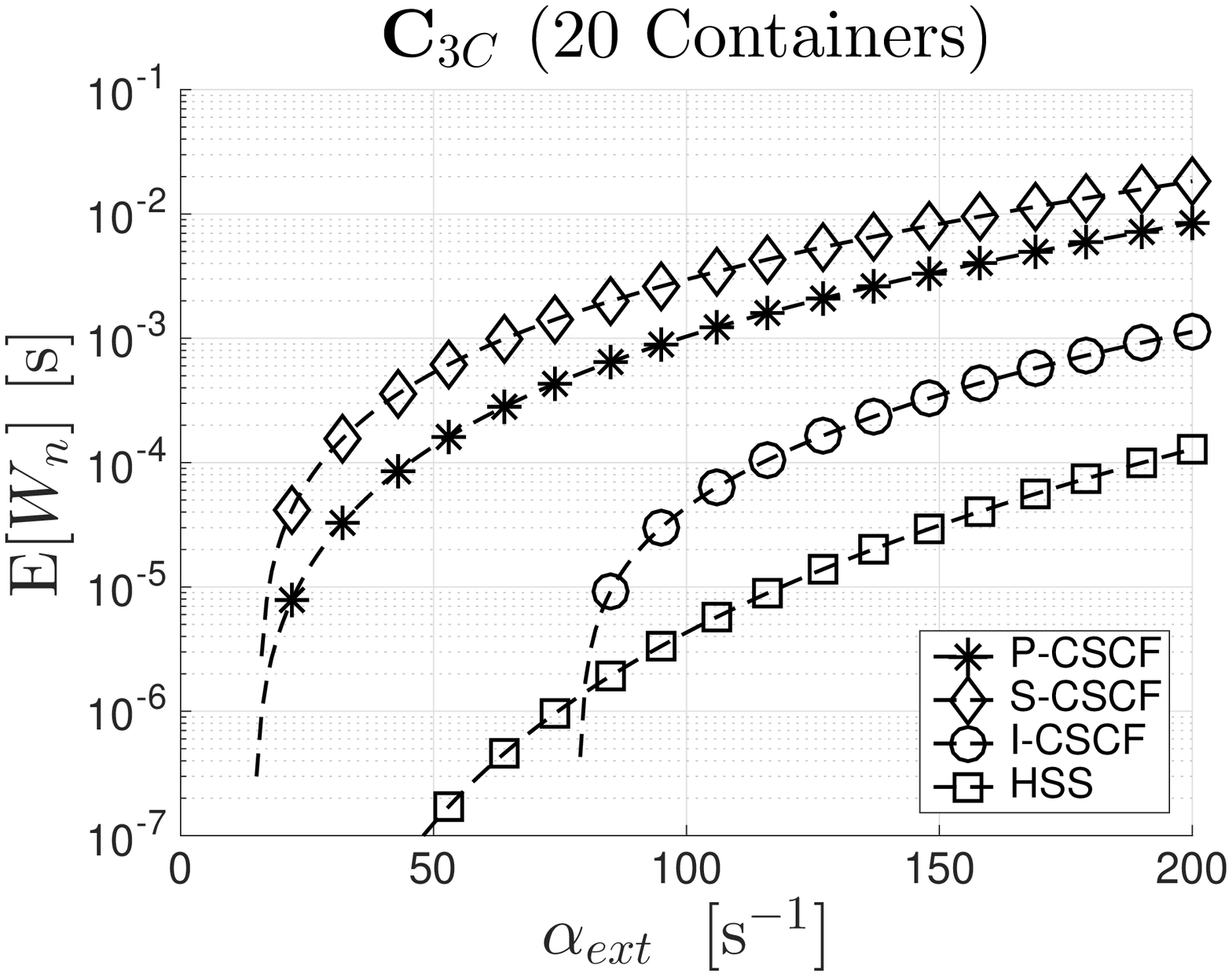}
	\end{minipage} \hspace{2pt}
	\LineSep
	\begin{minipage}[t]{0.23\textwidth}
		\includegraphics[width=\textwidth]{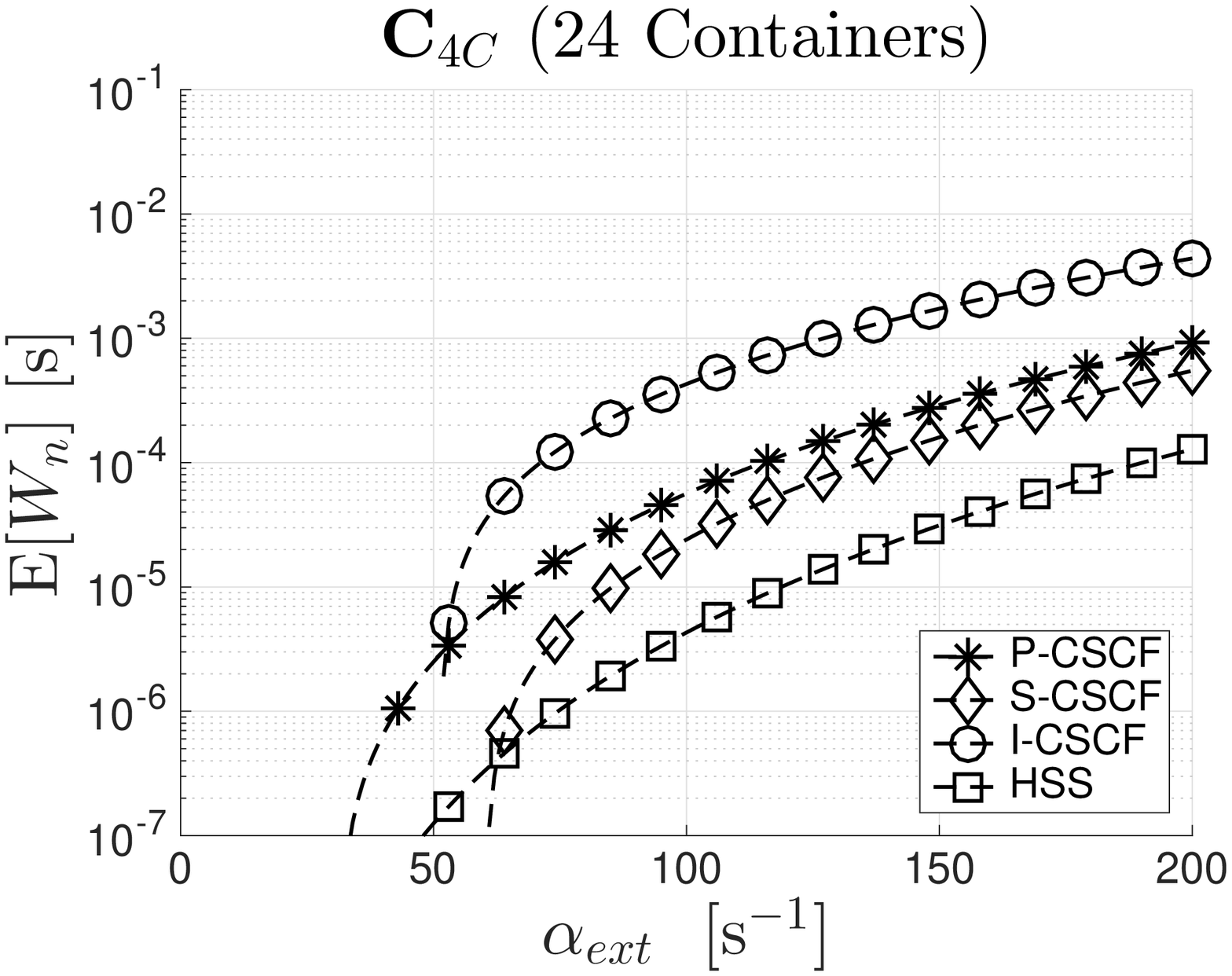}
	\end{minipage} \hspace{2pt}
	\begin{minipage}[t]{0.23\textwidth}
		\includegraphics[width=\textwidth]{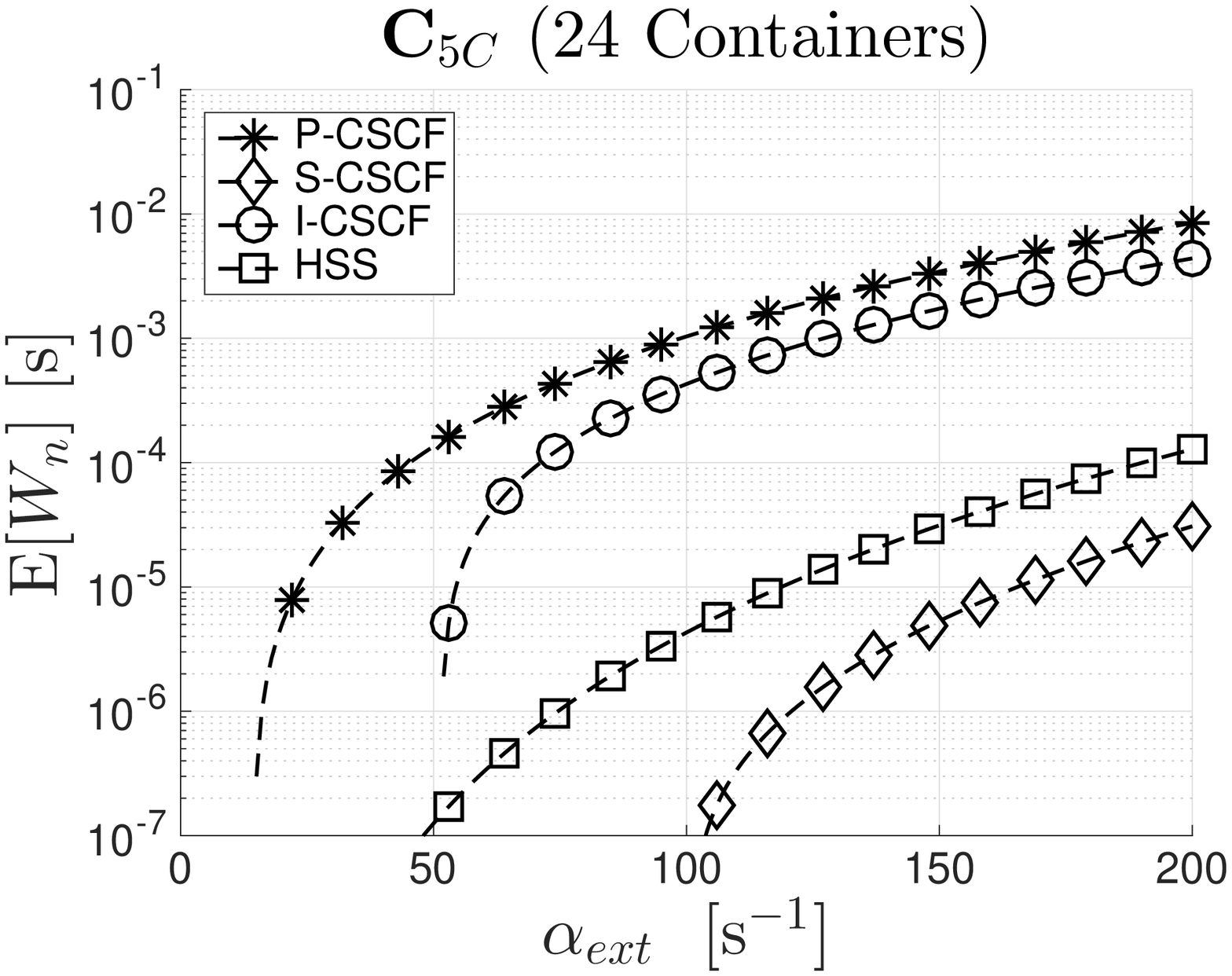}
	\end{minipage} \hspace{2pt}
	\begin{minipage}[t]{0.23\textwidth}
		\includegraphics[width=\textwidth]{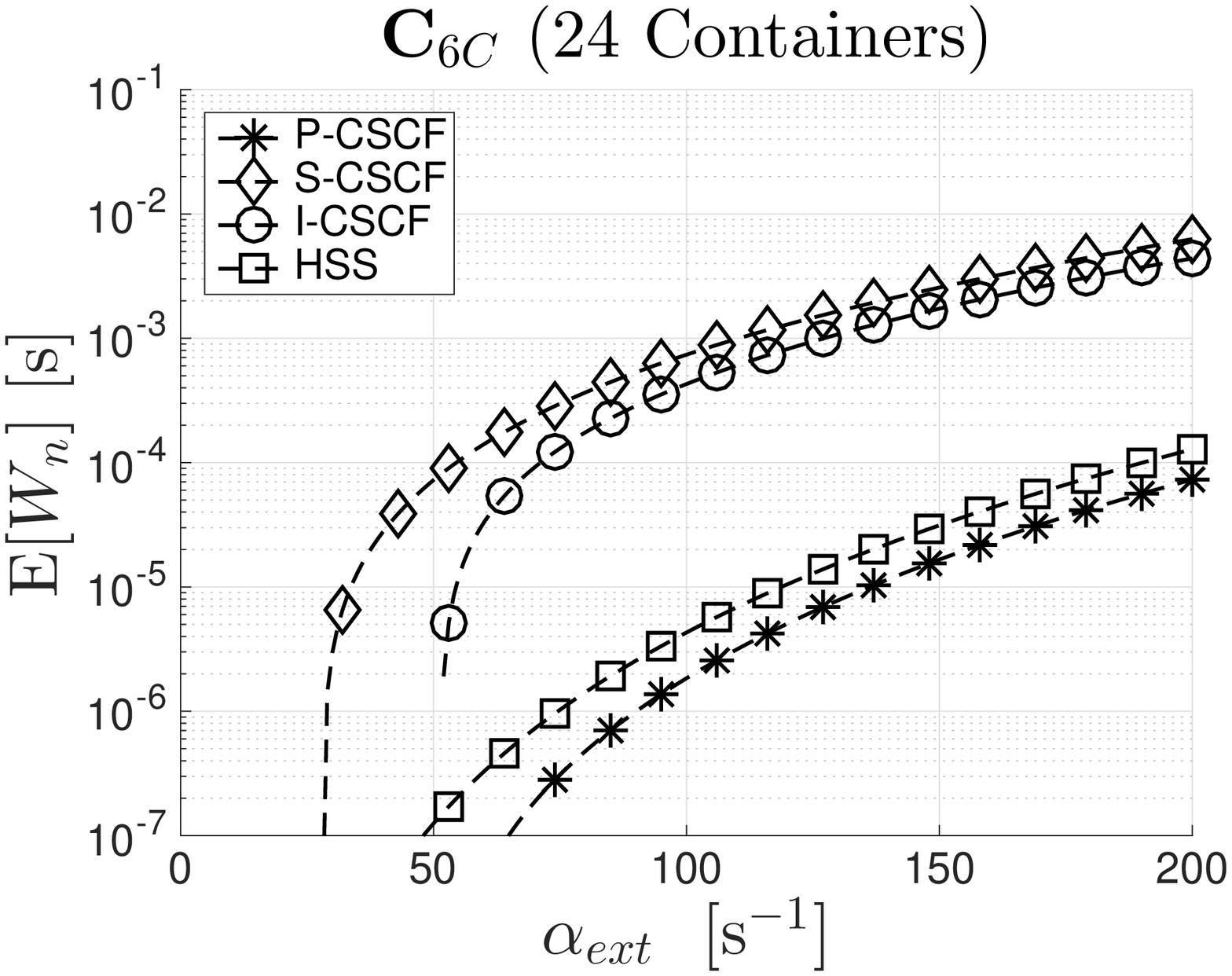}
	\end{minipage} \hspace{2pt}
	\begin{minipage}[t]{0.23\textwidth}
		\includegraphics[width=\textwidth]{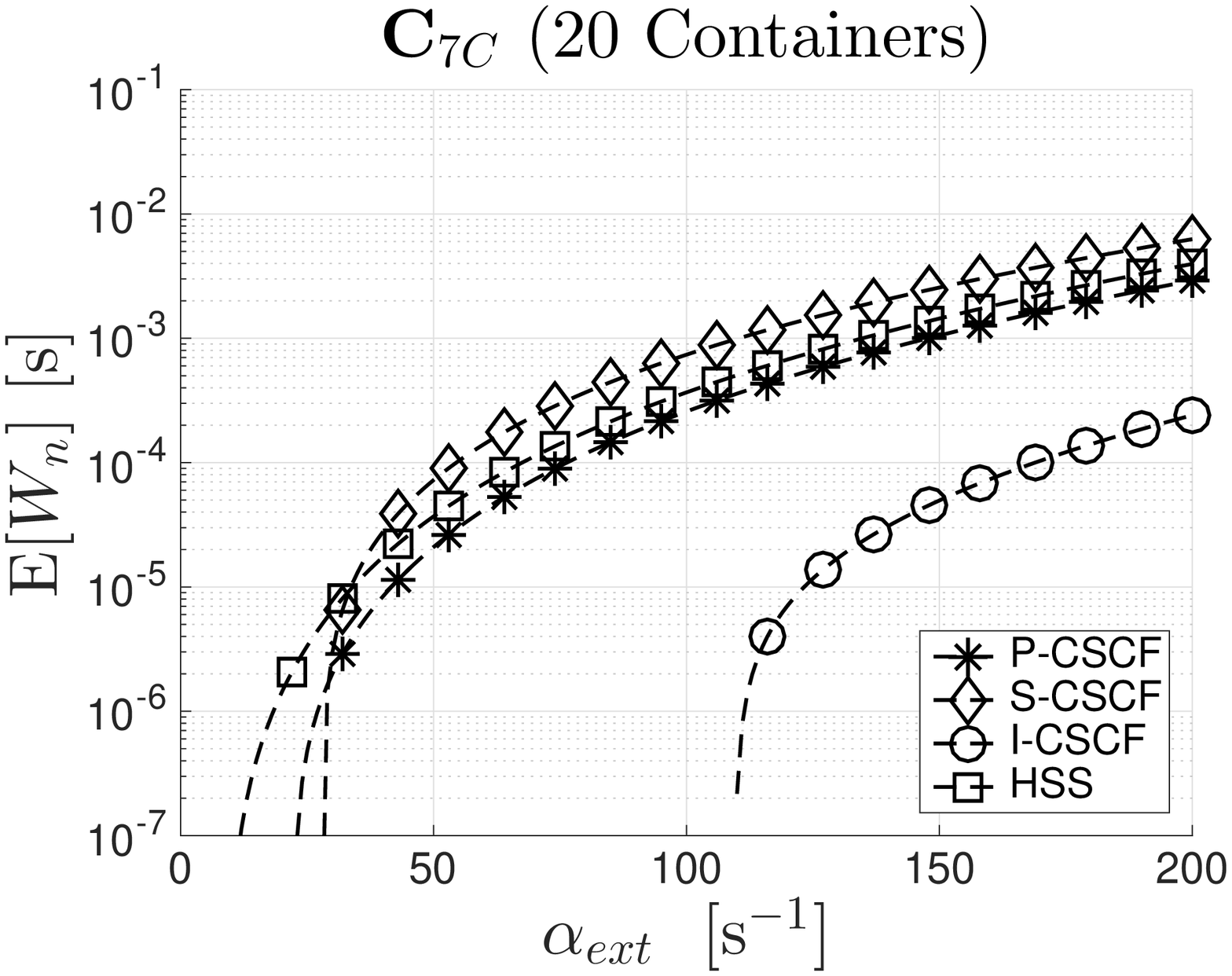}
	\end{minipage} \hspace{2pt}
	\caption{Co-located configurations: Mean waiting times per node $\mathbb{E}[W_n]$ for configurations from $C^*_C$ (top left panel) to $C_{7C}$ (bottom right panel).}
	\label{fig:all_w_coloc}
\end{figure*}
\begin{figure*}[t!]
	\centering
	\begin{minipage}[t]{0.23\textwidth}
		\includegraphics[width=\textwidth]{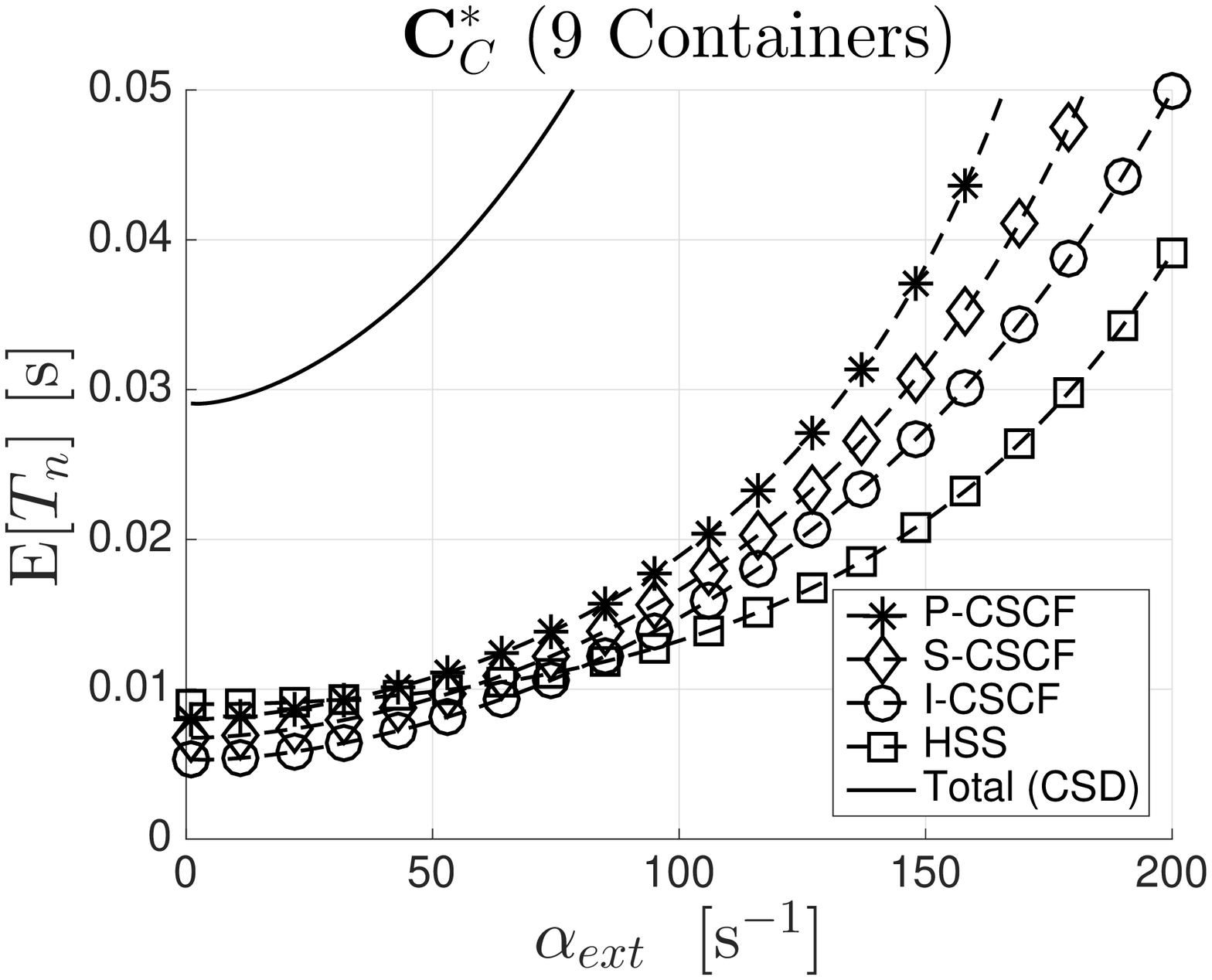}
	\end{minipage} \hspace{2pt}
	\begin{minipage}[t]{0.23\textwidth}
		\includegraphics[width=\textwidth]{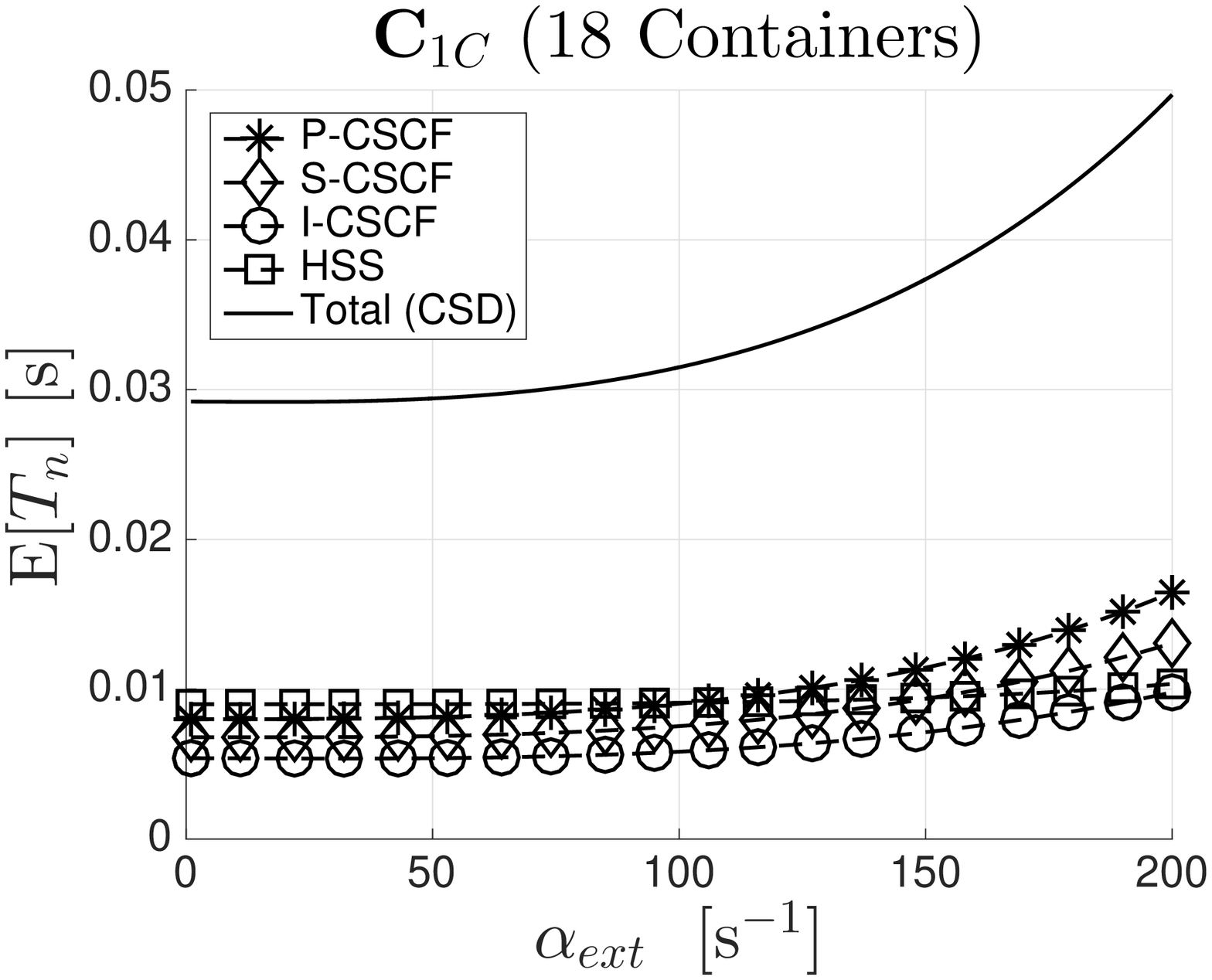}
	\end{minipage} \hspace{2pt}
	\begin{minipage}[t]{0.23\textwidth}
		\includegraphics[width=\textwidth]{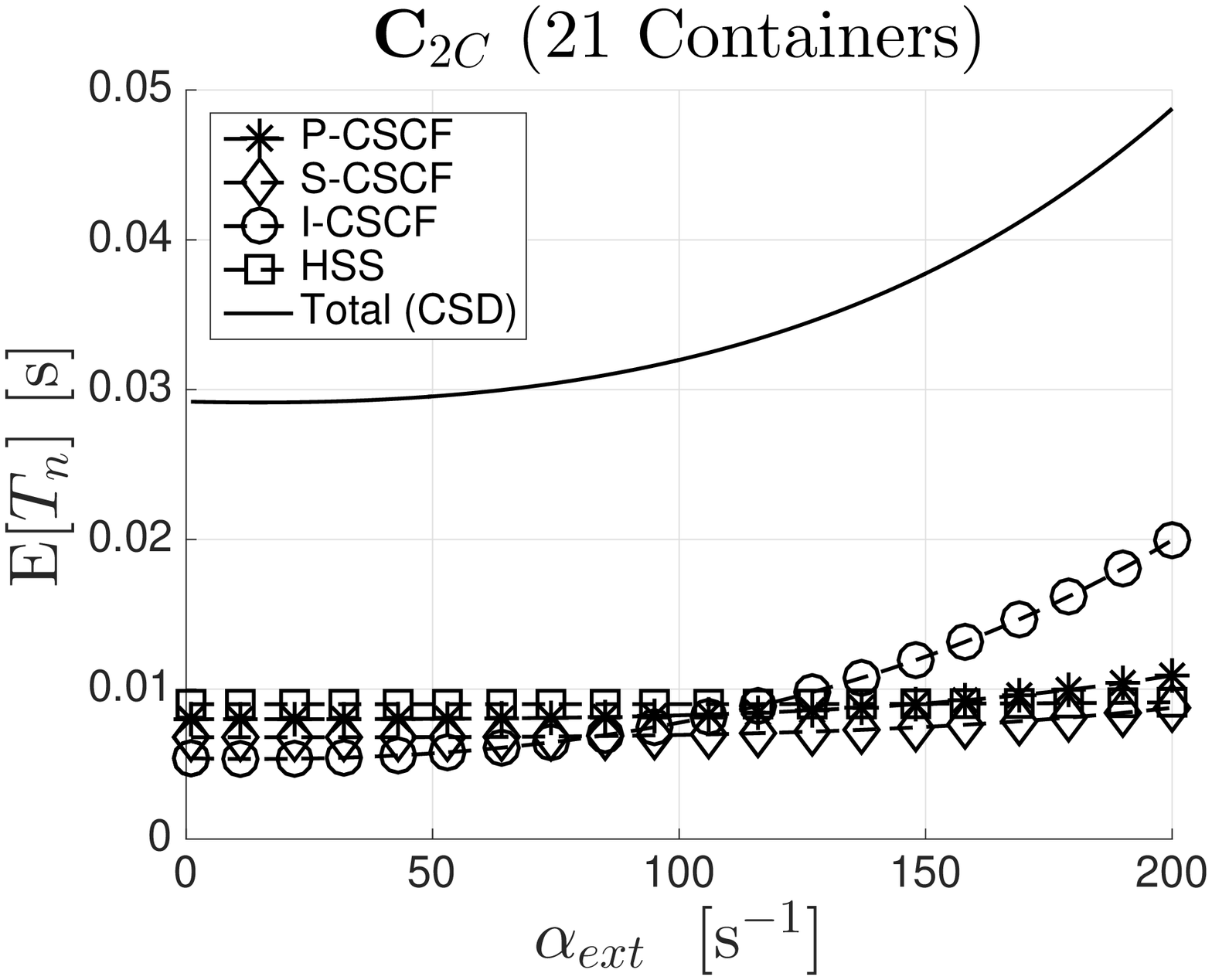}
	\end{minipage} \hspace{2pt}
	\begin{minipage}[t]{0.23\textwidth}
		\includegraphics[width=\textwidth]{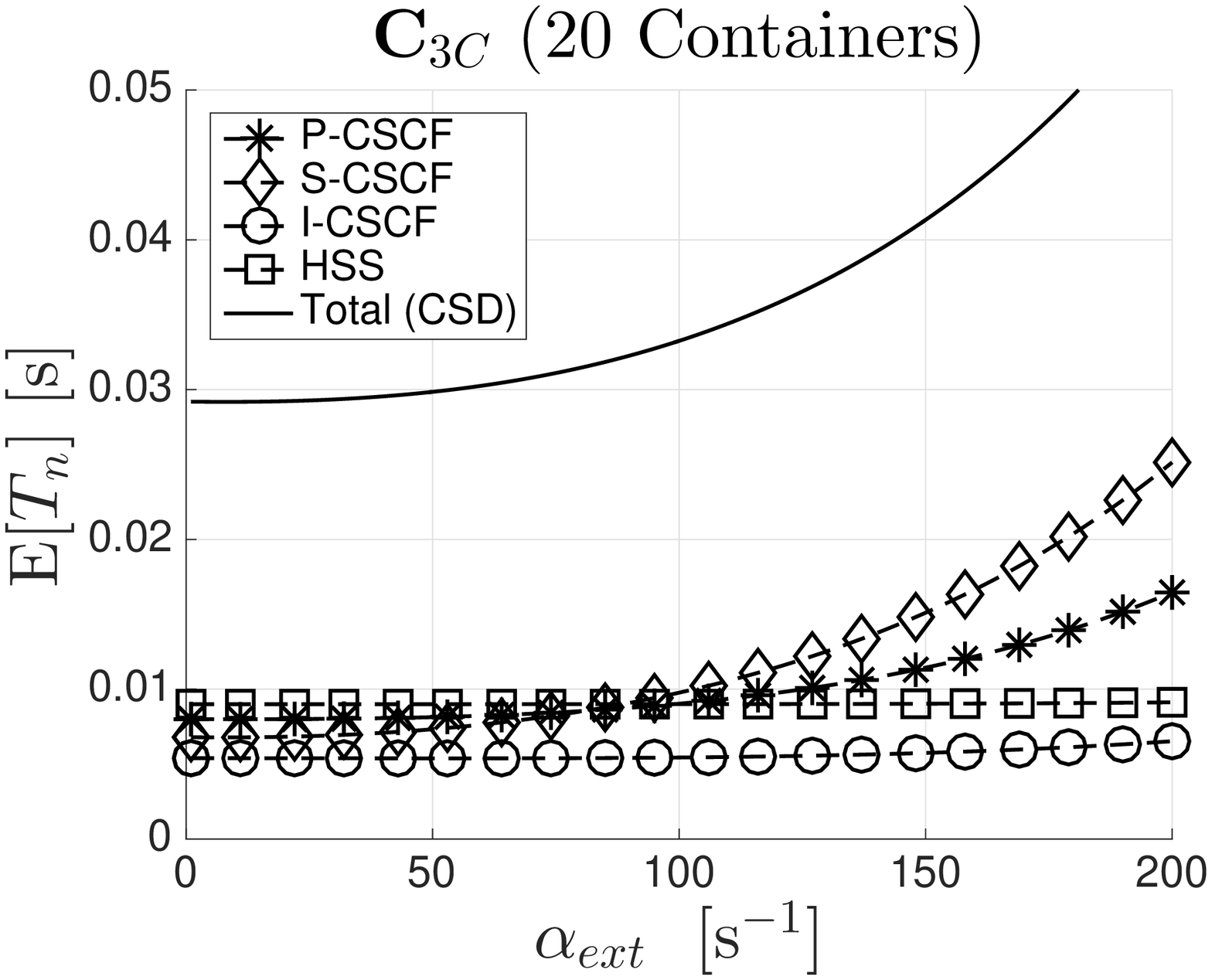}
	\end{minipage} \hspace{2pt}
	\LineSep
	\begin{minipage}[t]{0.23\textwidth}
		\includegraphics[width=\textwidth]{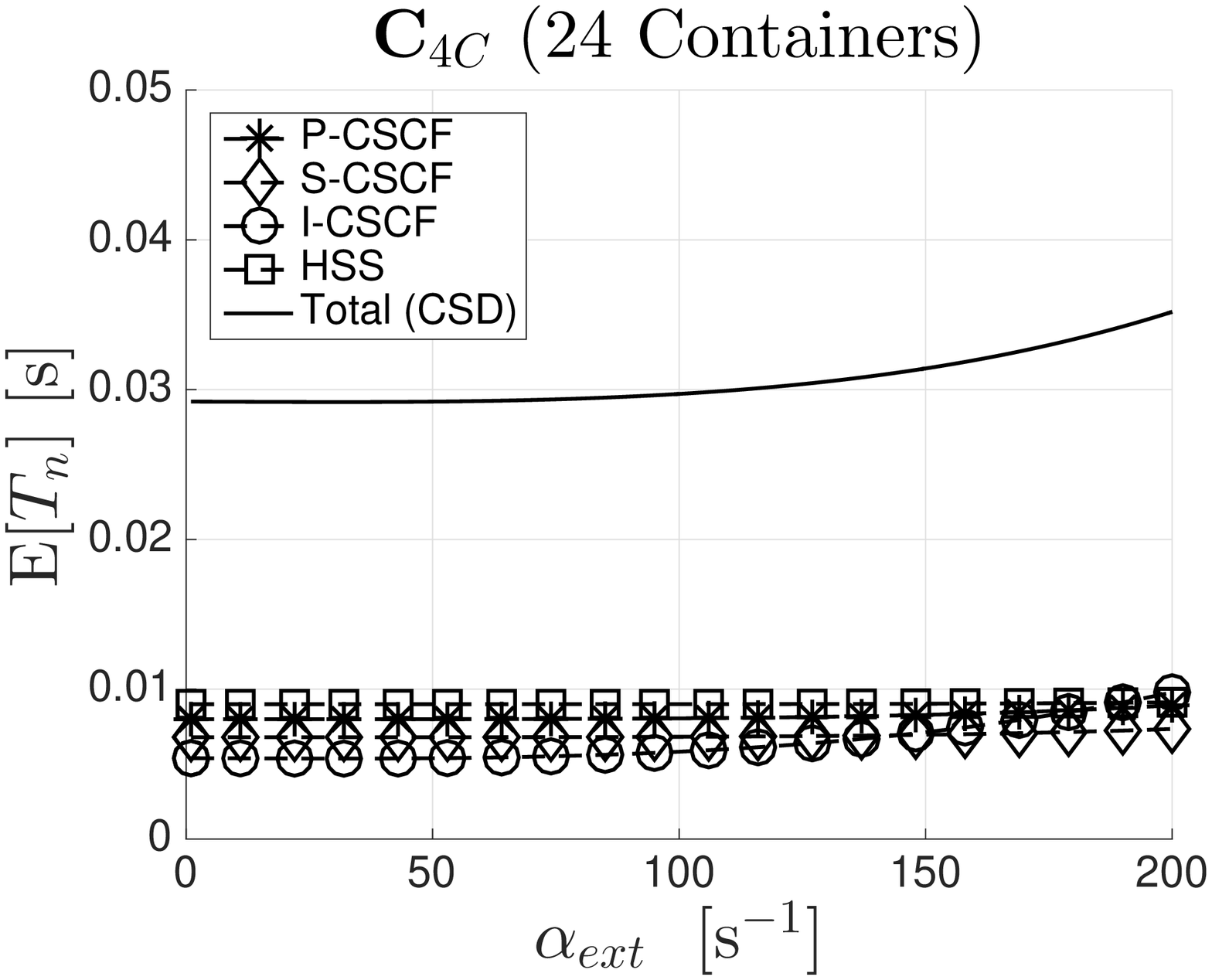}
	\end{minipage} \hspace{2pt}
	\begin{minipage}[t]{0.23\textwidth}
		\includegraphics[width=\textwidth]{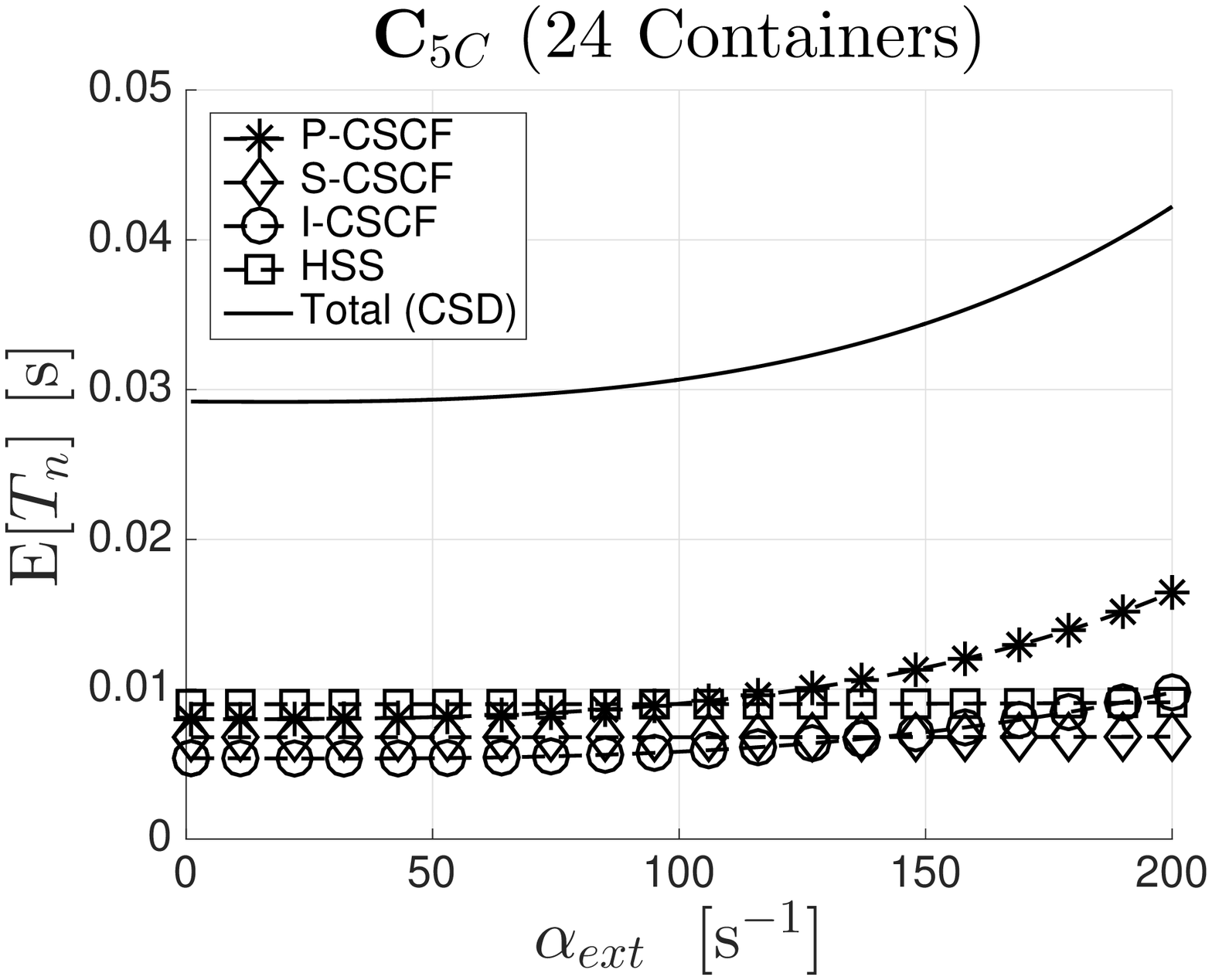}
	\end{minipage} \hspace{2pt}
	\begin{minipage}[t]{0.23\textwidth}
		\includegraphics[width=\textwidth]{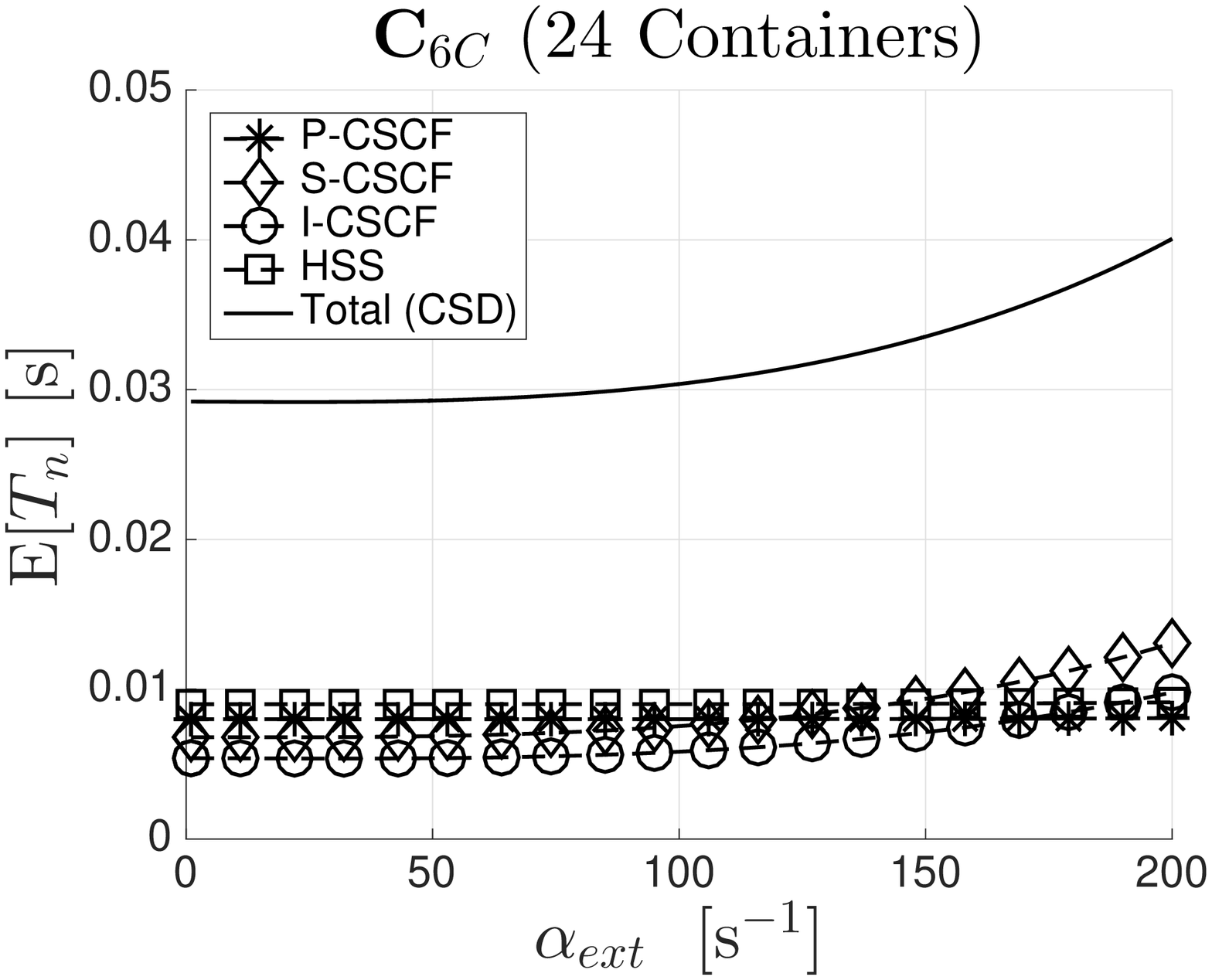}
	\end{minipage} \hspace{2pt}
	\begin{minipage}[t]{0.23\textwidth}
		\includegraphics[width=\textwidth]{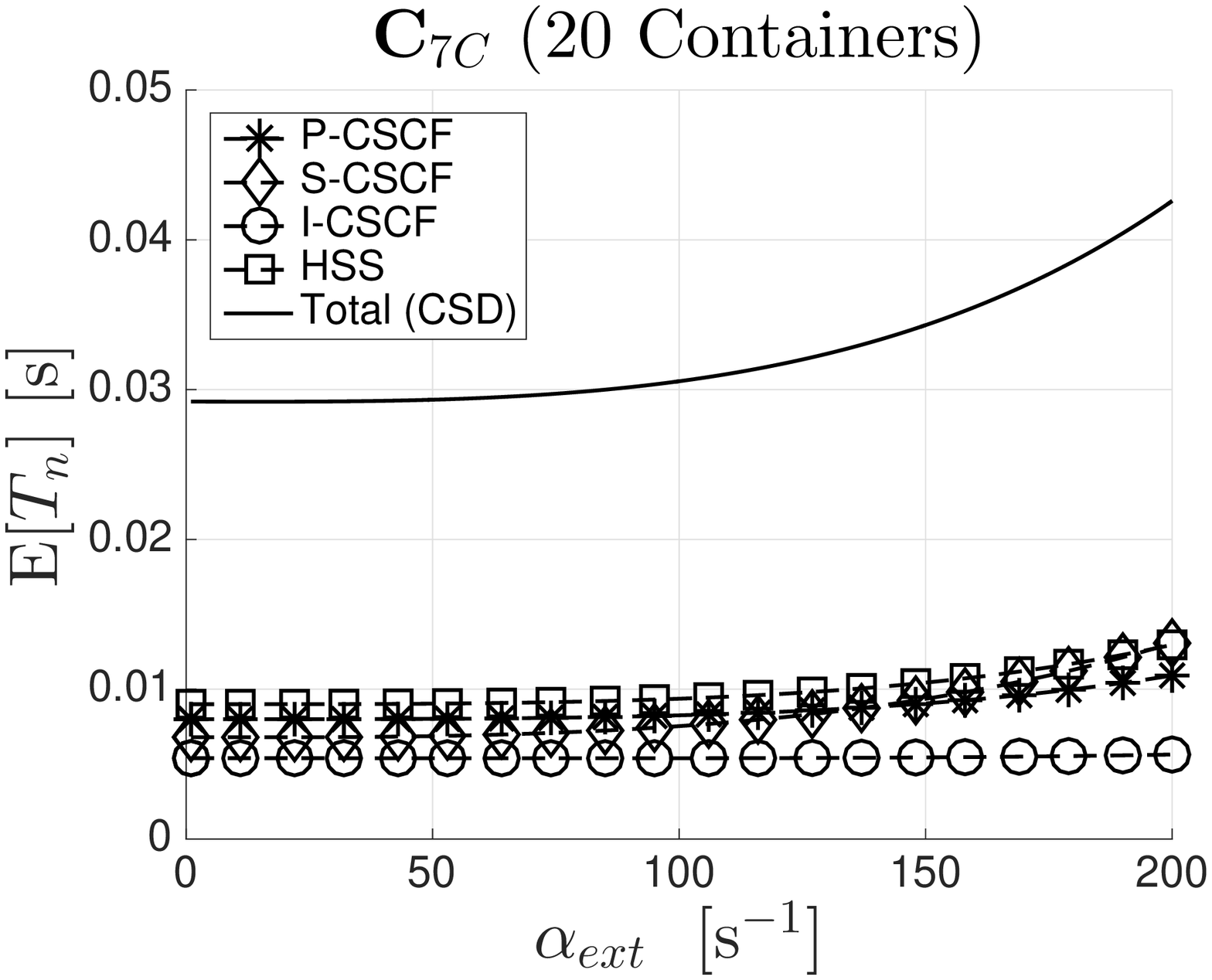}
	\end{minipage} \hspace{2pt}
	\caption{Co-located configurations: Mean response times per node $\mathbb{E}[T_n]$ for configurations from $C^*_C$ (top left panel) to $C_{7C}$ (bottom right panel).}
	\label{fig:all_t_coloc}
\end{figure*}

Now we present a specific analysis to reveal how a load variation (e.g. a load decrease) affects $\mathbb{E}[W_n]$ and $\mathbb{E}[T_n]$, and, then the whole CSD. Accordingly, the two panels of Figs. \ref{fig:all_w} and \ref{fig:all_t} report\footnote{Note that, for visualization comfort, we adopt a semi-log scale in the panel of Figs. \ref{fig:all_w} and a linear scale in the panel of Figs. \ref{fig:all_t}.}, respectively, the mean waiting time per node $\mathbb{E}[W_n]$ and the mean response time per node $\mathbb{E}[T_n]$ for all the homogeneous configurations presented in the Table \ref{tab:homotab}. 
The idea is to analyze the trend of such metrics before the external load reaches critical peaks ($\alpha_{ext}=200$ s$^{-1}$). 
Let us start by analyzing the panels of Fig. \ref{fig:all_w}. From the top left panel to the bottom right panel we report the configurations of Table \ref{tab:homotab} by preserving the ascending cost order (from $C^*_H$ to $C_{7H}$). We also report in parenthesis (near the configuration name) the total number of containers per softIMS configuration. At a first sight, we can observe, on average, a downward shift of curves by increasing the number of containers. See, for instance, the $C^*_H$ case ($9$ containers and curves lying between $10^{-2}$ and $10^{-1}$ in correspondence of the maximum external load) and the $C_{4H}$ case  ($23$ containers and curves lying between $10^{-4}$ and $10^{-2}$ in correspondence of the maximum external load). This is not particularly surprising since, as the number of containers grows, a softIMS configuration can simultaneously serve more sessions, thus the mean waiting times per node decrease. In contrast, it is interesting to notice how the allocation strategies play a crucial role to govern the behavior of the mean waiting times. See, for instance, the couple of configurations $C_{2H}$ and $C_{7H}$, where the same number of containers (amounting to $22$) does not imply the same behaviors of the curves (similar considerations hold for the couple of configurations $C_{4H}$ and $C_{6H}$ having the same number of containers amounting to $23$). This is basically due to the combined effect of different measured service times per node (see Table \ref{tab:params} for numerical values) coupled with a different distribution of containers per node. For example, curves are approximately equally spaced for $C_{7H}$ but not for $C_{2H}$, where the HSS curve is distant from the remaining curves. This stems from an ``aggressive'' container allocation policy adopted for HSS in $C_{2H}$ ($8$ containers, see Table \ref{tab:homotab}), which highly counterbalances its average service time being the lowest among the other nodes (see Table \ref{tab:params}).  
The trend of curves in Fig. \ref{fig:all_w} directly reflects into the panels of Fig. \ref{fig:all_t}, since $\mathbb{E}[W_n]$ and $\mathbb{E}[T_n]$ are related by virtue of (\ref{eq:tot_wait_time}).  In all the panels of Fig. \ref{fig:all_t}, we highlight the behavior of CSD (black continuous curve) resulting from the contribution of mean response time at single nodes as specified by (\ref{eq:csd}), where it is possible to observe a slower increase of the CSD curve as the  global number of containers grows. As regards single curves per node, the observed behavior directly results from the competing effect of the mean service times and the mean waiting times, here more accentuated by virtue of (\ref{eq:tot_wait_time}). For example, slower curves (e.g. $C_{2H}$, $C_{3H}$, $C_{7H}$) reveal a predominance of mean service time per node over the mean waiting times. In contrast, quickly growing curves (e.g. $C^*_H$, $C_{5H}$) indicate high mean waiting times, possibly related to an inefficient container allocation strategy.
Remarkably, when the orders of magnitude of mean waiting and service times are similar, it is possible to improve the overall CSD performance by: $i)$ increasing the number of containers compatibly with costs and with $c_{max}$ in the optimization problem (\ref{eq:optprob}) aimed at decreasing the mean waiting times, and/or $ii)$ increasing the computational resources (e.g. CPU, RAM, etc.) aimed at  accelerating the request processing which, in turn, means to decrease the service times. In contrast, when the orders of magnitude are considerably different, the only way to improve the CSD performance is to mitigate the effect of the slower variable between the mean waiting time and the mean service time.
	
A similar analysis can be performed by considering the co-located configurations presented in Table \ref{tab:colotab}. The results of such an analysis are reported in the panel of Figs. \ref{fig:all_w_coloc} and \ref{fig:all_t_coloc} for the mean waiting times and the mean response times, respectively, where the ascending cost order (from $C^*_C$ to $C_{7C}$) has been still preserved. For such co-located configurations, it is possible to derive similar considerations as already done for the homogeneous case presented in Figs. \ref{fig:all_w} and \ref{fig:all_t}. Also in this case we can notice that configurations with the same number of containers (see, e.g.,  $C_{4C}$, $C_{5C}$, $C_{6C}$ all hosting a total amount of $24$ containers) exhibit different behaviors in terms of mean waiting times and mean response times. As occurred for the homogeneous case, this is basically due to the different containers allocation policies across the nodes. For example, eight containers on the P-CSCF node for the configuration $C_{6C}$ result in a lower mean waiting time w.r.t. the configuration $C_{5C}$ where only four containers are hosted onto the P-CSCF node, in view of Eq. (\ref{eq:cosme}). Likewise, both configurations $C_{3C}$ and $C_{7C}$ host $20$ total containers but differently allocated on the various nodes, so that the mean waiting times and mean service times are differently distributed for the two configurations. In this case we also observe a jump of cost (see Table \ref{tab:colotab}) from $38$ (conf. $C_{3C}$) to $42$ (conf. $C_{7C}$) that directly comes from the fact that $C_{7C}$ uses more network replicas than $C_{3C}$ for the P-CSCF node ($4$ NRs \textit{vs.} $2$ NRs) and for the S-CSCF node ($4$ NRs \textit{vs.} $3$ NRs), respectively. Moreover, the high number of NRs coupled with a well balanced number of containers per NR makes the $C_{7C}$ configuration extremely robust, thus the steady-state availability amounts to six nines (see Table \ref{tab:colotab}).
	
In summary, the performability analysis can lead to these results: $i)$ independently from the configuration type (homogeneous/co-located), deploying more containers has a beneficial effect on the performance metric (CSD) since it allows to tackle its possible unexpected increase (e.g. due to a network congestion), at the price of a higher cost; $ii)$ the containers allocation strategies across NRs can affect the overall availability since some nodes may exhibit different levels of "robustness" (e.g. in terms of failures), thus a well-designed assignment of NRs/containers has to be planned per node; $iii)$ co-located deployments are more cheaper than homogeneous ones, since the shared infrastructure between I-CSCF and HSS allows to reduce the deployment costs. 
	
\section{Conclusions}
\label{sec:conclusions}

Network service chains represent the novel way of providing services by means of softwarized (e.g. virtualized and/or containerized) network nodes traversed according to a predefined path. A challenging use case is given by the IP Multimedia Subsystem, whose nodes are traversed in a chained fashion to provide multimedia services within $5$G networks. With reference to a softwarized IMS (softIMS) architecture (realized by means of a realistic testbed), we tackle two critical issues: $i)$ ensuring given performance levels (choosing the Call Setup Delay as performance metric) across the whole chain, and $ii)$ guaranteeing availability requirements (by means of smart redundancy strategies) to tackle the faults. The former issue is addressed through the queueing network formalism along with an optimal allocation problem whose numerical solution is guaranteed by an algorithm nicknamed OptCNT; the latter is treated through the Stochastic Reward Nets (SRN) methodology, where an algorithm dubbed OptSearchChain automates the solution of SRN schemes and seeks the optimal softIMS configurations (higher availability at minimum cost) through an exhaustive search with pruning. 
Moreover, we achieve a set of results useful to compare the performability of two popular containerized architectural deployments (homogeneous and co-located) in terms of: $i)$ the performance/costs trade-off associated to an increasing number of containers resulting into better performance (shorter Call Setup Delays) but at higher costs; $ii)$ the containers allocation strategy (namely, how many containers on which nodes) which has an impact both on the steady-state availability and on the mean waiting/response times.
	
The proposed novel assessment can be easily adapted to other chained-like structures (SDN-based networks, Traffic-Engineered Networks, virtualized mobile domains, etc.) provided that performance and availability metrics are obtainable. In particular, we foresee intriguing implications in the field of service chain provisioning automation. For instance, automatic procedures can be implemented in dedicated NFV management nodes (e.g. MANO - MANagement and Orchestration). Such procedures can be designed to dynamically add or remove softwarized network resources on the basis of performance requirements (e.g. a delay-based constraint across a virtualized mobile network) and/or availability demands (e.g. a critical Traffic-Engineered network which has to guarantee the challenging six nines availability requirement).  
Remarkably, the use of artificial intelligence (AI) in this field will open the doors to future $6$G scenarios which, according to the network scientists, will be characterized by smart resource management, intelligent network adjustments and automatic service provisioning.
\vspace{-3mm}
\section*{Acknowledgement}
The authors express their deep gratitude to Prof. Maurizio Longo for the stimulating discussions, and for the endless encouragement offered during the writing of this work.

\newpage

\begin{IEEEbiography}[{\includegraphics[width=1in,height=1.15in,clip,keepaspectratio]{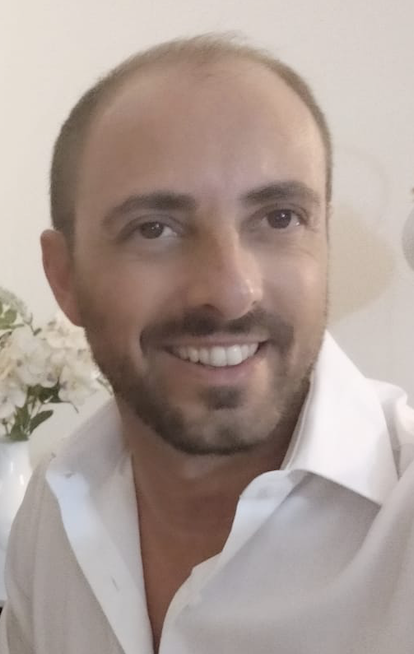}}] 
	{Mario Di Mauro} received the Laurea degree in electronic engineering from the University of Salerno (Italy) in 2005, the M.S. degree in networking from the University of L'Aquila (Italy) jointly with the Telecom Italia Centre in 2006, and the PhD. degree in information engineering in 2018 from University of Salerno.
	He is a Research Fellow with University of Salerno. His main fields of interest include: network availability, network security, data analysis for telecommunication infrastructures.
\end{IEEEbiography}

\vspace{-220 pt}
\begin{IEEEbiography}[{\includegraphics[width=1in,height=1.15in,clip,keepaspectratio,angle=270]{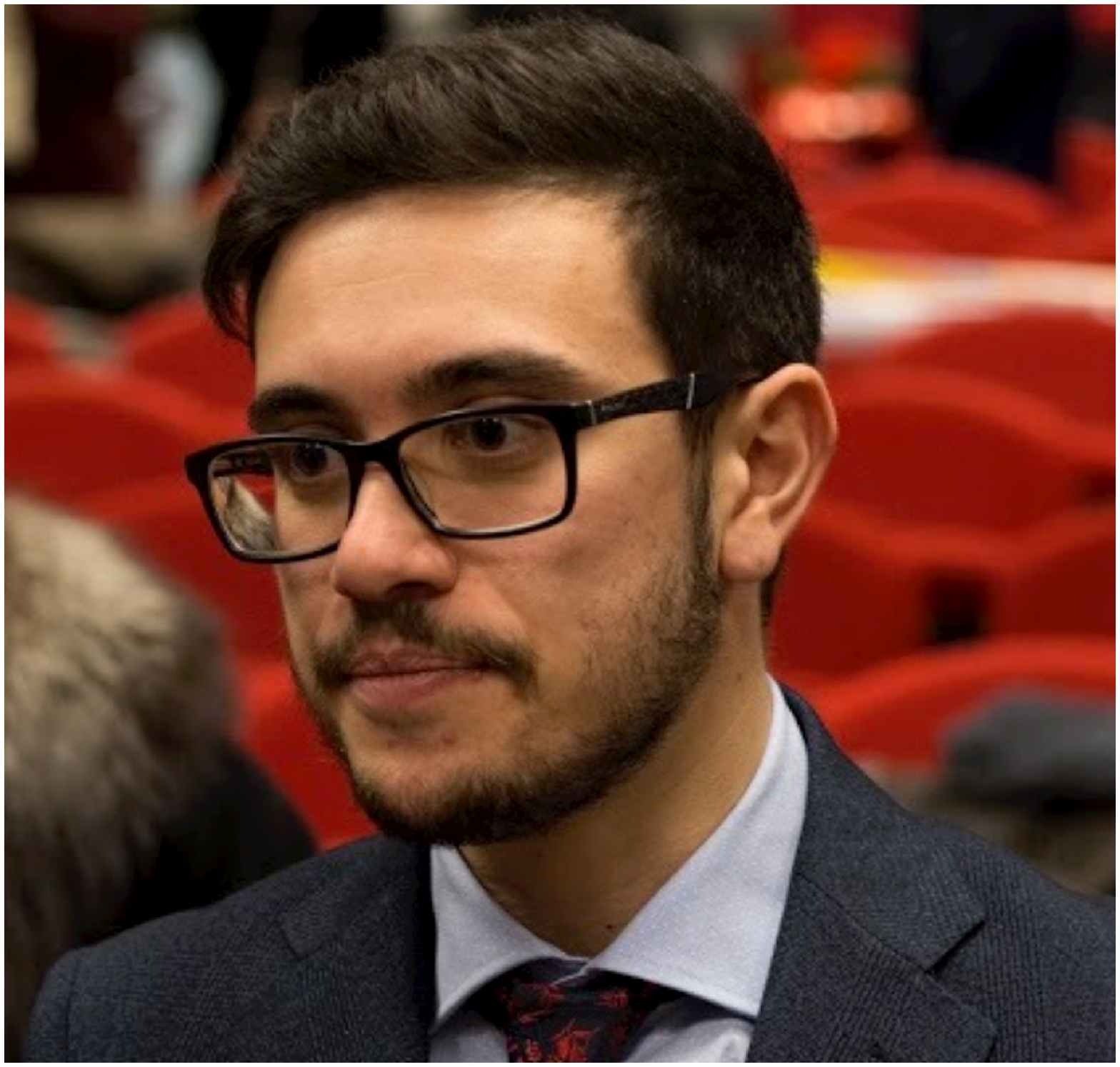}}] 
	{Giovanni Galatro} received the Laurea degree (summa cum laude) in information engineering from the University of Salerno (Italy) in 2018, and has been a visiting student at Dept. of Computer Science at Groningen University (Netherlands). In 2017 he got a scholarship with Telecommunication and Applied Statistics groups, focused on the availability analysis of modern telecommunication infrastructures. 
\end{IEEEbiography}

\vspace{-220 pt}
\begin{IEEEbiography}[{\includegraphics[width=1in,height=1.15in,clip,keepaspectratio]{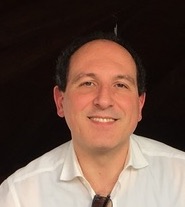}}]
	{Fabio Postiglione} is currently an Assistant Professor of Applied Statistics with the Dept. of Information and Electrical Engineering and Applied Mathematics (DIEM) at University of Salerno (Italy). He received his Laurea degree (summa cum laude) in Electronic Engineering and his Ph.D. degree in Information Engineering from University of Salerno in 1999 and 2005, respectively. His main research interests include degradation analysis, lifetime estimation, reliability and availability evaluation of complex systems (telecommunication networks, fuel cells), Bayesian statistics and data analysis. 
	He has co-authored over 100 papers, mainly published in international journals.
\end{IEEEbiography}

\vspace{-220 pt}
\begin{IEEEbiography}[{\includegraphics[width=1in,height=1.15in,clip,keepaspectratio]{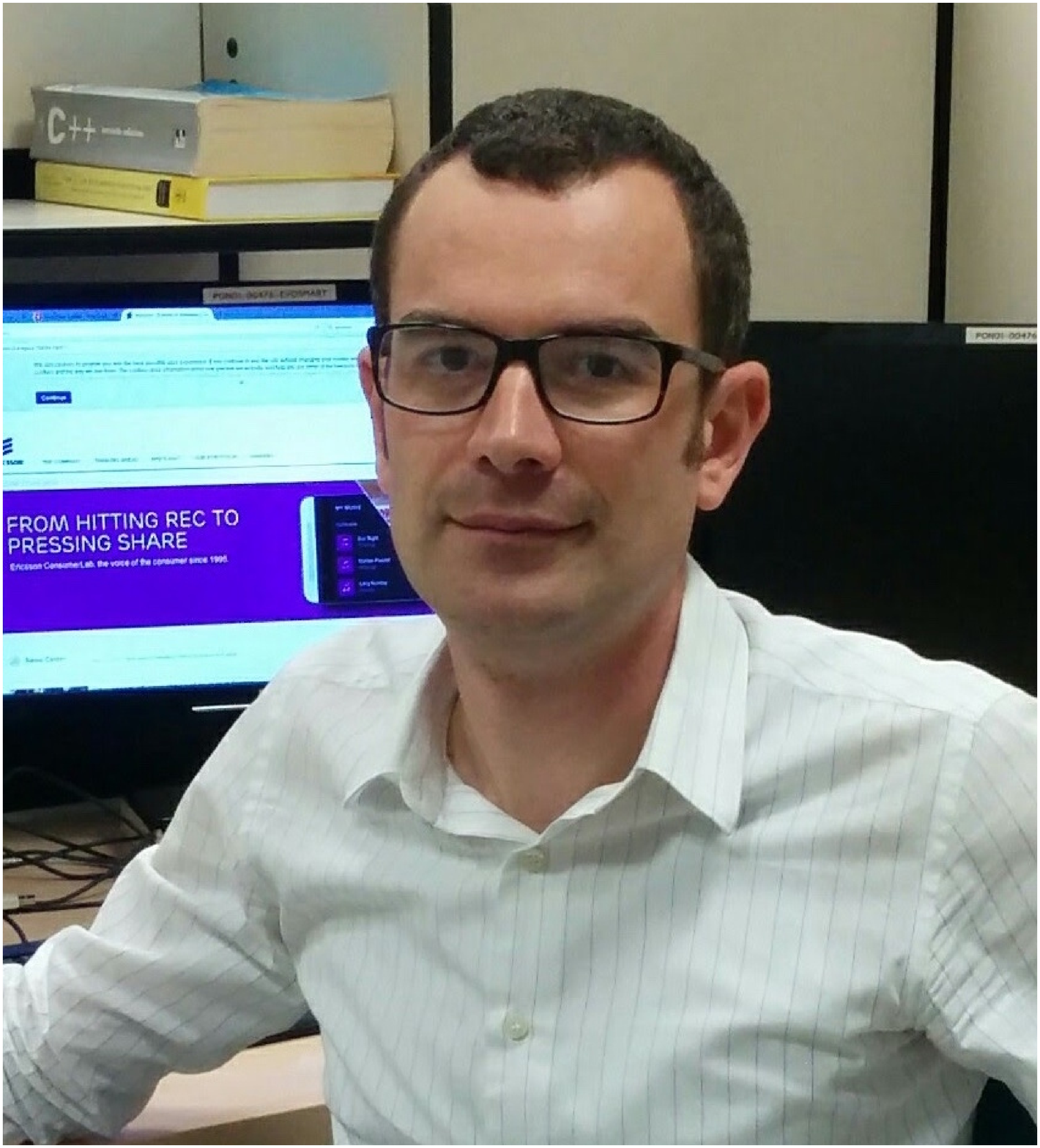}}] 
	{Marco Tambasco} received his Master's degree in Electronic Engineering from University of Salerno in 2010, and he is pursuing the PhD at the same University. He then joined CoRiTeL (Research Consortium on Telecommunications) and he is now an industrial researcher for Ericsson Telecommunication. Research interests include networks analysis and design, availability and security of cloud-based telecommunication systems (NFV/SDN).
\end{IEEEbiography}

\end{document}